\documentclass{aa}

\usepackage{xcolor}
\usepackage{amsmath}
\newcommand{\angstrom}{\textup{\AA}}
\usepackage{flexisym}
\usepackage{graphicx}
\usepackage{adjustbox}
\usepackage{booktabs}
\usepackage{tabularx}
\usepackage{multirow}
\usepackage[breaklinks]{hyperref}
\usepackage{float}

\usepackage{enumitem}

\newcommand\gaby[1]{\textcolor{blue}{#1}}

\newlength{\bulletwidth}
\newcommand{\mitem}{\setlength{\leftskip}{\leftmargin}\hspace*{-\labelsep}\hspace*{-4\bulletwidth}$\bullet$\hspace*{\labelsep}}
\newcommand{\mend}{\setlength{\leftskip}{0cm}}
\usepackage{txfonts}
\begin{document}

   \title{\textsc{AGNfitter-rx}: Modelling the radio-to-X-ray SEDs of AGNs}

   %\subtitle{I. Comparing torus and accretion disk models}
\author{L. N. Martínez-Ramírez\inst{1,2,3, 4} \and G. Calistro Rivera \inst{1}\thanks{ESO Fellow} \and Elisabeta Lusso \inst{5, 6} \and F. E. Bauer \inst{2,3} \and Emanuele Nardini \inst{6} \and  Johannes Buchner \inst{7} \and Michael J.I. Brown \inst{8} \and Juan C. B. Pineda \inst{9} \and Matthew J. Temple \inst{10, 11} \and Manda Banerji \inst{12} \and M. Stalevski \inst{13,14} \and Joseph F. Hennawi \inst{15, 16}
}% 

\institute{European Southern Observatory (ESO), Karl-Schwarzschild-Stra\ss e 2, 85748 Garching bei M\"unchen, Germany \\\email{ltmartinez@uc.cl} %gabriela.calistrorivera@eso.org
\and
Instituto de Astrofísica, Facultad de Física, Pontificia Universidad Católica de Chile Av. Vicuña Mackenna 4860, 7820436 Macul, Santiago, Chile
\and
Millennium Institute of Astrophysics (MAS), Nuncio Monseñor Sótero Sanz 100, Providencia, Santiago, Chile
\and
Max Planck Institut f\"ur Astronomie, K\"onigstuhl 17, D-69117, Heidelberg, Germany
\and 
Dipartamento di Fisica e Astronomia, Universitá di Firenze, Via G. Sansone 1, I-50019 Sesto Fiorentino, FI, Italy
\and 
INAF - Osservatorio Astrofisico di Arcetri, Largo E. Fermi 5, I-50125 Firenze, Italy
\and
Max Planck Institute for Extragalactic Physics, Gie\ss enbachstra\ss e, D-85741 Garching, Germany
\and
School of Physics \& Astronomy, Monash University, Clayton, VIC, Australia
\and 
Escuela de Física - Universidad Industrial de Santander, 680002 Bucaramanga, Colombia
\and 
Instituto de Estudios Astrofísicos, Facultad de Ingeniería y Ciencias, Universidad Diego Portales, Av. Ejército Libertador 441, Santiago 8320000, Chile
\and
Centre for Extragalactic Astronomy, Department of Physics, Durham University, South Road, Durham DH1 3LE, UK
\and
School of Physics and Astronomy, University of Southampton, Southampton SO17 1BJ, UK
\and
Astronomical Observatory, Volgina 7, 11060 Belgrade, Serbia
\and 
Sterrenkundig Observatorium, Universiteit Gent, Krijgslaan 281-S9, Gent B-9000, Belgium
\and
Physics Department, Broida Hall, University of California, Santa Barbara, CA 93106-9530, USA
\and
Leiden Observatory, Leiden University, PO Box 9513, NL-2300 RA Leiden, the Netherlands
}

  \abstract
  {
  We present new frontiers in the modelling of the spectral energy distributions (SED) of active galaxies by introducing the radio-to-X-ray fitting capabilities of the publicly available Bayesian code \textsc{AGNfitter}.  
  The new code release, called \textsc{AGNfitter-rx}, models the broad-band photometry covering the radio, infrared (IR), optical, ultraviolet (UV) and X-ray bands consistently, using a combination of theoretical and semi-empirical models of the AGN and host galaxy emission. This framework enables the detailed characterization of four physical components of the active nuclei: the accretion disk, the hot dusty torus, the relativistic jets/core radio emission, and the hot corona; alongside modeling three components within the host galaxy: stellar populations, cold dust, and the radio emission from the star-forming regions. Applying \textsc{AGNfitter-rx} to a diverse sample of 36 AGN SEDs at $z \lesssim 0.7$ from the AGN SED ATLAS, we investigate and compare the performance of state-of-the-art torus and accretion disk emission models on fit quality and inferred physical parameters. We find that clumpy torus models that include polar winds and semi-empirical accretion disk templates including emission line features significantly increase the fit quality in 67\% of the sources, by effectively reducing by $2\sigma$ fit residuals in the $1.5-5 \hspace{1mm}\mu \rm m$ and $0.7 \hspace{1mm}\mu \rm m$ regimes.
  We demonstrate that, by applying \textsc{AGNfitter-rx} on photometric data, we are able to estimate inclination and opening angles of the torus, consistent with spectroscopic classifications within the AGN unified model, as well as black hole mass estimates in agreement with virial estimates based on H$\alpha$.
  We investigate wavelength-dependent AGN fractions across the spectrum for Type 1 and Type 2 AGNs, finding dominant AGN fractions in radio, mid-infrared and X-ray bands, in agreement with empirical methods for AGN selection. The wavelength coverage and the flexibility for the inclusion of state-of-the-art theoretical models make \textsc{AGNfitter-rx} a unique tool for the further development of SED modelling for AGNs in present and future radio-to-X-ray galaxy surveys.
  
  }

   \keywords{galaxies: active, galaxies: nuclei, quasars: general, methods: statistical
               }

   \maketitle
%
%-------------------------------------------------------------------

\section{Introduction}

The accretion of matter onto the central supermassive black hole (SMBH) is the key process responsible for the luminous emission observed across the entire electromagnetic spectrum in active galactic nuclei (AGN) \citep[][]{blandford1990active, peterson1993reverberation, urry1995unified, netzer2015revisiting, padovani2017active}. 
This emission is multi-scale, spanning sub-parsec to kiloparsecs sizes, and often difficult to spatially distinguish from the host galaxy components. Thus, a composite spectral energy distribution (SED) is observed, which can be challenging when interpreting the physical properties of both the galaxy and its AGN to understand more complex processes such as black hole growth and the co-evolution of the galaxies and AGNs. 
The emission of AGNs spans the electromagnetic spectrum from radio to X-rays or even up to $\gamma$-rays in the case of blazars \citep[e.g., ][]{ackermann2011radio}. The optically thick gaseous accretion disk around the SMBH produces a multi-colour black body spectrum from optical to UV energies.
Additionally, it includes a soft X-ray excess and a hard X-ray tail produced by Compton up-scattering of disk photons in the hot optically thin corona and its reprocess flux \citep[e.g., ][]{kubota2018physical}. 
Obscuring dust surrounding the disk in the shape of a torus is responsible for absorbing optical/UV photons to re-radiate them into the IR regime with a predominant bump in the MIR. In Compton-thick systems ($\log \rm N_{\rm H} \geq 24$, this high column density dust also down-scatters X-rays due to Compton recoil, and produces a strong iron $\rm K_{\alpha}$ line and a reflection spectrum due to the inner wall of the torus \citep{matt2003changing, comastri2004compton}. Finally, the large-scale and powerful jets with bright shock ends release radio emissions originating in the interactions of relativistic electrons or positrons with the magnetic fields present in the jets.

Distinct physical processes taking place in AGNs leave their unique imprints and dominate in different energy regimes within the SED. The most widely used method to unravel the physical information contained in the multiwavelength emission of AGNs is SED-fitting with a combination of emission models for the dominant photometric components of active galaxies \citep[e.g., ][]{berta2013panchromatic, ciesla2015constraining, calistrorivera16, leja2018hot, Yang2020, thorne22}. 
As deeper photometric surveys of increasing spectral and spatial coverage and resolution, as well as more advanced theoretical models emerge, many of the parameterized emission models need to be updated and validated. 
An example is the smooth dusty torus models which are not able to reproduce the observed silicate emission line in some Seyfert 2 and need over dust-sublimation temperatures to be consistent with observations \citep{tanimoto2019xclumpy}. Also, extinction produced by dust along the polar direction, on scales up to hundreds of parsecs \citep{asmus2016subarcsecond, honig2019redefining, stalevski2019dissecting, calistro2021multiwavelength}, not accounted for in the traditional torus models. 
Similarly, simple theoretical accretion discs models do not consider observed high-velocity winds \citep{nardini2015black,tombesi2015wind}, predict UV power law index steeper than observed \citep{davis2007uv} and are still not conclusive about the mechanisms driving soft X-rays \citep{kubota2018physical}.

The depth of information extracted from data in the SED-fitting approach, however, is limited by the dimensionality of the parameter space, a product of both the number of photometric bands and model complexity. A complex parameter space involves dealing with the existence of degeneracies and correlations between parameters \citep{calistrorivera16}, which can hinder physical interpretations when not considered. 
Even when accounted for through Bayesian methods, degenerate models in the IR-to-UV can limit the accuracy of the information obtained through the SED fitting significantly.

To overcome some of these degeneracies, we can take advantage of the information in the radio and X-ray bands. Although these regimes have not been commonly included in galaxy SED fitting codes, their relevance as tracers of AGN activity makes them crucial to further inform the multiwavelength modelling of AGN SEDs. 
On, the one hand, radio emission is a powerful tracer of both star formation and AGN activity, as it is unaffected by dust. Moreover, the radio emission originating in cosmic-ray electrons from star formation in the galaxy is correlated to the galactic IR luminosity \citep{helou1985thermal, de1985spiral}, offering an avenue to disentangle the origin of the radio emission.
Thus, the remaining radio spectra associated with the AGN emission can give us insights into the feedback mechanisms and the outflow geometry and structure \citep{panessa19, silpa2022quasar,  CR24}. On the other hand, the high-energy X-ray emission exhibits great penetrating power and is thus almost unaffected by obscuration. Unlike radio fluxes, the galaxy's contribution in X-rays due to binary stars and star formation in highly star-forming galaxies is very low compared to that of the AGN.

In this paper, we present an extended version of the publicly available \textsc{AGNfitter} code \citep{calistrorivera16} released as \textsc{AGNfitter-rx}. \textsc{AGNfitter-rx} is a Python code designed to fit the radio-to-X-ray SED of AGNs and their host galaxies. 
One of the key features of this release is the introduction of new libraries of theoretical and empirical emission models and improved flexibility and user-friendly customizability, which opens up endless possibilities to tackle the SED-fitting task. \textsc{AGNfitter-rx} thus works as an interface between the observational and theoretical modelling community, as its flexibility allows the easy addition of new templates and comparisons between existing and new models.
While AGN components have been included recently in SED-fitting codes \citep[eg.,][]{leja2018hot, Yang2020, thorne22}, most of these have a smaller wavelength coverage and are all focused on inferring the potential impact of the presence of an AGN on the galaxy parameters, AGN fractions or AGN identification. 
In addition to these tasks, \textsc{AGNfitter-rx} is further tailored as a tool to characterize the AGN physics, and to robustly infer the physical parameters associated with the multiwavelength emission in the AGN.
Besides, despite the large diversity of existing theoretical models for both the torus and the accretion disk, only a few studies have focused on comparing the performance of different models in SED fitting \citep{garcia2022torus, gonzalez2019exploring, esparza2021dust,cerqueira2023coronal}.

This paper is divided as follows.
In Section \ref{sec:agnfitter1.0} we introduce the new functions which allow for the flexible inclusion of new physical models and new instrument filters.
In section \ref{sec:models} we introduce the new library of physical models which are already included in the new version of the code.
In section \ref{sec:brownAGN} we demonstrate the use of \textsc{AGNfitter-rx} by applying the algorithm on a sample of nearby AGNs, to compare the capabilities of different state-of-the-art torus and accretion disk models in reproducing the photometric data.

We adopt a concordance flat $\Lambda$-cosmology with $\rm H_{0}= 67.4 \text{ km s}^{-1} \text{Mpc}^{-1}$ , $\Omega_{m} = 0.315$, and $\Omega_\Lambda = 0.685$ \citep{aghanim2020planck}.

\section{The \textsc{AGNfitter-rx} release}\label{sec:agnfitter1.0}

\textsc{AGNfitter-rx} is built upon the code \textsc{AGNfitter} \citep{calistrorivera16} which models the spectral energy distribution of galaxies and AGNs. The first version of the code models the emission from the sub-mm to the UV ($11 <\log \frac{\nu}{\rm Hz}<16$) with 4 physical components: the accretion disk, the hot circumnuclear dust, the stellar populations and the cold dust emission in star-forming regions \citep[see detailed description in][]{calistrorivera16}.
\textsc{AGNfitter-rx} expands this coverage to lower and higher frequencies ($8 <\log \frac{\nu}{\rm Hz}<20$), by introducing two additional physical components to model several orders of magnitude in frequency in the radio regime, as well as the X-ray emission.

Radio models, as well as the torus, accretion disk, stellar populations and cold dust models, consider different levels of contribution, making it possible to model diverse populations spanning from radio-quiet to radio-loud AGNs. In particular, studying radio-quiet AGNs is interesting as it has recently been reported by \cite{kang2022x} that this population seems to have a higher high-energy spectral break compared to radio-loud AGNs. The high-energy cutoff suggests a coronae with high temperature and small opacity in AGNs with low radio emission. Considering also that X-rays are now part of the code, it opens the possibility to study this possible correlation between radio and X-ray emission now.

As \textsc{AGNfitter}, \textsc{AGNfitter-rx} explores the parameter space using a Bayesian Markov Chain Monte Carlo method, based on the \textsc{emcee} code \citep{foreman2013emcee}. 
Additionally, it now includes an alternative Bayesian methodology implementing the nested sampling Monte Carlo algorithm \textsc{ultranest} \citep{buchner2014statistical, buchner2019collaborative}. Through the random sampling of the parameter space, the code recovers posterior probability density functions (PDFs) of the physical parameters driving the multiwavelength emission. 
Achieving high computational efficiency is particularly crucial, as the size of the parameter space is directly predicated upon the user's choice of models. Moreover, the Bayesian approach allows for the introduction of prior knowledge on parameter distributions (see Section \ref{subsec:priors}) which is useful e.g. to consistently enforce the energy balance among otherwise independent components such as the optical-UV attenuated stellar emission and cold dust emission \citep{da2008simple}.
New significant developments in the structure of the code have increased the flexibility and customization capability of the model library, filter library and priors, as described below.  

 \begin{figure*}[h!]
    \centering
    \includegraphics[width = 0.76\linewidth]{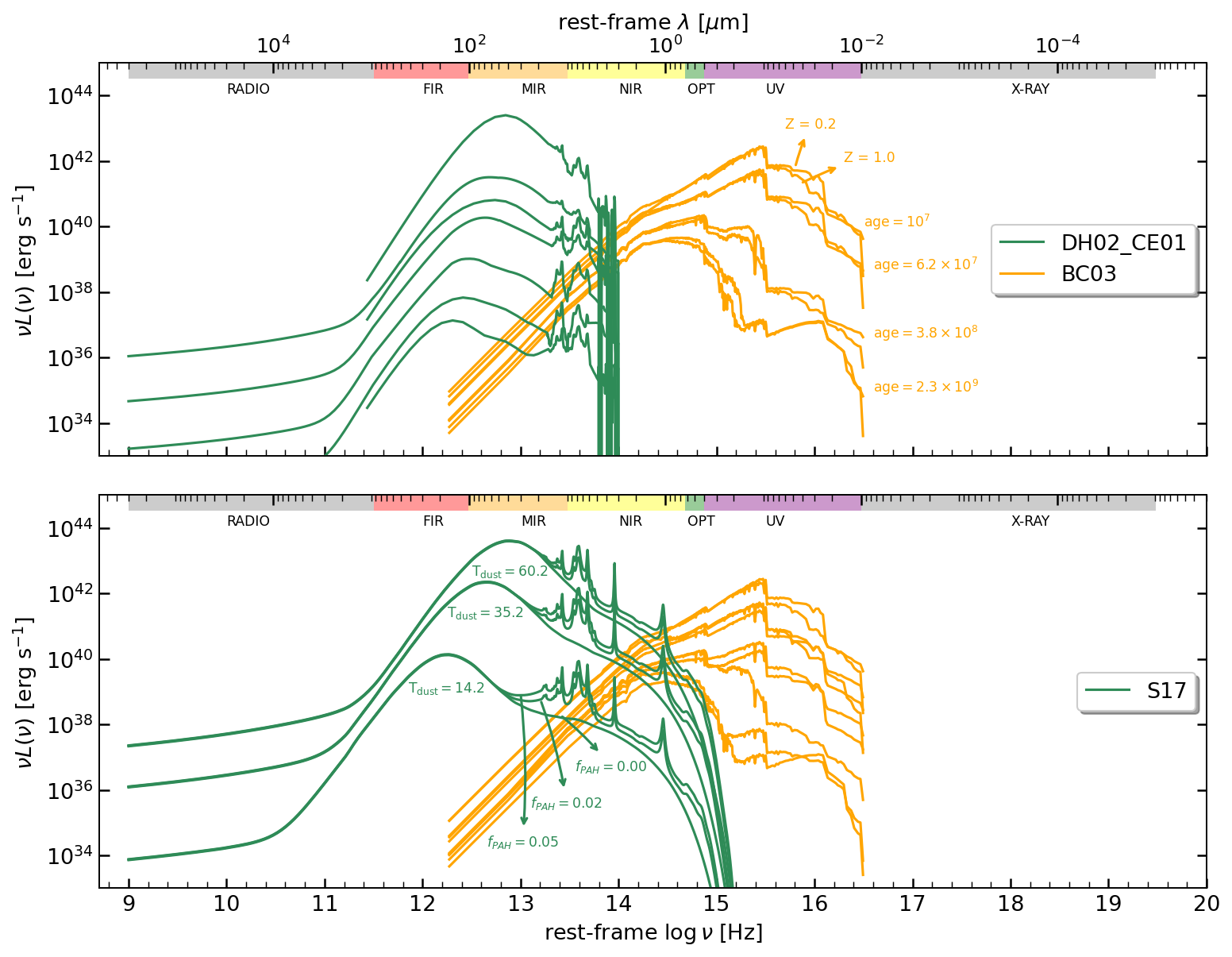}
    \caption{Examples of the model templates included in \textsc{AGNfitter-rx} for the physical components responsible for the emission of the host galaxy and its star-forming regions. At radio frequencies the host galaxy SED is dominated by the synchrotron emission associated with the diffusion of cosmic rays from supernova remnant (SNR) and pulsar wind nebula (PWN) acceleration sites through the galaxy ISM and the infrared emission is dominated by the cold dust emission (green lines) both associated with regions of high star-formation. The NIR-optical-UV emission is dominated by the stellar emission (yellow lines). The upper panel shows a subsample of semi-empirical templates by \cite{chary2001interpreting} and \cite{dale2002infrared} for the emission from star-forming regions and the stellar population synthesis models by \cite{bruzual2003stellar} for SFH of $0.05\times 10^{9}$ yr, and different ages and metallicities. The lower panel shows theoretical models for cold dust by \cite{schreiber2018dust} for increasing dust temperatures and $f_{\mathrm{PAH}}$ smoothly joined to the radio emission estimated with the FIR-radio correlation (eq. \ref{eq: IR-rad correlation}).}

    %synchrotron emission associated with the diffusion of cosmic rays from supernova remnant (SNR) and pulsar wind nebula (PWN) acceleration sites  through the galaxy ISM. (which generally coincide with star-forming regions) typically one applies a kernel smoothing to IR maps to model the radio. Typical diffusion scales are ~1kpc I think. This statement appears in a few places.
    \label{fig:SED_SB+GALmodels}
\end{figure*}

\subsection{Flexibility for adding new filters and models} 
\textsc{AGNfitter-rx} includes a compilation of 182 published filter curves from the most widely used telescopes.
The user can add new filters by providing the corresponding telescope responses.
However, for most X-ray CCD cameras and interferometric radio and sub-mm data, photometric filters are not provided by the observatories, given the complex telescope response curves. In these cases, when photometric filters are not available, a boxcar or Gaussian-like functions can be defined and included.

\textsc{AGNfitter-rx} is designed to allow the user to add customary models for each of the physical components.
The models must be entered as a Python dictionary containing a grid of templates.
Each template corresponds to a spectral flux density as a function of the rest-frame frequency and should be defined by a unique set of parameter values. 
The combination of all loaded model parameters ultimately defines the parameter space of the total model.
This development allows us to include and compare simultaneously several competing physical models in this study.

\subsection{Flexibility for including priors}\label{subsec:priors}

One of the main advantages of \textsc{AGNfitter-rx} is the flexibility for the users to include pre-defined or customary informative priors, based on ancillary information of the source. Optional priors which are pre-defined in the code include:

\vspace{2mm}

\mitem Lower and upper limits to the fractional galaxy contribution in the UV/optical can be estimated based on predictions from redshift-dependent galaxy luminosity functions.

\mitem Energy balance between the dust attenuation to the stellar emission in the optical/UV and the reprocessed emission by cold dust in the infrared (MIR and FIR). Three optional energy balance priors are available, allowing for a flexible set-up in which optical-UV dust-absorbed emission sets a lower limit to the infrared emission, a more restrictive condition with cold dust to attenuated stellar luminosities ratio given by a  broad Gaussian centred at 1, and no energy balance at all. 
In contrast to other codes where energy balance is a fixed property, the flexibility to include or drop energy balance is particularly relevant for sources where energy balance potentially breaks down.
These sources include dusty star-forming galaxies at intermediate and high redshifts where cold dust emission and rest-frame UV emission have been observed to be often spatially disconnected \citep{CR18, buat19}.  %Franz comment: Two comments here: 1) a fourth option might be interesting, wherein you apply a fraction obscuration. This might more explicitly address the example that Gaby references (it cannot fully address it, since the SF in and out of the dust could have unique SFHs, and hence distinct stellar continua). 2) energy balance should also apply to the AGN? And in the same sense as comment 1, there are obscured AGN which have very high scattered blue light fractions (see recent papers by Roberto Assef), so it might make sense to have fractional obscuration by the torus (or an additional scattered component tied to the main UV disk light)

\mitem New priors on the X-ray emission have also been included, which are described in detail in Section \ref{subsec:accretiondisk}.

\mend
 
\section{Physical models from the radio to the X-rays} \label{sec:models}

In this section, we describe the different physical components to model AGN and galaxy SEDs in the radio-to-X-ray regime, which in aggregate constitute the total physical model fit by \textsc{AGNfitter-rx}. 
For each physical component, several state-of-the-art models are included in the code for the user to implement,  fit and compare. 
We include a large range of theoretical, semi-empirical and empirical models, and note that the user can also easily include self-customised models.
\textsc{AGNfitter-rx} can therefore be used as a tool for model testing, as well as to compare the suitability of different sets of models for a given source or galaxy sample.

\begin{figure*}%[h!]
\begin{minipage}{\linewidth}
    \centering
    \includegraphics[width = 0.76\linewidth]{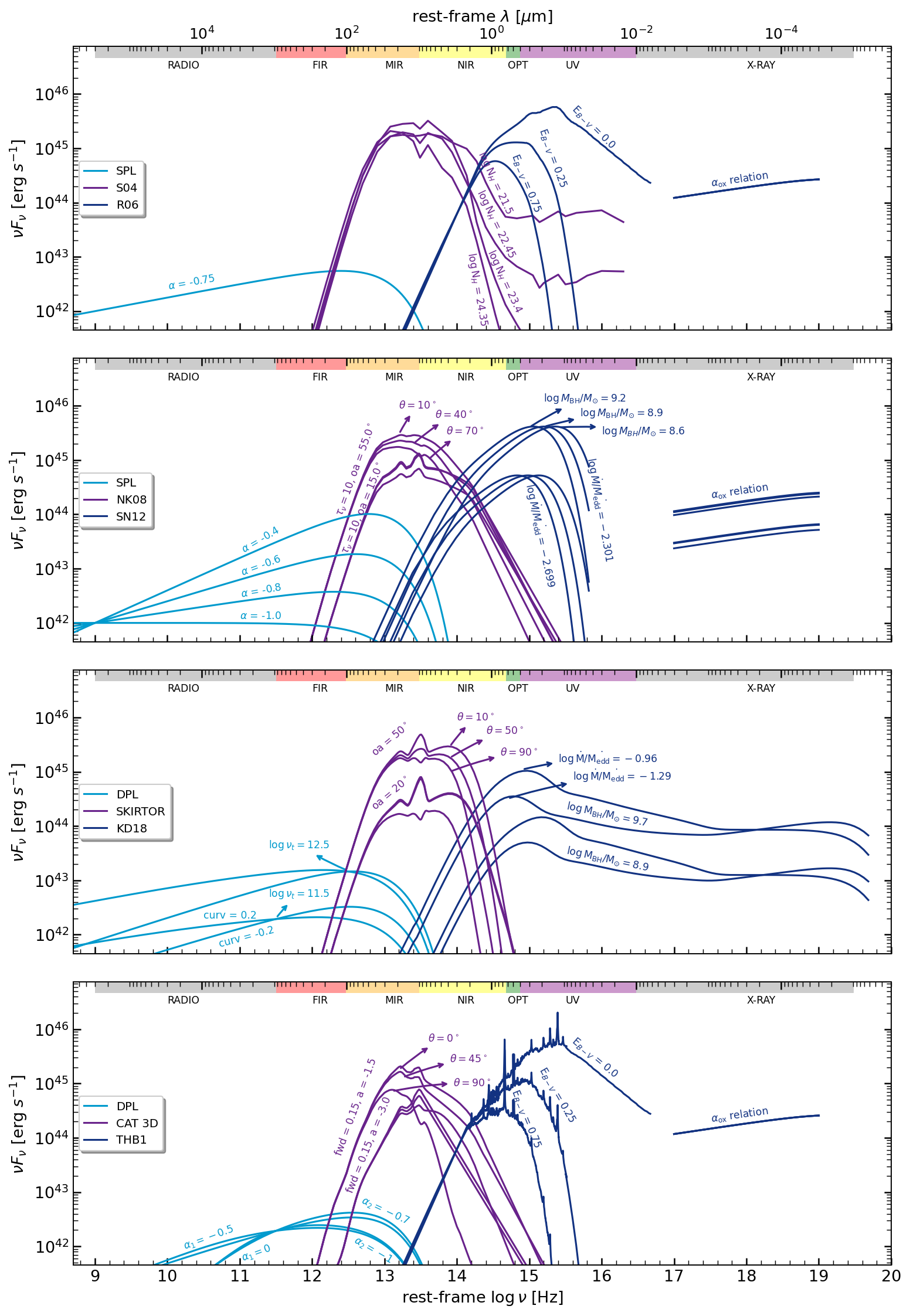}
    \caption{Examples of the model templates included in \textsc{AGNfitter-rx} for three distinct physical components of the nuclear emission (AGN) in active galaxies. At radio frequencies, the nuclear SED is dominated by the synchrotron emission (light blue curves) associated with jets and core emission at $\sim 10^{-1}- 10^{6}$ parsec scales. The mid-infrared emission is dominated by the hot dusty `torus' emission (purple curves) arising at $\sim 10$ parsec scales. The optical-UV-X-ray emission is dominated by the accretion disk and X-ray corona emission (dark blue curves) at $\sim 10^{-2}$ parsec scales. The first panel shows a simple cutoff power law (SPL) in radio, a subsample of torus templates by \cite{Silva2004} with different column densities,  
    and the accretion disk model by \cite{richards2006sloan} with an increasing reddening parameter. The second panel presents SPL emission with different slopes, torus models by \cite{nenkova2008agnII} for different inclination angles and opening angles, and optical depth of $10$, and accretion disk models by \cite{slone2012effects} for diverse black holes masses and accretion rates. The third panel shows a double power law (DPL) with $curv$ and $\log \nu_t$ as free parameters; torus templates by \cite{Stalevski2016} with optical depth of $3$ and diverse inclination and opening angles; and accretion disk models by \cite{kubota2018physical} defined by black holes masses and accretion rates. The fourth panel shows DPL with $\log \nu_t = 11.5$ Hz and different values of $\alpha_1$ and $\alpha_2$; torus model by \cite{honig2017dusty} depending on inclination angles, index of cloud distribution and cloud fractions; and the accretion disk template by \cite{temple2021modelling}.}

 % Comment from Franz: in all cases but KD18, the X-ray emission appears fixed despite strong changes in MBH and L/LEDD. this cannot be correct. It should follow alpha_ox vs LUV relation, plus account for any obscuration (adopting some standard NH = Av * 2.2e21
    \label{fig:AGNtemplates}
\end{minipage}
\end{figure*}

\subsection{Radio - Synchrotron SF and AGN models}\label{subsec:radio}

The two sources of radio emission considered in the model are the synchrotron emission originating in star-forming regions of the host galaxy and the synchrotron emission from the AGN. 

The radio emission from star-forming regions in the host galaxy is the product of the interaction of hot plasma electrons from supernova remnants and cosmic rays with the magnetic fields of the galaxy. 
This interaction produces predominantly non-thermal synchrotron emission across the radio regime, as well as thermal Bremsstrahlung radiation, which becomes predominant at frequencies above 30 GHz \citep{condon92}. 

To account for the radio emission from the host galaxy we include the semi-empirical starburst templates by \cite{dale2002infrared} which are already extended up to radio by using the empirically calibrated IR-radio correlation \citep[e.g.,][]{helou1985thermal, de1985spiral} with $\rm q_{\rm IR} = 2.34$, $\sigma = 0.26$ dex and assuming $90 \%$ of the $\rm \mathit{L}_{\rm 1.4 GHz}$ is given by the synchrotron emission of the cosmic ray electrons in the galaxy and the remaining $10 \%$ by thermal bremsstrahlung radiation produced in the star-forming regions. Both components were modelled by power law functions $L_{\nu} \propto \nu^{-\alpha}$ with $\alpha = 0.1$ for the thermal component and $\alpha = 0.8$ for the non-thermal one. Note that of the DH02\_CE01 model only the \cite{dale2002infrared} templates extend to radio while the \cite{chary2001interpreting} templates do not (see top panel of Fig. \ref{fig:SED_SB+GALmodels}).

Furthermore, we apply the total IR version of the same IR-radio correlation \citep[e.g.,][]{bell2003estimating} to the recently added S17 cold dust templates (see \ref{sub:cold_dust_models}). This relation establishes a link between the total rest-frame IR luminosity from the cold dust in star-forming regions and the emission at 1.4 GHz ($L_{1.4 \rm GHz}$ in erg s$^{-1}$ Hz$^{-1}$) :

\begin{equation}
\hspace{15mm}
    \rm q_{\rm IR} = \log\left( \frac{\mathit{L}_{\rm IR}}{(3.75 \times 10^{12} \rm Hz) \mathit{L}_{\rm 1.4 GHz}}\right),
\label{eq: IR-rad correlation}
\end{equation}

\noindent where $\mathit{L}_{\rm IR}$ is the integrated luminosity between $8$ and $1000 \hspace{1mm}\mu$m (in erg s$^{-1}$). We adopted the value of $\rm q_{IR} = 2.64 \pm  0.26$ from \cite{bell2003estimating} and undertook a conservative assumption of $\text{q}_{\text{IR}} = 2.64 + \sigma $ to include sources with faint radio fluxes. The current implementation is in the range of values found by \cite{molnar2021non} for $\rm \mathit{L}_{\rm IR}$ lower than $10^{7}$ [L$_\odot$]. The nonlinearity of the infrared-radio correlation, due to its dependence on redshift \citep{smith2014temperature,calistro2017lofar, molnar2021non} and stellar mass \citep{delvecchio2021infrared}, is being currently explored and there is still no consensus on the behaviour of $q_{\rm IR}$.

% Molnar et al. 2021 found q = 2.54 \pm 0.01, 0.1 dex lower than Bell 2003
%we may want to consider a more modern (slightly non-linear) assessment (?) => https://ui.adsabs.harvard.edu/abs/2021MNRAS.504..118M/abstract or at least say that the current implementation remains consistent with newer results?
To extend the host galaxy SED to radio frequencies, we assumed contribution percentages to the $\rm \mathit{L}_{\rm 1.4 GHz}$ of $90 \%$ and $10 \%$ for non-thermal and thermal components, in line with the values presented by \cite{condon92} and \cite{dale2002infrared}. Power laws with $\alpha = 0.1$ for the thermal \citep{dale2002infrared, condon92}, and $\alpha = 0.75$ for the non-thermal \citep{baan2006radio}  emissions were computed from $1$ to $10$ GHz, and then smoothly joined to IR cold dust emission.

Radio emission from AGNs can originate from several physical processes, and establishing its exact origin or the respective contributions is an active field of research \citep[e.g.][]{panessa19}.
AGN processes that produce radio emission are relativistic radio jets of diverse scales (from sub-galactic to giant FRI or FRII structures), shocks from AGN outflows, as well as the X-ray corona \citep{panessa19}.
Independent of these assumptions, \textsc{AGNfitter-rx} implements four different functional forms of increasing complexity to describe the radio emission related to the AGN (light-blue curves in Fig. \ref{fig:AGNtemplates}) following \cite{azadi2020disentangling}.
While these models are not linked to a particular physical origin, they nonetheless provide useful information on the shape of the SED and the physical conditions of the synchrotron emission. 
These properties can then be with physical parameters obtained from the multiwavelength modelling to establish relations which can further push our understanding of the nature of the radio emission.
The choice of the complexity of the radio AGN model depends on the availability of radio data points. 

\vspace{2mm}

\mitem \textit{Simple power law (SPL):} If the observed SED has 1 or 2 radio bands, \textsc{AGNfitter} automatically assumes a power law defined by a slope fixed at $-0.75$ or fit over the range $[-2.0, 1.0]$, respectively.

\mitem \textit{Double power law (DPL):} For SEDs with additional ($\geq3$) radio measurements, this formulation takes into account the transition from optically thin (lobes) to optically thick (cores and hot spots) region at a transition frequency $\nu_t$. The optically thin is given by a synchrotron-aged population of electrons while the optically thick is due to absorption of radio photons by synchrotron self-absorption. This complex emission is described as follows: \begin{equation}
    \hspace{15mm}
    \text{L}_{\nu} \propto \left(\frac{\nu}{\nu_t}\right)^{\alpha_1} \left[1- \text{exp} \left(-\left(\frac{\nu_t}{\nu}\right)^{\alpha_1-\alpha_2}\right)\right] \text{e}^{-\frac{\nu}{\nu_{\rm cutoff}}}
\end{equation} When 3 bands are available $\alpha_1 = -0.75$, the slope difference $curv = \alpha_1 - \alpha_2$ takes a value in the range $[-0.5, 0.8]$ and $\nu_t$ in the interval $[10^7, 10^{13}]$ Hz. Whereas, if there are more measurements $\alpha_1$ and $\alpha_2$ are free parameter with values in the ranges $[-1.0, 1.0]$ and $[-1.0, 0.0]$, respectively.

\mend

\noindent We note that, in all cases, energy due to the ageing of the electron population is accounted for by an exponential term with a cutoff frequency $\nu_{\rm cutoff} = 10^{13}$ Hz. As discussed in \cite{polletta2000far}, it is difficult to find the energy cutoff from the radio data because it relies on a drop in the spectra while the actual cutoff may be located at lower frequencies. As such we select a value consistent with the limits of $10^{10}$ and $10^{14}$ Hz used in \cite{azadi2020disentangling} and one order of magnitude lower than the upper limits found by \cite{polletta2000far}.
With these prescriptions and considering the component normalization, the minimum number of parameters possible to fit radio fluxes is 1 (i.e. normalization parameter RAD) and the maximum number is 4 (i.e. $\alpha_1, \alpha_2, \nu_t, \rm RAD$).

\subsection{Infrared - Cold dust models} \label{sub:cold_dust_models}

The emission of the cold dust from the star-forming regions in the host galaxy is modeled in \textsc{AGNfitter-rx} using two optional sets of templates: a compilation of the semi-empirical libraries by \cite{chary2001interpreting} and \cite{dale2002infrared} or the theoretical set of SEDs by \cite{schreiber2018dust}, hereafter denoted DH02\_CE01 and S17, respectively.

\vspace{2mm}

\begin{itemize}
    \item \textbf{DH02\_CE01:} A detailed description of DH02\_CE01 library, which was included in the first version of the code, can be found in \citet{calistrorivera16}. The upper panel of Fig.\ref{fig:SED_SB+GALmodels} presents a subsample of SEDs of the library.

    \item \textbf{S17: } The cold dust SED templates of \cite{schreiber2018dust} consist of two independent components: the dust continuum and the mid-infrared emission line spectra from complex molecules. The first one is produced by big ($>0.01 \hspace{1mm} \mu \rm m$) and small ($<0.01 \hspace{1mm} \mu \rm m$) silicate and carbonate grains with "quasi-black body like" emissivity in the MIR-to-FIR range. The second component corresponds to the emission of lines between $3.3 -12.3 \hspace{1mm} \mu \rm m$ due to the characteristic vibrational and rotational modes of polycyclic aromatic hydrocarbons (PAHs). Each component is modelled by 150 templates defined by two parameters: the dust temperature and the mass fraction of the PAH component ($f_{\mathrm{PAH}}$), which weights the contribution of PAHs to the total cold dust spectrum. The total spectrum is found by adding the SED of dust and PAHs for each possible combination of $T_{\rm dust}$ and $f_{\mathrm{PAH}}$, resulting in a flexible model consisting of 9600 templates. The flexibility of this model can be particularly useful to capitalize on new detailed observations of PAH emission in high-$z$ galaxies and AGN with the James Webb Space Telescope (JWST). A subsample of templates is plotted in the lower panel of Fig.\ref{fig:SED_SB+GALmodels}. For more details on this model, the reader may refer to \citet{schreiber2018dust}.
\end{itemize}

\hspace{-5mm}Thus, the number of free parameters is 2 (i.e. log IRlum and the normalization parameter SB) when choosing the DH02\_CE01 model and 3 when choosing the S17 (i.e. $\rm T_{\rm dust}$, fracPAH and SB).

\subsection{Infrared - AGN torus models}

To model the emission from the nuclear hot dust and gas, commonly referred to as the AGN torus, we implement four different models: the semi-empirical model by \cite{Silva2004} and three theoretical models by \cite{nenkova2008agnII}, \cite{Stalevski2016} and \cite{honig2017dusty} hereafter S04, NK08, SKIRTOR and CAT3D-Wind, respectively. 
The theoretical models assume the existence of a population of ("clumpy") dust clouds, resulting in lines of sight with probabilistic opacities that depend on the cloud spatial and density distribution. 
%mass, size

\vspace{2mm}
\begin{itemize}
\item \textbf{S04:}  These semi-empirical templates consider the emission of a smooth distribution of dust. They were included in the first version of \textsc{AGNfitter} (see \cite{calistrorivera16} for a detailed description) and are presented in the upper panel of Fig. \ref{fig:AGNtemplates}.

\item \textbf{NK08:} The CLUMPY model by \cite{nenkova2008agnII} considers dust clouds with a standard Galactic composition ($53 \%$ silicates and $47 \%$ graphite) and column density of $10^{22}-10^{23}$ cm$^{-2}$  spread out in an axisymmetrical torus, internally limited by the dust sublimation region. The cloud distribution follows a Gaussian profile in the polar angle and a power law profile in the radial coordinate. 

\item \textbf{SKIRTOR:} The two-phase model by \cite{Stalevski2016} consists of a torus composed of high-density clumps immersed in a smooth distribution of low-density dust. Templates were calculated with the continuum radiative transfer code SKIRT \citep{baes2015skirt, camps2015skirt, camps2020skirt} by considering the geometry of a flared disk truncated the dust sublimation region. The dust distribution follows a power law profile with an exponential cut-off as a function of radius and polar angle, while the clumps scale radius of $0.4$ pc and a density profile given by a standard smoothing kernel with compact support\footnote{See \url{https://www.particleworks.com/technical_column_vol3_02_en.html} for brief explanation.}, are generated randomly. 

\item \textbf{CAT3D:} The CAT3D-Wind model \citep{honig2017dusty} considers dust and gas around AGNs as an ensemble of an inflowing clumpy disk and outflowing wind. Similar to NK08, the distribution of clouds with optical depth $\tau_V = 50$ and radius of $0.035$ times the graphite grains sublimation radius at $1500$ K (r$_{\text{ sub}} \sim 1$ pc), is given radially by a power law but in the vertical direction follows a Gaussian distribution. A unique feature of this model is the inclusion of a polar wind component which is embedded in a hollow cone and, due to the sublimation in the innermost region of the torus, is only composed of big grains of dust.

\end{itemize}

\noindent In the last three cases we include simplified versions of the models, since the original ones have 6, 6 and 8 fitting parameters, respectively. These large sets of SEDs are computationally demanding and may lead to overfitting of the data. Furthermore, in some cases, the spectral resolution of the SEDs may be insufficient to notice the effect of each parameter, which would cause the parameters to be unable to converge. For that reason, we reduced the number of templates by averaging all the models with respect to all but the 3 most relevant parameters (discussed through private communication with the authors of the models). We show in Section \ref{subsec:toruscomparison} that the reduction of the parameter space in these models is an approach to produce templates suitable for fitting. Those simplified templates deal with limitations relying on the existence, sensitivity, bandwidth, and spectral resolution of the rest-frame MIR data.

For NK08 and SKIRTOR are the inclination angle, opening angle and optical depth of the torus, while for CAT3D they are the inclination angle, index of the radial power law of the cloud distribution and fraction of clouds in the wind compared to the torus plane. As a result, our final SED grid consists of 1080, 400 and 378 templates for NK08, SKIRTOR and CAT3D, respectively, of which a subsample is plotted in the different panels of Fig. \ref{fig:AGNtemplates}. As a consequence of the previous model reductions and considering the normalization parameter, a minimum of 2 (i.e. the normalization parameter TO and $\log \rm N_{\rm H}$ or incl) and a maximum of 4 parameters (i.e. TO, incl, oa and $\tau_\nu$ or TO, incl, a and fwd, more details in table \ref{Tabla: parametros finales}) are required to model the torus emission.

\begin{table*}[ht]

\begin{adjustbox}{max width=\textwidth}
\centering
\begin{tabular}{@{}ccccc@{}}
\toprule
\multicolumn{1}{c}{\textbf{Component}}           &  \multicolumn{1}{c}{\textbf{Model}}                                              & \multicolumn{1}{c}{\textbf{Parameter}} & \multicolumn{1}{c}{\textbf{Description}}                                                & \textbf{Range}    \\ \midrule
\multicolumn{1}{c}{\multirow{5}{*}{Galaxy}}     & \multicolumn{1}{c}{\multirow{5}{*}{BC03_metal}}                                                 & \multicolumn{1}{c}{$\tau$}               & \multicolumn{1}{c}{Time scale of the exponential SFH  {[}log year{]}}                   & {[}0.05, age($z$){]} \\  
\multicolumn{1}{c}{}   & \multicolumn{1}{c}{} & \multicolumn{1}{c}{age}               & \multicolumn{1}{c}{Galaxy age {[}log year{]}}                                      & {[}7, age($z$){]}    \\  
\multicolumn{1}{c}{}  & \multicolumn{1}{c}{}   & \multicolumn{1}{c}{Z}             & \multicolumn{1}{c}{Metallicity {[}$Z_\odot${]}}                                              & {[}0.2, 2{]}      \\
\multicolumn{1}{c}{} & \multicolumn{1}{c}{}  & \multicolumn{1}{c}{E(B-V)$_{\text{gal}}$}            & \multicolumn{1}{c}{Galaxy reddening}                                         & {[}0,0.6{]}         \\ 
\multicolumn{1}{c}{}  & \multicolumn{1}{c}{}  & \multicolumn{1}{c}{GA}                & \multicolumn{1}{c}{Galaxy normalization}                                          & {[}-10, 10{]}     \\ \midrule
\multicolumn{1}{c}{\multirow{4}{*}{Cold dust}}  &   \multicolumn{1}{c}{DH02_CE01}  & \multicolumn{1}{c}{log IRlum [$L_\odot$]}           & \multicolumn{1}{c}{Parametrization of starburst templates}                                                    & {[}7, 15{]}     \\ 
\multicolumn{1}{c}{}  & \multicolumn{1}{c}{\multirow{2}{*}{S17}}                                                 & \multicolumn{1}{c}{$\text{T}_{\text{dust}}$}             & \multicolumn{1}{c}{Dust temperature {[}K{]}}                                  & {[}14.24, 42{]}   \\
\multicolumn{1}{c}{}  & \multicolumn{1}{c}{}  & \multicolumn{1}{c}{fracPAH}           & \multicolumn{1}{c}{PAHs fraction}                                                    & {[}0, 0.05{]}     \\ 
\multicolumn{1}{c}{}  & \multicolumn{1}{c}{DH02_CE01, S17} & \multicolumn{1}{c}{SB}                & \multicolumn{1}{c}{Starburst normalization}                                          & {[}-10, 10{]}     \\  \midrule
\multicolumn{1}{c}{\multirow{3}{*}{Torus}}            & \multicolumn{1}{c}{S04}                                                  & \multicolumn{1}{c}{log N$_{\text{H}}$}                & \multicolumn{1}{c}{Torus column density {[}log cm$^{-2}${]}}                    & {[}21, 25{]}    \\
\multicolumn{1}{c}{}   &    \multicolumn{1}{c}{NK08, SKIRTOR, CAT3D}   &  \multicolumn{1}{c}{incl}              & \multicolumn{1}{c}{Torus inclination angle {[}$^\circ${]}}                       & {[}0, 90{]}      \\ 
\multicolumn{1}{c}{}   &    \multicolumn{1}{c}{\multirow{2}{*}{NK08, SKIRTOR}}   &  \multicolumn{1}{c}{oa}              & \multicolumn{1}{c}{Opening angle {[}$^\circ${]}}                       & {[}10, 80{]}      \\ 
\multicolumn{1}{c}{}   &    \multicolumn{1}{c}{}   &  \multicolumn{1}{c}{$\tau_\nu$}              & \multicolumn{1}{c}{Optical depth at $9.7 \mu \rm m$}                       & {[}10,300{] and [3,11]}      \\ 
\multicolumn{1}{c}{}   &    \multicolumn{1}{c}{\multirow{2}{*}{CAT3D}}   &  \multicolumn{1}{c}{a}              & \multicolumn{1}{c}{index of cloud dist. }                       & {[}-3, -1.5{]}      \\ 
\multicolumn{1}{c}{}   &    \multicolumn{1}{c}{}   &  \multicolumn{1}{c}{fwd}              & \multicolumn{1}{c}{cloud fraction }                       & {[}0.15, 2.25{]}      \\ 
\multicolumn{1}{c}{}  & \multicolumn{1}{c}{S04, NK08, SKIRTOR, CAT3D}  & \multicolumn{1}{c}{TO}                & \multicolumn{1}{c}{Torus normalization}                & {[}-10, 10{]}     \\ \midrule
\multicolumn{1}{c}{\multirow{4}{*}{\begin{tabular}[c]{@{}c@{}}Accretion \\ disk\end{tabular}}} & \multicolumn{1}{c}{\multirow{2}{*}{SN12, KD18}}  & \multicolumn{1}{c}{log M$_{\text{BH}}$}  & \multicolumn{1}{c}{Black hole mass {[}log $M_\odot${]}}                         & {[}7.4, 9.8{] and [}6.0, 10.0{]}    \\ 

\multicolumn{1}{c}{}  & \multicolumn{1}{c}{}& \multicolumn{1}{c}{log $\dot{m}$} & \multicolumn{1}{c}{Accretion rate{[}$\log \dot{M}/ \dot{M}_{\rm edd}${]}}                  & {[}-4, 2.49{]} and {[}-0.15, 0.0{]}    \\ 
\multicolumn{1}{c}{} & \multicolumn{1}{c}{R06, THB21, SN12, KD18} & \multicolumn{1}{c}{E(B-V)$_{\text{bbb}}$}  & \multicolumn{1}{c}{BBB reddening}                                         & {[}0,1{]}         \\ 
\multicolumn{1}{c}{} &  \multicolumn{1}{c}{R06, THB21} &\multicolumn{1}{c}{BB} & \multicolumn{1}{c}{BBB normalization}                                          & {[}-10, 10{]}     \\ \midrule
\multicolumn{1}{c}{\multirow{2}{*}{Hot corona}}  & 
\multicolumn{1}{c}{\multirow{2}{*}{R06, SN12, THB21}} &
\multicolumn{1}{c}{$\Delta \alpha_{\rm ox}$} & \multicolumn{1}{c}{ $\alpha_{\rm ox}-L_{2500 \angstrom }$ dispersion} & {[}-0.4, 0.4{]} \\ 

\multicolumn{1}{c}{}  & \multicolumn{1}{c}{}& \multicolumn{1}{c}{$\Gamma$} & \multicolumn{1}{c}{Photon index}  & {[}0, 3{]}  \\\midrule
\multicolumn{1}{c}{\multirow{5}{*}{Radio from AGN}} &
\multicolumn{1}{c}{SPL}  &
\multicolumn{1}{c}{$\alpha$} &
\multicolumn{1}{c}{Slope of synch. {[}log $\nu${]}} & {[}-2, 1{]}  \\
\multicolumn{1}{c}{} &
\multicolumn{1}{c}{\multirow{4}{*}{DPL}} &
\multicolumn{1}{c}{curv}   & \multicolumn{1}{c}{Difference in slopes} & {[}-0.5, 0.8{]} \\
\multicolumn{1}{c}{} &
\multicolumn{1}{c}{}  &
\multicolumn{1}{c}{$\nu_{t}$}   & \multicolumn{1}{c}{Transition frequency {[}log $\nu${]}} & {[}7, 13{]} \\
\multicolumn{1}{c}{} &
\multicolumn{1}{c}{}  &
\multicolumn{1}{c}{$\alpha_1$}   & \multicolumn{1}{c}{Slope of synch. aged {[}log $\nu${]}} & {[}-1, 1{]} \\
\multicolumn{1}{c}{} &
\multicolumn{1}{c}{}  &
\multicolumn{1}{c}{$\alpha_2$}   & \multicolumn{1}{c}{Slope of synch. self-abs {[}log $\nu${]}} & {[}-1, 0{]} \\
\multicolumn{1}{c}{} &
\multicolumn{1}{c}{SPL, DPL}  &
\multicolumn{1}{c}{RAD}   & \multicolumn{1}{c}{Radio normalization} & {[}-10, 10{]}   \\ \bottomrule
\end{tabular}
\end{adjustbox}
\caption{Description, name and values of the \textsc{AGNfitter-rx} fitting parameters. The value  age($z$) correspond to the age of the Universe at the observing time, according the assumed cosmological model $\Lambda$CDM with $H_{0}=70 \rm kms^{-1}\text{ Mpc}^{-1}$, $\Omega_{\text{m}}= 0.3$ and $\Omega_{\Lambda}= 0.7$.}

\label{Tabla: parametros finales}
\end{table*}

\subsection{NIR/Optical/UV - Stellar population models}

The stellar emission templates were built employing the stellar population synthesis as in the first version of \textsc{AGNfitter}. 
We took the single stellar population (SSPs) models from \cite{bruzual2003stellar} defined for ages from $10^7$ to $10^{10}$ yr and metallicities ranging from $0.004$ to $0.04$. This formalism includes a prescription for thermally pulsing stars in the asymptotic giant branch.
To calculate the composite stellar population models we include the initial mass function by \cite{chabrier2003galactic} and an exponentially decreasing star formation history (SFH). To account for a wider variety of stellar populations, 10 different characteristic times for the decay of SFH were considered with values between $0.05$ to $11$ Gyr. Unlike the first version of \textsc{AGNfitter}, metallicity is now a free parameter of the SSPs.
\textsc{AGNfitter-rx} builds the stellar emission library using the \textsc{smpy} code \citep{duncan2015powering}, which gives the user the flexibility to include new templates with different star formation histories if desired. With new updates, the number of parameters required to fit the galaxy emission model is 4 or 5 (i.e. the normalization parameter GA, $\tau$, age, Z, E(B-V)$_{\rm gal}$) if we include metallicity.

%metallicities other than solar are now considered allowing to count the spectral evolution of SSPs, different atmosphere models and a better relative estimation of the number of red and blue supergiants. 
%This is mainly due to the transition of oxygen-rich to carbon-rich stars caused by the carbon dredge-up during the thermally pulsing regime of the asymptotic giant branch. Higher metallicities increase the minimum initial mass necessary to form carbon star and therefore, decreases the ratio of carbon-rich to oxygen-rich stars. Also, since star evolve at lower temperatures and luminosities, increasing metallicities redden the SSP colors and increase the mass-to-light ratio. 

\subsection{Optical/UV  - Accretion disk models} \label{subsec:accretiondisk}

One of the main strengths of \textsc{AGNfitter} in comparison to other SED-fitting codes is the superior modelling of the BH accretion disk emission and its reddening due to dust attenuation.
\textsc{AGNfitter-rx} further improves on this by including four libraries for the detailed modelling of the accretion disk emission\footnote{We use "accretion disk emission" to mean "UV and optical emission from the region inside the dust sublimation radius, including the broad line region.}: a modified version of the semi-empirical templates by \cite{richards2006sloan} and \cite{temple2021modelling} denoted as R06 and THB21, respectively; and the theory-built sets of templates by  \cite{slone2012effects} and \cite{kubota2018physical}, denoted as SN12 and KD18, respectively. 

\vspace{2mm}

\begin{itemize}

\item \textbf{R06:} The big blue bump model by \cite{richards2006sloan} is based on observed quasar spectra and a NIR blackbody tail. The library is composed of one SED which is then modified by different levels of dust reddening following the Small Magellanic Clouds (SMC) dust attenuation law (upper panel of Fig. \ref{fig:AGNtemplates}), and was implemented in the first version of the code (for more details see \citealt{calistrorivera16}).

\item \textbf{SN12:} The emission model by \citet{slone2012effects} is based on an $\alpha$-disk model (optically thick and geometrically thin accretion disk), and the inclusion of relativistic temperature corrections, the effect of viscous dissipation and the comptonization of the radiation in the disk atmosphere. We include a simplified model by assuming a zero-spin black hole and windless disk, resulting in a set of 108 templates (some of them presented in the upper central panel of Fig. \ref{fig:AGNtemplates}). The templates are parametrized by the mass of the SMBH taking values between of $\log \rm M_{\rm BH} = [7.4, 9.8]$ in intervals of $\Delta \log \rm M_{\rm BH} = 0.3$ and the accretion rate with values in the range $\dot{\rm M}/\dot{\rm M}_{\rm edd} = [0.0, 310.116]$ in variable intervals in the range $\Delta \dot{\rm M}/\dot{\rm M}_{\rm edd} = [0.001, 154.69]$.

%\gaby{We note the effect of this components is negligible?}.
% SN12 The original model considers winds, which change the accretion rate ($\dot{M}$) locally; relativistic temperature corrections, the effect of viscous dissipation and the comptonization of the emitted radiation in the disk atmosphere.

\item \textbf{KD18:} The theoretical model AGNSED \citep{kubota2018physical} considers the emission of an outer Novikov-Thorne accretion disk, an inner warm Comptonising region and a hot corona, all of them radially delimited. We adopted a simplified version which implied setting all the parameters except the black hole mass and the accretion rate to their typical values and assumed a second comptonisation component as a source of the soft X-ray excess  \citep[for more details see][]{kubota2018physical}\footnote{Python code to generate the templates in \url{https://github.com/arnauqb/qsosed}}.  The final set is composed of 225 SEDs defined by black hole masses between $\log \rm M_{\rm BH} = [6, 10]$ in intervals of $\Delta \log \rm M_{\rm BH} = 0.286$ and accretion rates in the range $\log \dot{\rm M}/\dot{\rm M}_{\rm edd} = [-1.5, 0]$ in intervals of $\Delta \dot{\rm M} = 0.107$.  A subsample of the library is shown in the two lower panels of Fig. \ref{fig:AGNtemplates}.

\item \textbf{THB21:} The accretion disk library by \citet{temple2021modelling} is a set of composites empirically-derived SED of luminous quasars. The templates include a broken power-law that models the accretion disk, a blackbody dominating from $1$ to $3 \hspace{1mm} \mu \rm m$ due to hot dust; and account for both broad and narrow emission lines as well as additional continuum emission from blended line emission.

To soften the edges of the SED in \textsc{AGNfitter-rx}, we add the low ($\nu < 10^{14.15}$ Hz) and high frequency ($\nu > 10^{15.5}$ Hz) tails of the R06 model by normalizing at the transition frequencies. The final template appertains to a zero redshift and non-reddened quasar with an average emission line template, and it is shown in the last panel of Fig. \ref{fig:AGNtemplates} after applying the \textsc{AGNfitter} reddening prescription.

\end{itemize}

\noindent As mentioned in the R06 description, the different levels of extinction produced by dust are accounted for by applying the SMC reddening law \citep{prevot1984typical} to the non-reddened templates of all the aforementioned accretion disk models. The needed parameter in this prescription is the reddening $\text{E(B-V)}_{\text{BBB}}$ with values between $0$ and $1$.
In summary, to fit the accretion disk component, the user needs between 2 (i.e. the normalization parameter BB and $\text{E(B-V)}_{\text{BBB}}$) and 3 (i.e. $\log \rm M_{\rm BH}$, $\dot{\rm M}/\dot{\rm M}_{\rm edd}$ and $\text{E(B-V)}_{\text{BBB}}$) parameters.

%On one hand the values taken by the parameters were: spin of 0, an electron temperature in the hot corona of 100 keV and 0.2 keV in the warm component, a photon index of the warm component of 2.5, a luminosity of the corona of 0.02 times the Eddington luminosity and a reflection albedo of 0.3.

\subsection{X-rays  - corona and star formation} \label{subsec:accretiondisk-xray}

For optical-UV accretion disk models which do not include X-rays, we add an X-ray model. Including the X-ray emission in AGN SED fitting is crucial as X-ray fluxes may be used to provide additional information to constrain accretion disk emission and break potential degeneracies between the blue accretion disk emission and galaxies with young/blue stellar populations in the optical-UV region. The X-ray emission is produced in the hot corona in the vicinity of the black hole, containing information related to the accretion process and the black hole properties. Considering the above, we used the empirical correlation $\alpha_{\rm ox}- L_{2500 \text{\AA}}$ \citep{lusso2017quasars,just2007x} which connects the coronae emission at $2$ keV with the  emission of the AGN at $2500 \hspace{1mm} \text{\AA}$:

\begin{equation}
\hspace{18mm}
    \alpha_{\rm ox} = -0.137 \log (L_{2500 \text{\AA}}) + 2.638,
\label{eq: alpha_ox}
\end{equation}

\noindent where $\alpha_{\rm ox}$ is the slope of the SED between $2500 \hspace{1mm} \text{\AA}$ and $2$ keV given by:

\begin{equation}
\hspace{18mm}
    \alpha_{\rm ox} = -0.3838 \log (L_{2500 \text{\AA}}/L_{2\text{keV}}).
\label{eq: slope_UV_RX}
\end{equation}

\noindent Based on the $2500 \hspace{1mm} \text{\AA}$ flux, we computed the expected $\alpha_{\rm ox}$ from the relation. Then, we included a grid of dispersion values of $\Delta \alpha_{\rm ox}$, with respect to the empirical relation, to calculate the corresponding 2 keV fluxes for each accretion disk template. The dispersion of the relation is a fitting parameter taking a maximum value of $\Delta \alpha_{\rm ox} = |0.4|$ reported by \cite{lusso2016tight} for low-quality data and sources with variability. Once the X-ray normalization is fixed, the SEDs were extended from 2 keV to more energetic X-rays by assuming a power law with a photon index of $1.8$ and an exponential cut-off at $300$ keV as assumed in \cite{Yang2020}. When at least 2 X-ray measurements are available, the photon index is a free-fit parameter that takes values from $0.00$ to $3.00$ in intervals of $\Delta \Gamma = 0.15$. Note that obscuration is not accounted for, therefore this effective power law carries information about both obscuration and potential accretion rate estimates. Also, $\alpha_{\rm ox}-L_{2500 \text{\AA}}$  relation is valid to constrain soft X-ray emission only for type-1, radio-quiet, non-BAL\footnote{BAL: Broad absorption line} AGNs \citep{strateva2005soft}. Similarly, the SED of Blazars is likely modelled with synchrotron emission from radio to X-rays, therefore $\alpha_{\rm ox}-L_{2500 \text{\AA}}$  relation is not providing reliable physical interpretation for those objects.

As part of the postprocessing analysis, we compute the host galaxy contribution to the X-rays due to the emission of high-mass X-ray binaries (HMXBs) and the hot ionized interstellar medium (ISM) by using the \cite{mineo2014x}:

\begin{equation}
\hspace{18mm}
L_{0.5-8 \text{keV}} = (4.0 \pm 0.4) \times 10^{39} \text{SFR}
\label{eq: host_gal_RX}
\end{equation}

\noindent where $L_{0.5-8 \text{keV}}$ is the luminosity between $0.5$ and $8$ keV in erg s$^{-1}$ and SFR the star formation rate in M$_\odot$yr$^{-1}$. For this purpose, we used the IR SFR computed from the \cite{murphy2011calibrating} calibration for the integrated $8-1000 \hspace{1mm}\mu \rm m$ luminosity of the cold dust emission:

\begin{equation}
\hspace{18mm}
    \left(\frac{\rm SFR_{\rm}}{\rm M_\odot  \rm yr^{-1}}\right) = 3.88 \times 10^{-44} \left(\frac{\it L_{\rm IR}}{\rm erg \hspace{1mm} \rm s^{-1}}\right).
\end{equation}

\noindent Although we selected the most robust SFR estimator, it is worth noting that \textsc{AGNfitter-rx}, as in its initial release, also delivers as output optical/UV SFR estimated from the stellar population emission (see \cite{calistrorivera16} for more details). We included the host galaxy X-ray emission derived from the IR SFR in the estimation of the AGN fraction.

In table \ref{Tabla: parametros finales} a summary of the model fitting parameters in \textsc{AGNfitter-rx}, including names, values and descriptions, is presented. The \textsc{AGNfitter} code \citep{calistrorivera16}, including the \textsc{AGNfitter-rx} code release, are publicly available as open-source code in Python 3 in \url{https://github.com/GabrielaCR/AGNfitter}.

\begin{figure*}[ht!]
\centering
    \includegraphics[trim={0 1.57cm 0 0},clip, width = 0.8\linewidth]{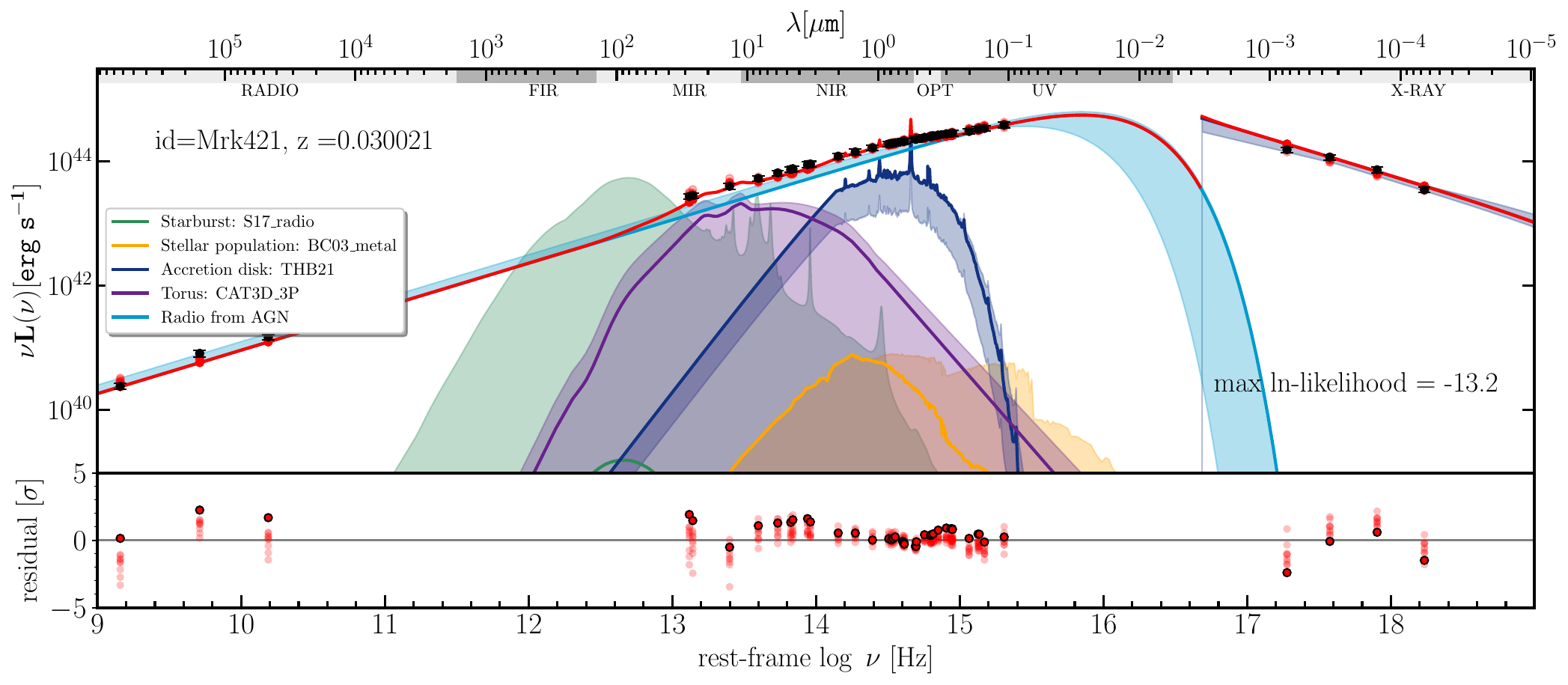}
    \includegraphics[trim={0 1.57cm 0 1.55cm},clip, width = 0.8\linewidth]{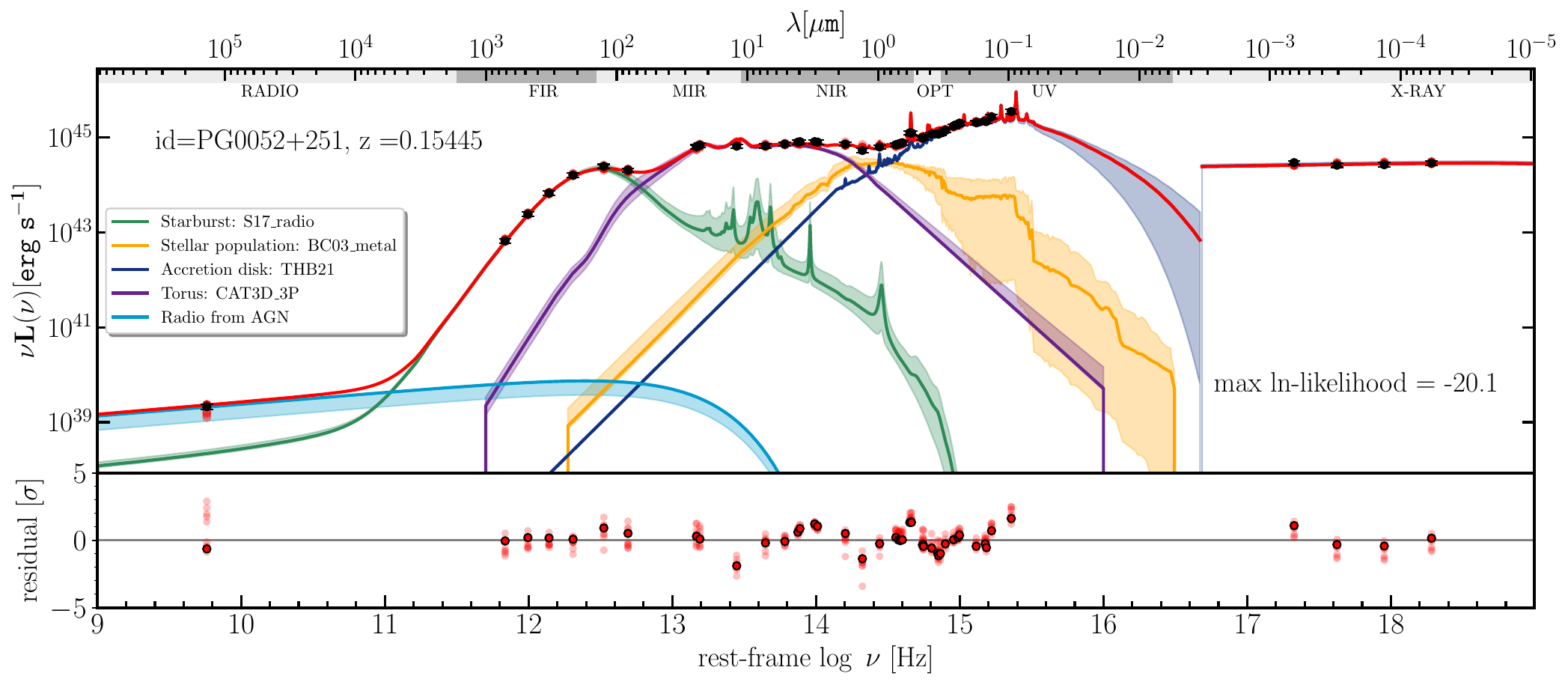}
    \includegraphics[trim={0 1.57cm 0 1.55cm},clip, width = 0.8\linewidth]{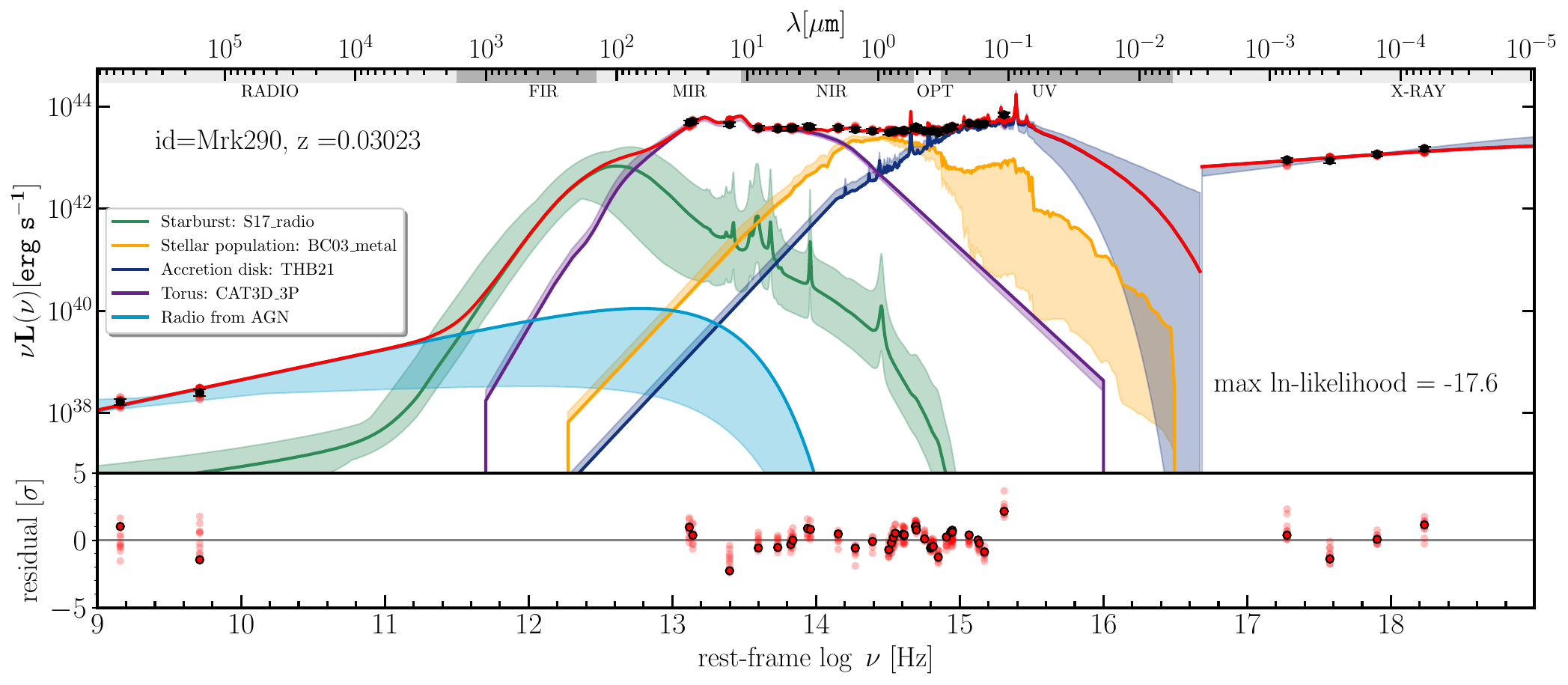}
    \includegraphics[trim={0 0 0 1.55cm},clip, width = 0.8\linewidth]{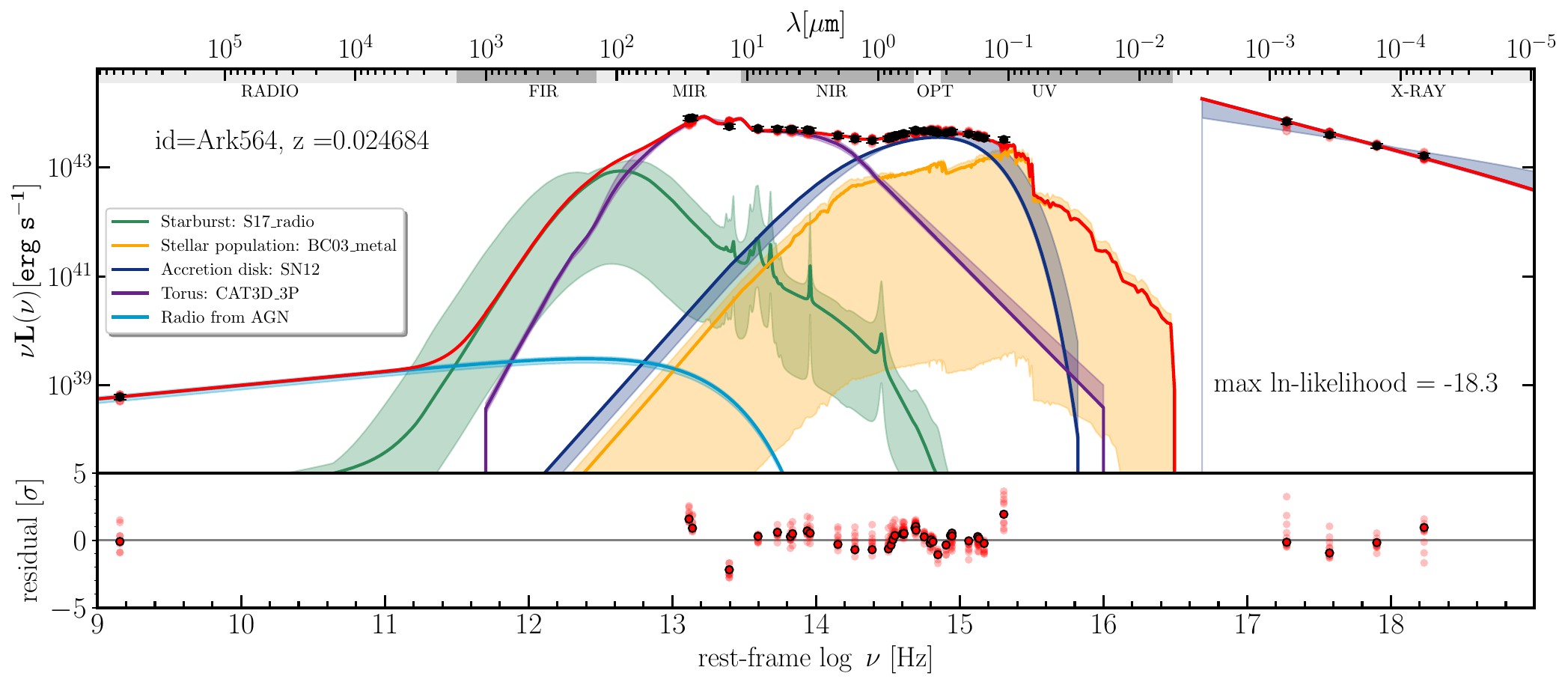}
    \caption{Examples of the best SED fittings of the different AGN types: the blazar Mrk 421 in the top panel, the Seyfert1 PG0052+251 in the second panel, the Seyfert 1.5 Mrk 290 in the third panel, and the Seyfert 2 Ark 564 in the bottom panel. The yellow and green solid curves show the emission from the stellar population and cold dust of the galaxy, respectively. Purple, dark blue and light blue solid curves show the torus, accretion disk, and radio AGN emission models, respectively. 10 models constructed from combinations of parameters randomly selected from the posterior PDFs (hereafter referred to as realizations) are plotted as a shaded area and the residuals of each fit realization are presented in the graphs below each source. A flexible energy balance prior was assumed so that the cold dust IR emission is at least comparable to the dust-absorbed stellar emission.}
    \label{fig: SEDfitting_1}
\end{figure*}

\begin{figure*}[ht!]
\centering
    \includegraphics[trim={0 1.57cm 0 0},clip, width = 0.8\linewidth]{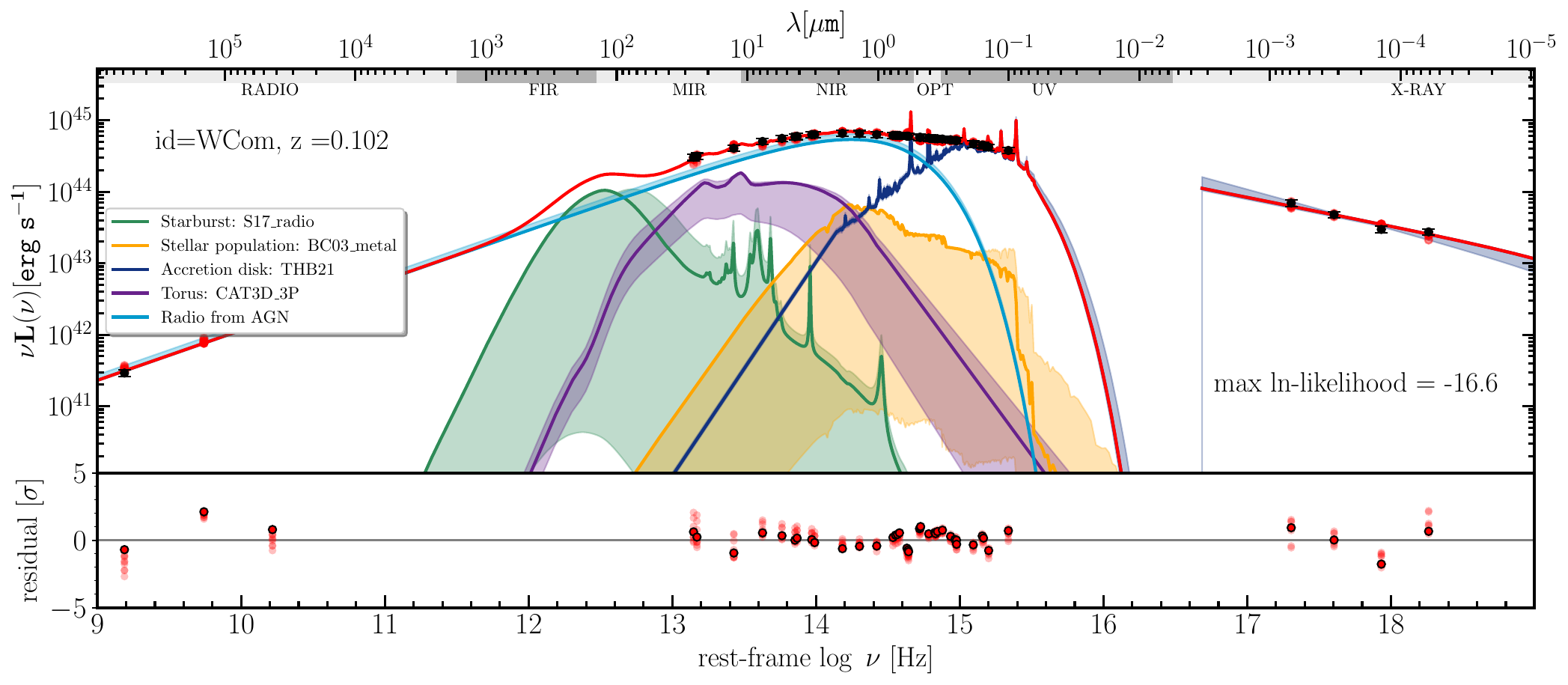}
    \includegraphics[trim={0 1.57cm 0 1.55cm},clip, width = 0.8\linewidth]{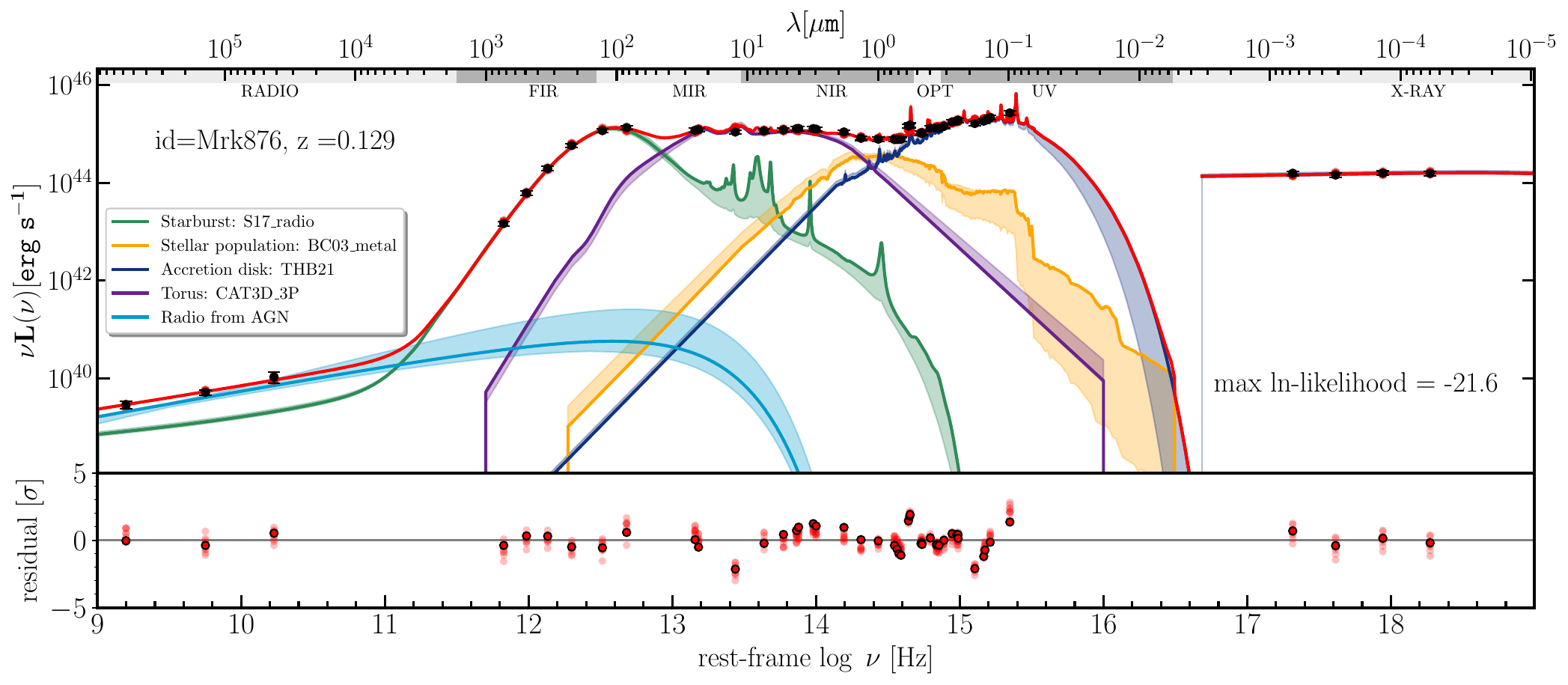}
    \includegraphics[trim={0 1.57cm 0 1.55cm},clip, width = 0.8\linewidth]{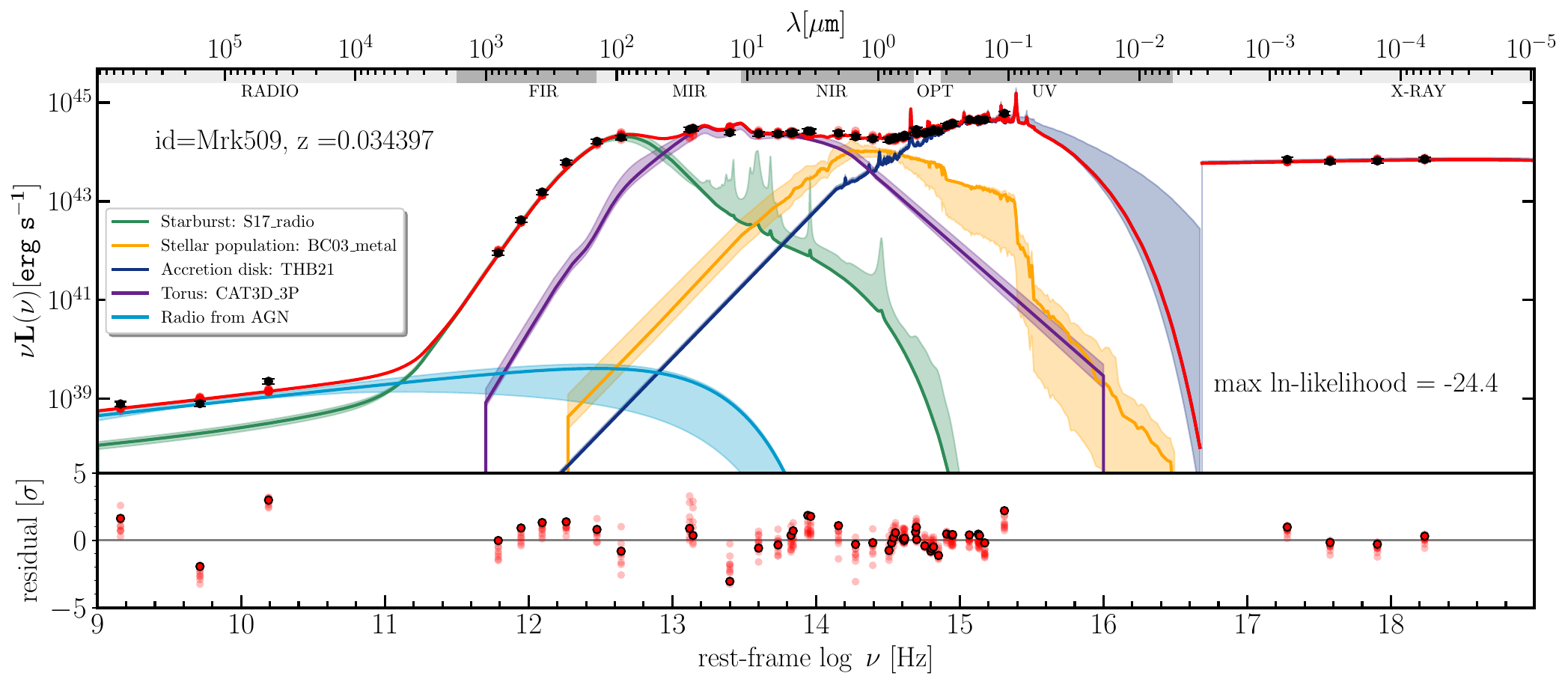}
    \includegraphics[trim={0 0 0 1.55cm},clip, width = 0.8\linewidth]{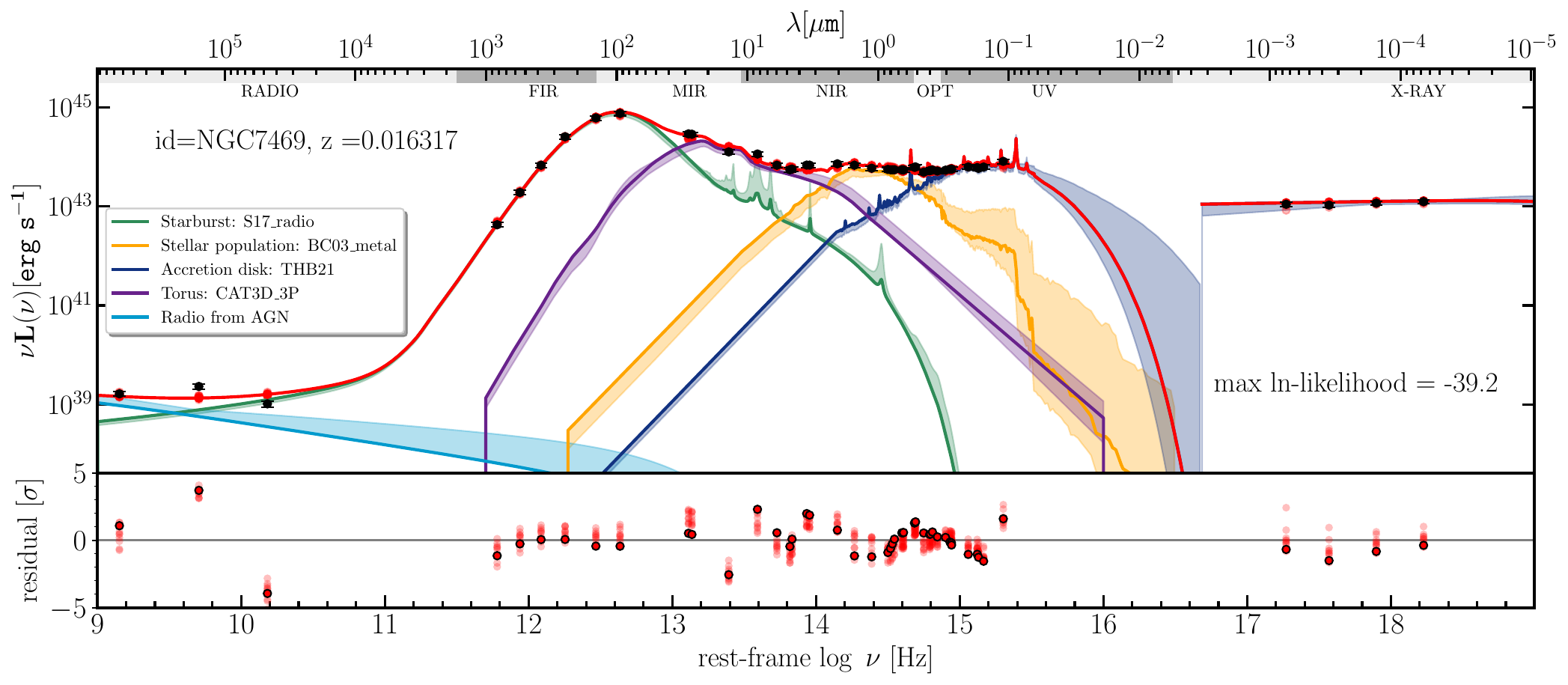}
    \caption{Continued from Figure \ref{fig: SEDfitting_1}. Examples of the best SED-fittings for the blazar WCom, the Seyfert 1 Mrk876 and the Seyferts 1.5 Mrk509 and NGC7469.}
    \label{fig: SEDfitting2}
\end{figure*}

\section{Data: radio-to-X-ray SEDs of local AGN}\label{sec:brownAGN}

To test the capabilities of \textsc{AGNfitter-rx}, we fit the SEDs of a sample of 36 nearby active galaxies selected from the AGN SED ATLAS by \citet{brown19}. This sample was chosen as it comprises a well-characterized and diverse set of local AGNs, with well-constrained photometric measurements with a high signal-to-noise ratio. To derive these synthetic SEDs, individual spectra were combined and scaled by a multiplicative factor to match aperture photometry measured from images using circular apertures. The scaling factors aim to relieve the effects of different instruments, extraction apertures and AGN variability producing discontinuities on the SEDs.  The resulting photometric random uncertainties of the synthetic SEDs are extremely small compared to the errors in the photometric calibrations, scatter introduced by AGN variability and sources of systematic error since the AGNs in our sample are very bright. The photometric data is a compilation of available observations from GALEX, Swift, SDSS, PanSTARRS, Skymapper, 2MASS, WISE, Spitzer and  Herschel, covering a broad wavelength range from the UV to the IR. For more details on the photometry, we refer the reader to \citet{brown19}. 

According to the spectroscopic classification by \citet{veron2010catalogue}, this sample includes 8 Seyfert 1, 5 Seyfert 1n (Seyfert 1s with narrow Balmer lines), 6 Seyfert 1.2, 8 Seyfert 1.5, 1 Seyfert 1.9, 2 Seyfert 2, 1 Seyfert 1i (Seyfert 2s with reddened broad line region), 1 LINER and 4 BL Lacertae objects.

To test the new \textsc{AGNfitter} developments, we extended the spectral coverage of the \citet{brown19} SEDs by including archival photometric data in the radio and X-ray regimes.
In the radio regime, we added 1.4, 5 and 15 GHz radio data from OVRO, Parkes, VLBA, VLA, Green Bank 140-ft, 100-m Effelsberg, 26-m Peach Mountain radio telescopes, obtained through the NASA/IPAC Extragalactic Database\footnote{\url{https://ned.ipac.caltech.edu/}}. These bands were chosen because they were available for most AGNs (32 out of 36) in the sample. %\gaby{hadn't we also added the low-freq radio data?}

In the X-rays, we added the 0.5-1 keV, 1-2 keV, 2-4.5 keV and 4.5-12 keV bands from the XMM-Newton Serendipitous Source Catalogue 4XMM-DR11 \citep{webb2020xmm}. As explained by \cite{Yang2020}, there are no standard transmission curves for X-ray bands due to the high sensibility of measurements to observing conditions and source characteristics. Hence, we follow the procedure by \cite{risaliti2019cosmological} to calculate monochromatic fluxes at effective area-weighted mean energies (henceforth called pivot points) and include delta Dirac filters centred at those energies.

We got the pivot energies 0.76, 1.51, 3.22 and 6.87 keV for bands 2, 3, 4 and 5 of 4XMM-Newton, respectively, through private communication with the authors of the catalog. We computed correction factors ($\rm f_{\rm corr}$) to consider the effect of Galactic absorption while correcting for the difference between $\rm N_{\rm H}$ at the source location and the catalog assumed value. Factors were computed based on the reported and reference fluxes from \cite{webb2020xmm} for $\Gamma = -1.42$ and $N_H = 1.7 \times 10^{20} \text{ cm}^{-2}$, and the column densities from the gas map of the Python package gdpyc\footnote{Gas and Dust Python Calculator}. Thus, the corrected monochromatic fluxes at the  pivot points were estimated as follows:

\begin{equation}
\hspace{15mm}
    \text{F}_{\nu_2-\nu_1(\rm corr)} = \nu^{(1+\Gamma)} \frac{\text{F}_{\nu_2-\nu_1} (2+\Gamma)}{\nu_2^{2+\Gamma} - \nu_1^{2+ \Gamma}} \rm f_{corr},
\label{eq: flux_pivot_energy}
\end{equation}
where $F_{\nu_2-\nu_1}$ is the catalog reported flux within a given band, $\nu_1$ is lower and $\nu_2$ the upper-frequency limits of each band and a power law function with $\Gamma = -1.42$ (used to derive the fluxes in the 4XMM catalogue) is assumed to model the X-ray emission. By definition, the monochromatic flux at the pivot energy does not depend on the photon index value assumed. The above procedure was possible for $29$ of $36$ AGNs in the sample with available X-ray data.

Our final data set comprises $36$ galaxies, from which $25$ have radio-to-X-ray coverage with up to $49$ photometric data points, $7$ have radio-to-UV coverage with up to $45$ photometric data points, and 4 have IR-to-X-ray coverage with up to $46$ photometric data points. Out of the $36$ galaxies, $22$ has FIR coverage. As photometry was not acquired at the same time, variability may be affecting our analysis. However, the trade-off is having very good wavelength coverage and high-quality photometry to perform the fittings.
The extended photometric catalogues are published as supplementary data to this publication \gaby{link}.

\section{Results}\label{sec: results}

In Figs. \ref{fig: SEDfitting_1} and \ref{fig: SEDfitting2} we present examples of the fitting results that include Seyfert 1, Seyfert 1.5, Seyfert 2 AGN and blazars. The different curves represent the emission of the physical components corresponding to the most probable model (solid colour curves) and 10 realizations randomly selected from the posterior PDFs (shaded area). 
The line colors are coded similarly as in Figs. \ref{fig:SED_SB+GALmodels} and \ref{fig:AGNtemplates}.
The total SEDs are depicted as red lines and the observed photometry as black-filled circles with error bars. The lower part of each panel presents the residuals estimated as the difference between observed and model fluxes divided by the quadrature-combined errors of the photometry for each band. The residuals of the best model and 10 random realizations are shown as red circles with outlined and shaded areas, respectively. The annotated maximum likelihood value is the highest value achieved by the live points in the nested sampling fitting process.

We employed the nested sampling method to perform the fits, which relies on a population of live points to explore parameter space. The iterative process removes the point with the lowest likelihood and searches for a new, independent point on ellipsoids around the remaining live points. The volume of live point decreases while the likelihood values increase. The process runs until the live point population is so small that does not contribute any probability mass.

This process takes approximately $25$ minutes per source to fit a maximum of $19$ parameters. We were able to concurrently run up to $4$ sources, leveraging \textsc{AGNfitter-rx} parallelization. While the computational time aligns with that of MCMC, the methods exhibit distinct convergence criteria. In emcee, convergence relies on assessing the effective number of independent samples within each chain, as measured by the integrated autocorrelation time. In contrast, UltraNest employs a convergence criterion based on the weight of the live points, since once this weight becomes negligible, it no longer improves the results, even if the iterations persist.

The set-up parameters were: 400 as the minimum number of live points, 400 effective samples, and 20 steps for the slice sampler.

It can be seen that different AGN classes require slightly different sets of models to optimize the fitting of the observed SED. This is not surprising as the AGNs in our sample are quite diverse. However, we can exploit the flexible capabilities of \textsc{AGNfitter-rx} to investigate general trends of the different models included in the public version of the code and characterise how the results depend on the choice of the model.  
We also investigate the AGN physical components with the largest variety of models in the code, the torus and the accretion disk. For this purpose, we compare the performance of different models associated with a given component, while keeping fixed models for the other physical components. Isolating the effects of the different torus and disk models is of course complex due to dissimilar wavelength coverage and overlap with other physical components, such as stellar emission and cold dust, for instance.

For this purpose, we first set the R06 model for the disk, S17 for the cold dust, BC03 with metallicity for the stellar population and test the different hot dust models: 1 and 3 parameter versions of NK0, SKIRTOR, CAT3D; and the empirical S04. Once the best torus model was found, we proceeded to evaluate the different templates of the accretion disk using this model and maintaining S17 and B03 as cold dust and stellar population models. In all the runs, we assumed a flexible prior to constraining the cold dust to be at least as luminous as the dust-absorbed stellar emission.

The primary statistical outputs of the code are the distribution of logarithmic likelihood (log $\mathcal{L}$) for 100 random model realizations, its corresponding expected fluxes, the residuals compared to the data and the model logarithmic marginal likelihood (log Z from now on evidence). Beyond the quality of the fit informed by the likelihood and residuals, our Bayesian approach allows us to estimate and analyze posterior probabilities of the models given the observed SED P(M|D). According to the Bayes theorem, it is define as:

\begin{align}
    \hspace{25mm} \rm P(M|D) & = \frac{\rm P (D|M) P(M)}{\rm P(D)}
\label{eq: posterior_odds}
\end{align}

where P(D|M), P(D) and P(M) corresponds to the likelihood, the evidence and the prior probability of the models, respectively. Since the sampling process outputs log $\mathcal{L}$ and log Z, we can compute log-posterior probabilities by assuming all the models are a priori equally probable. Due to the evidence term, posterior odds contain information of the probability of a given model to generate the observed SED by considering all combinations of parameters, enabling a robust model comparison.

\subsection{Comparing torus models}\label{subsec: comp torus}

%posterior PDFs cummulative sum of live point weights > 1e-4

\begin{figure*}[ht!]
    \centering
    \includegraphics[width = 0.43\linewidth]{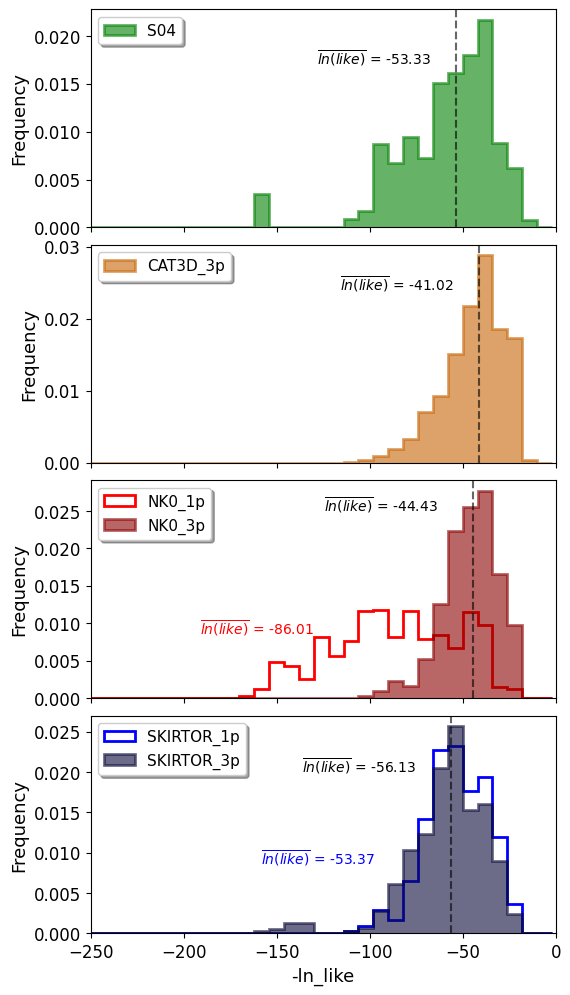}
    \includegraphics[width = 0.539\linewidth]{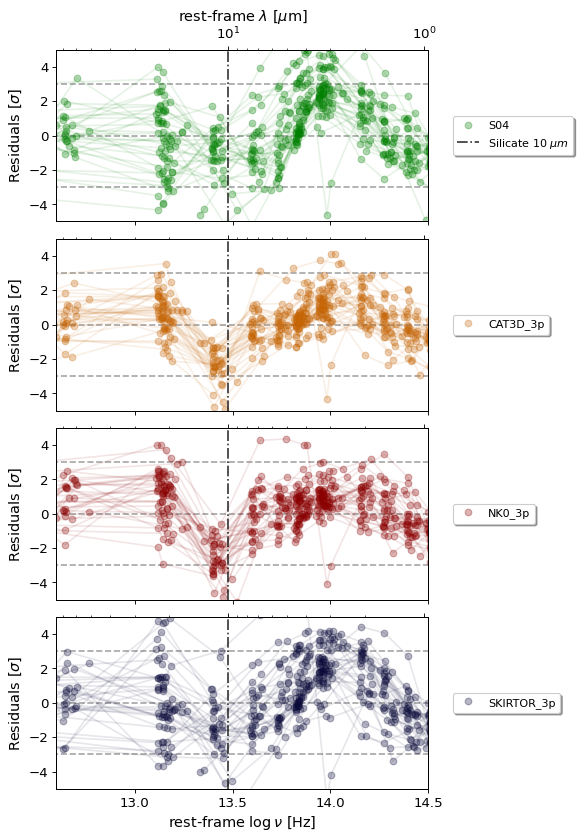}
    \caption{Performance analysis of different torus models in SED-fitting: the homogeneous S04 (upper panel), the clumpy and windy CAT3D (central upper panel), the clumpy NK08 (central lower panel) and the two-phase SKIRTOR model (lower panel). The left panel shows the histograms of the logarithm of the likelihood values corresponding to each model and the dashed line in each plot indicates the median value of the distribution. Residuals from the best fits for each galaxy of the sample are shown in the right panel. The plot shows the existence of templates that manage to capture some SED features (high density of points around zero) while systematically failing in modeling specific regions of the observed SED (high dispersion). The dash-dotted line indicates the 10 $\mu \rm m$ feature and the dashed lines the limits of $0 \sigma$, $3\sigma$ and $-3\sigma$. }
    \label{fig: TorusM_likeli_residuals}
\end{figure*}

The first indicator that we use to compare the torus emission models is the likelihood, which is a measurement of the goodness of the fit. The likelihood distributions for $100$ combinations of parameters randomly selected from the posterior PDFs (hereafter referred to as realizations) of the fits of the 36 galaxies in the sample are presented in the left panel of Fig. \ref{fig: TorusM_likeli_residuals}, for each of the torus models. We considered both relatively-extended and simplified versions of the models, with 3 and 1 fitting parameters, respectively. Since the likelihood values are on a logarithmic scale, it implies all the histograms have a broad distribution. This feature is mainly due to the diversity in the AGN sample and highlights that some SEDs are better modeled than others. 

The panels of Fig. \ref{fig: TorusM_likeli_residuals} show how increasing model complexity may not necessarily imply a higher quality of the fit, given our current data. Although the NK08 and SKIRTOR models are driven by the same physical parameters and are equally simplified, NK08 has a very large increase in likelihood between the 1-parameter and 3-parameter models, reaching a difference of the order of $\approx {10}^{40}$. In appendix \ref{fig: TO_alphas} is presented the spectral slope space covered by our torus models given by $\alpha = -\log(F_\nu (\lambda_2)/F_\nu(\lambda_1))/\log(\lambda_2 / \lambda_1)$  with $\alpha_{\text{MIR}}$ and $\alpha_{\text{NIR}}$ corresponding to $[\lambda_2, \lambda_1]$ equal to $[14, 8]$ $\mu \rm m$ and $[6, 3]$ $\mu \rm m$; and the estimates of the rest-frame NIR and MIR spectral slope of our sample of AGNs based on Spitzer and WISE photometry. The plot shows how similar the areas covered by SKIRTOR with $3$ and $1$ parameters are in contrast to the different regions covered by NK08 with $3$ and $1$ parameters. This discrepancy in the $\alpha_{\text{MIR}}$-$\alpha_{\text{NIR}}$ space might be responsible for the considerable likelihood increase from 1-parameter to 3-parameter templates. Additionally, NK08 1-parameter covers a lower range of $\alpha_{\text{MIR}}$ compared to S04 and SKIRTOR 1-parameter model, while most of the sources of our sample spread in wide range of low $\alpha_{\text{MIR}}$ values which explains why the average NK08 is not performing so well. Consistently with the likelihood analysis, we found CAT3D and NK08 3-parameter are the closest models to the spectral slope space covered by our sample.

The histograms show that the more complex NK08 model results in the best overall distribution of fits, while the simplified version fails to capture important features of the SEDs. The set of templates in the simplified NK08 model does not exhibit the characteristic silicate 10 $\mu \rm m$ feature in absorption, only as an emission line, which changes its strength with the inclination angle of the torus. Meanwhile, the simplified version of the two-phase model SKIRTOR presents a transition between emission and absorption of the 10 $\mu \rm m$ line allowing to fit type 1, type 2 and intermediate AGNs such as those found in our sample. It raises the possibility of exploring another procedure for averaging NK08 SEDs in the future that yields a wide variety of silicate $10$ $\mu \rm m$ features, likely the main driver of $\alpha_{\text{MIR}}$.

The inclusion of the polar wind component improves the quality of the fit, as can be seen in the central upper panel with the CAT3D model. 
The likelihood distribution is one of the narrowest and the one with the highest likelihood values. $25$ out of $36$ active galaxies achieved the maximum likelihood with CAT3D as the torus model in the fitting. These results suggest that using CAT3D with 3 parameters (i.e. incl, a, fwd) to model the torus gives the best result on average if we have a large enough number of photometric bands, and SKIRTOR with 1 parameter (i.e. incl) otherwise. Given that our sample comprises predominantly Sy1 AGNs ($\sim 83 \%$), this finding is in agreement with that reported by \cite{garcia2022torus, gonzalez2019exploring} where the clumpy disk + wind model produces the best fit for Sy1. Favorable physical conditions and viewing angle of type 1 AGNs favor them for detecting IR dusty polar outflows.
Similarly, recent SED studies of red quasars at intermediate redshifts \citep{calistro2021multiwavelength} suggest that the presence of dusty winds could potentially explain the IR excess at $2-5$ $\mu \rm m$ in QSO SEDs. 
Specifically, the NIR bump can be explained by the presence of a hot graphite dust heated by the AGN \citep{garcia2017infrared}, reaching temperatures up to $1900$ K and thus producing emission peaking in the NIR. However, alternative scenarios such as a contribution from direct emission from the accretion disk \citep{hernan2016near, gonzalez2019exploring}, host galaxy or a compact disk of host dust \citep{honig2013dust, tristram2014dusty} have been proposed. 

In the right panel of Fig. \ref{fig: TorusM_likeli_residuals} we plot the combined residuals estimated as the difference between the observed data and the resulting models of SED fitting as a function of frequency for the complete sample.
The purpose of this exercise is to search for general structure and trends for the different hot dust models which could inform us of their advantages and limitations in modelling specific frequencies of the IR regime.
Around 10 $\mu$m ($\log {\nu}{\rm Hz} \sim 13.5$), NK08 and CAT3D present negative residuals which means that most of the time, templates with emission features were fitted while the data suggest absorption features or no features, that is, the predicted fluxes are consistently overestimated by the models. This limitation may have effects on the uncertainties in the estimated inclination angle. 

Moving towards shorter wavelengths, a positive bump is observed which implies an underestimation of fluxes between approximately 1.5 $\mu \rm m$ and 5 $\mu \rm m$ ($\log \frac{\nu}{\rm Hz} \sim 14$). In S04 the bump has the highest amplitude meaning the highest overestimation, while in CAT3D and NK08 with 3 parameters the lowest ones. The behaviour is consistent with the likelihood distributions and, due to the high amount of photometric bands that lie in this region, this bump largely determines the quality of the fit. This infrared excess, which is not recovered by models, has been reported previously in quasars and Seyfert galaxies \citep{mor2009dusty, burtscher2015obscuration, temple2021exploring}, and according to some studies it could be attributed to hot dusty winds not considered in most torus models \citep{calistro2021multiwavelength, honig2019redefining}.  This bump can potentially affect the properties of the host galaxies, making them brighter and redder than they would otherwise be.

Alternative analysis as the Akaike Information Criterion (AIC), founded on information theory, balances between fit quality and the complexity of the model by including an over-fitting penalisation term. AIC measures the loss of information in our understanding of the process that generates the data.  The criteria are useful in the current comparison to assess if fitting simplified or complex torus models has a significant statistical effect. The model with the least loss of information, i.e. the lowest AIC, is favored. It is defined by the number of model parameters (k) and the maximum likelihood ($\mathcal{L}$) as follows:

\begin{equation}
    \hspace{30mm}
    \rm AIC = - 2 \ln \mathcal{L} - 2k
\label{eq: AIC}
\end{equation}

In table \ref{table_torus_model_comp} we present the median of the distribution of maximum likelihoods for our sample of AGNs, its corresponding median logarithmic evidence, the median logarithmic posterior odds and the median AIC criteria. As observed in likelihood histograms of the realizations, the CAT3D model is the most favorable according to its high likelihood, high evidence and low AIC values. Since the evidence corresponds to a likelihood function integrated over all parameter space, it quantifies how well the high diversity of CAT3D templates in general explains the observed data.  As the evidence increases, the posterior probability decreases, which means that, within the enormous possible combination of parameters, the specific producing the maximum likelihood is probably not the only one. The previous might be a consequence of potential model degeneracy which results in a slightly lower posterior probability ($46.54 \%$) for CAT3D compared to the highest ($50.07 \%$) value. 

\begin{table}[ht!]
\begin{tabular}{lllll}
\hline
Model       & ln $\mathcal{L}$ & ln Z    & log P(M|D) & AIC    \\
\hline
S04       & -48.63 & -96.65 & 50.07 $\%$ & 257.94 \\
NK08\_1p   & -76.93 & -124.16 & 48.82$\%$ & 388.26\\
SKIRTOR\_1p & -47.9 & -96.76 & 47.39$\%$ & 254.6 \\
NK08\_3p & -37.86 & -87.82 & 49.86$\%$ & 212.35 \\
SKIRTOR\_3p & -50.88 & -100.24 & 47.79$\%$ & 272.3 \\
CAT3D\_3p   & -34.38 & -80.83 & 46.54$\%$ & 196.34\\
\hline
\end{tabular}
\caption{Results of the torus model comparison based on the median values of different criteria:  the maximum logarithm of the likelihood, the logarithm of the evidence, the posterior probability and the Akaike information criterion (AIC), computed by assuming all models have the same prior probability.}
\label{table_torus_model_comp}
\end{table}

S04 presents the maximum mean posterior odds with $\sim 4 \%$ higher probability than CAT3D, despite the lower likelihood values. Fig. \ref{fig: TO_alphas} shows how the slope space covered by S04 is limited to a well-defined track, providing a limited variety of templates. Therefore, the S04 template producing the maximum likelihood, although it does not produce a high-quality fit, is the one that does the best within the set of possible S04 templates, i.e. its combination of parameters is the most likely. 

In all the cases the Bayes factor $\rm B_{ij} = \rm Z_i/\rm Z_j$ give a factor of more than $10^{2}$ with strong evidence of the CAT3D model ($\log \rm Z_i = -80.83$) concerning the other models ($\log \rm Z_j$) according to the \cite{jeffreys1961international} scale. For the 1-parameter torus models the analysis is not conclusive since SKIRTOR presents a slightly higher likelihood and AIC, but lower evidence and posterior than S04. However, the Bayes factor $\rm B_{SKIRTOR, S04} = 10^{-0.11} \approx 0.78$ means a weak evidence supporting S04. Finally, it is worth highlighting that the increase in the parameter number in NK08 and SKIRTOR seems to have a subdominant effect on AIC compared to the likelihood improvement.

\subsection{Comparing accretion disk models}\label{subsec: comp disk}

\begin{figure*}[ht!]
    \centering
    \includegraphics[width = 0.48\linewidth]{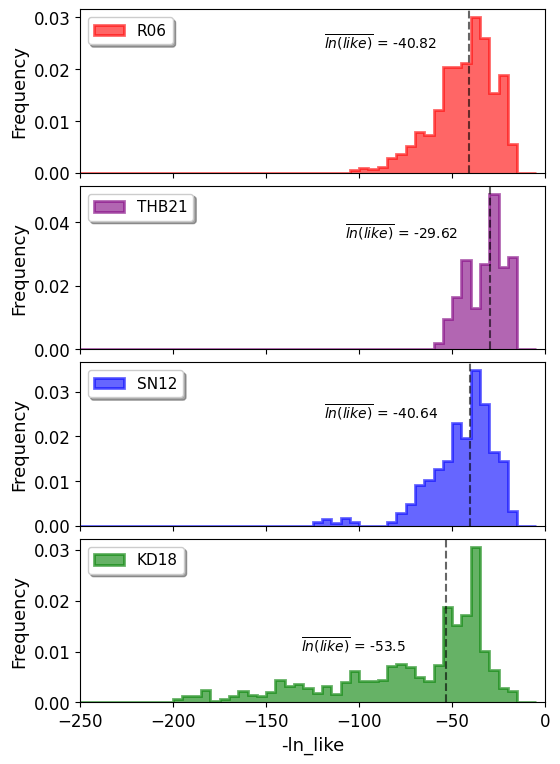}
    \includegraphics[width = 0.44\linewidth]{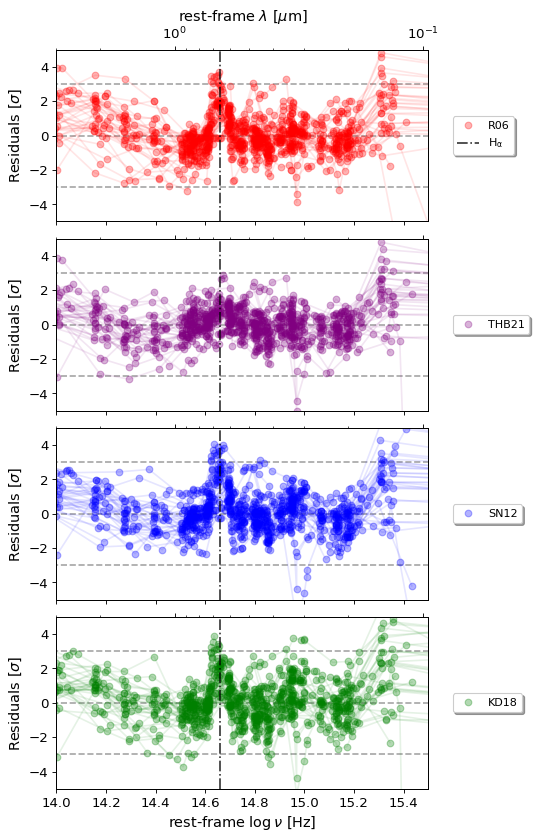}
    \caption{Performance analysis of different accretion disk models in SED-fitting: the semi-empirical R06 (upper panel), the semi-empirical with emission lines THB21 (central upper panel), the theoretical $\alpha$-disk SN12 (central lower panel) and the theoretical 3-component KD18 model (lower panel). The left panel shows the histograms of the logarithm of the likelihood values corresponding to each model and the dashed line in each plot indicates the median value of the distribution. Residuals from the best fits for each active galaxy of the sample are shown in the right panel. The effect of overlapped emission lines, such as the doublet of [\ion{N}{ii}] $\lambda\lambda$ 6549,6585 \AA \hspace{1mm} and  \ion{H}{$\alpha$} $\lambda$ 6563 \AA, is large, reducing residuals to about zero.
    The dash-dotted line indicates the 0.65 $\mu \rm m$ broad emission lines feature and the dashed lines the limits of $0 \sigma$, $3\sigma$ and $-3\sigma$.}
    \label{fig: BB_likelihood}
\end{figure*}

Analogously to the previous comparison, we started by studying the likelihood distributions in the left panel of Fig. \ref{fig: BB_likelihood}. In this case, the semi-empirical models R06 and THB21 reached higher likelihood values and narrower distributions than the theoretical models SN12 and KD18. THB21 produces the best fitting and is the unique model that considers emission lines in the spectrum. This result is informative since it shows that high-resolution spectral features of the broad AGN emission lines can have a high impact on the photometry \citep[e.g., ][]{schaerer09, shim11, marshall2022fresh}. 
In fact, the mean value of the likelihood distribution differs by a factor of $\approx 10^{10}$ from that of R06 even though both models were calibrated following a similar methodology with quasars and the redshift range of the data sets is the same. 
The limitation of these semi-empirical models, in comparison to the theoretical models discussed below, is that they do not provide any estimates of the physical properties of the black hole and accretion disk beyond SED luminosity or reddening.

Conversely, theoretical models have more difficulties in reproducing the shape of SEDs and this may be due to the limitations we still have to fully understand the emission of accretion disks. It is interesting to study if, despite the lower likelihood values, these models provide acceptable estimates for example of the black hole masses (see Section \ref{subsec:bhcomparison}).  In this way, we will be able to evaluate in which cases it is convenient for the user to select one type of model or another, given their aimed scientific question.

In appendix \ref{fig: BB_alphas} is presented the spectral slope space covered by our accretion disk models given by $\alpha = -\log(F_\nu (\lambda_2)/F_\nu(\lambda_1))/\log(\lambda_2 / \lambda_1)$  with $\alpha_{\text{ox}}$ and $\alpha_{\text{B-V}}$ corresponding to $[\lambda_2, \lambda_1]$ equal to $[2500, 8.21]$ $\AA$ and $[4420, 5400]$ $\AA$; and the estimates of the rest-frame spectral slopes of our sample of AGNs based on GALEX, SWIFT and XMM-Newton photometry. The plot shows how the SN12 model using the flexible $\alpha_{\text{ox}}- \textit{L}_{2500\AA}$ nicely overlaps our sources with both low $\alpha_{\text{ox}}$ and low $\alpha_{\text{B-V}}$ while KD18 recovers some sources with $\alpha_{\text{B-V}}$. For our sample in particular, the parameter space covered by KD18 with BH mass and accretion rate as free parameters is not enough to mimic the spectral emission of some sources. Almost half of the sources outside the KD18 parameter space in Fig.\ref{fig: BB_alphas} account for the extended low probability tail of the probability distribution in Fig.\ref{fig: BB_likelihood}. The lack of an X-ray obscuration prescription may be playing a role here, since absorption produces changes in the observed spectral slope, while the KD18 model only accounts for intrinsic photon index changes driven by the accretion rate.

The residuals in the right panel of Fig. \ref{fig: BB_likelihood} show a pronounced and narrow peak around $\lambda \approx 0.7 \hspace{1mm}\mu$m in all the models except THB21. This underestimation of the flux appears to be caused by the overlap of several emission lines such as the doublet of [\ion{N}{ii}] $\lambda\lambda$ 6549,6585 \AA \hspace{1mm} and  \ion{H}{$\alpha$} $\lambda$ 6563 \AA, as suggested in previous works \citep[e.g., ][]{schaerer09, shim11}. In support of this, the errors in the THB21 model, which accounts for the emission lines, are very close to 0 and less dispersion of the points is observed, which is evidence of a systematic good fit. 
Besides, the FUV region contains a bump in all the models resulting from poor modeling of the soft excess as a consequence of the little good data in this region. 

For a proper model comparison, we calculated the posterior odds (equation \ref{eq: posterior_odds}) and the AIC criteria (equation \ref{eq: AIC}) as in the torus model comparison presented in the previous section. Table \ref{table_bb_model_comp} summarizes the median distribution of maximum likelihoods for our sample of AGNs, its corresponding median logarithmic evidence, the median logarithmic posterior odds and the median AIC criteria. In this case, all the evaluated criteria agree with the likelihood distribution of the realizations (Fig. \ref{fig: BB_likelihood}) in pointing to THB21 as the preferred accretion disk model. Despite R06 and THB21 being single templates without fitting parameters other than the normalization, reddening and the same X-ray $\alpha_{\rm ox}$ recipe; the effect of the broad emission lines on the likelihood and evidence is significant. The Bayes factor of $\approx 10^{5.1}$ also represents strong evidence of THB21 when compared with R06.

The theoretical SN12 model has similar results as R06 with just a smaller posterior but the Bayes factor $10^{0.24} \approx 1.74$ suggests very strong evidence over R06. The AIC values show that for a similar fit quality, the increase of 1 fitting parameter of SN12 compared to R06 is not critical. Although the complexity of the KD18, both likelihood and evidence suggest templates are not diverse enough to produce high-quality fits, in particular at the X-rays regime. The inclusion of a third free parameter such as the spin, the spectral index of the warm Comptonisation component or the dissipated luminosity of the corona can improve the performance.

\begin{table}[ht!]
\centering
\begin{tabular}{lllll}
\hline
Model & log $\mathcal{L}$ & log Z   & log P(M|D) & AIC    \\
\hline
R06   & -34.70        & -81.06 & 48.27 $\%$  & 197.81 \\
THB21 & -26.14        & -75.96 & 48.73 $\%$  & 158.4 \\
SN12  & -34.21        & -80.82 & 45.62 $\%$  & 197.54 \\
KD18  & -48.99        & -100.49 & 47.66 $\%$  & 265.61\\
\hline
\end{tabular}
\caption{Results of the accretion disk model comparison based on the median values of different criteria:  the maximum logarithm of the likelihood, the logarithm of the evidence, the posterior probability and the Akaike information criterion (AIC), computed by assuming all models have the same prior probability.}
\label{table_bb_model_comp}
\end{table}

\subsection{Comparing model combinations}\label{subsec: comp model comb}

The best-fit results are produced using combinations of CAT3D and THB21 models for $67 \%$ of the sources. When analyzing type 1 and 2 AGNs independently, this success rate translates into $71\%$ and $40\%$, respectively. However, there is not a sufficiently large sample of type 2 AGNs to obtain conclusive results. The $67\%$ reaches a value of $81 \%$ when also considering combinations of CAT3D with other disk models such as R06, SN12 or KD18. The number of parameters required for these best fits is $19$ consisting of $5$ parameters associated with stellar, $3$ with cold-dust, $4$ with torus, $2$ with accretion disk, $2$ with X-ray and $3$ with radio AGN models. This is not a problem for our data set as we have $49$ photometric bands with a minimum of $40$ valid data points allowing convergence of the live points exploring the parameter space. However, if poorer sampling of the SEDs is available, then the simplified versions of the models might be required.

Current and upcoming surveys, such as JWST surveys and the Vera C. Rubin Observatory Legacy Survey of Space and Time (LSST), are poised to effectively sample the critical frequency ranges found in the residuals for the different emission models. The short and long wavelength channels of NIRCam on JWST are particularly well-suited for in-depth studies of the overlapping doublet of [\ion{N}{ii}] and \ion{H}{$\alpha$} emission lines in AGNs up to z $\sim 6.5$, as well as the torus NIR excess ( $3-4$ $\mu$m restframe) in local AGNs. Additionally, MIRI imaging filters on JWST are sensitive to the crucial MIR torus peak and the silicate line in low-z AGNs, and they also capture the NIR excess in AGNs at wide redshift ranging up to z$~6$. The r, i, z, and y bands of LSST will enable the detection of the emission lines feature for AGNs at z $< 0.5$. These capabilities highlight the significant potential of current and near-future surveys to probe our torus and accretion disk models in the high redshift Universe.

\begin{figure*}[ht!]
    \centering
    \includegraphics[width = 18.0 cm]{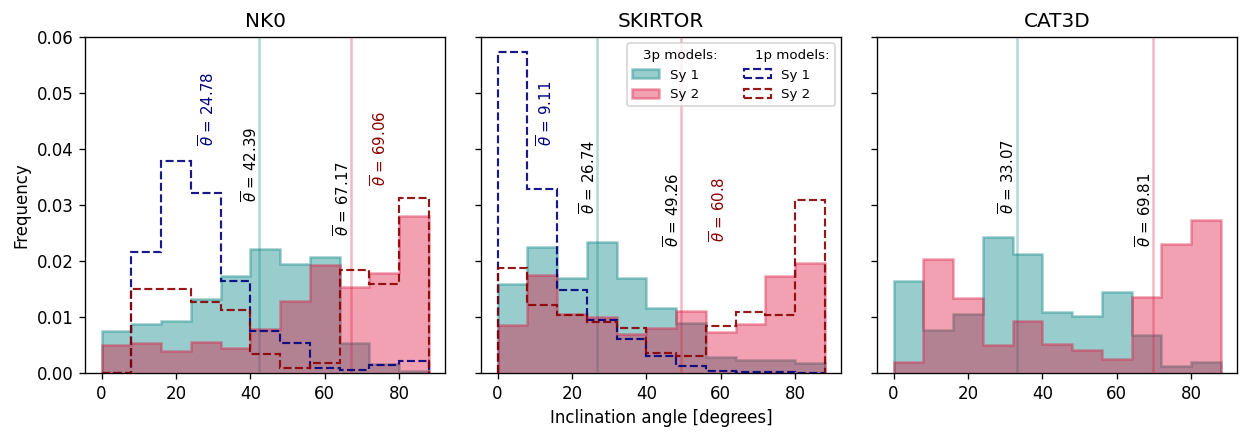}
    \caption{Normalized histogram of the total distribution of inclination angle estimates by the NK08 (left panel), SKIRTOR (central panel) and CAT3D (right panel) model for the 36 galaxies of the sample classified as Seyfert 1 (light blue) and Seyfert 2 (magenta). The dotted contours represent the distribution of the angles for the 1-parameter while the solid histograms for the 3-parameter models. The solid vertical lines highlight the median values of the distributions.}
    \label{fig: TorusM_theta}
\end{figure*}

When analyzing the morphology of the SEDs from Figs. \ref{fig: SEDfitting_1} and \ref{fig: SEDfitting2} we can see how blazars present a highly decreasing spectrum in X-rays with $\Gamma$ values around $-2.69$ and a high-energy emission of the accretion disk and the radio jet.  Seyfert 1 AGNs present a high energy emission of the disk too but a lower energy radio emission and a less steep X-ray power law ($\Gamma \approx -2$). On the contrary, in Seyfert 1.5 and 2 we observe a nuclear hot dust luminosity comparable to or slightly higher than that of the disk. 

An important aspect we found evidence for, is how the X-ray fluxes help to break the degeneracy between the galaxy and AGN contribution to the UV, as can be seen for example in the SED fit of Ark 564. 
It can also be noted that despite the lack of information in the FIR in some cases such as Mrk 290 and Ark 564, it is possible to roughly estimate the emission of cold dust using the energy balance prior to the stellar emission.

\section{Discussion}\label{sec:discussion}

%Doore, F/error match S/N of the spectra, calibration uncertainties given by the corresponding instrument user handbook + 10% systematic effects
%Leja minimum 5% error in the photometry to reflect systematics in models and systematic effects in the  measured photometry
The spectrum-like quality of this SED catalog results in extremely small photometric uncertainties as can be seen in Figs. \ref{fig: SEDfitting_1}, \ref{fig: SEDfitting2} and \ref{fig: SEDfitting3} to \ref{fig: SEDfitting9}. Despite point spread function and point source coincidence loss corrections were applied to observed photometry used to create the synthetic SEDs in \cite{brown19}, no additional photometric calibration uncertainties were added to the final SEDs. A flux-to-error value matching the signal-to-noise ratio of the spectra, calibration uncertainties given by the instruments user handbook and/or fixed percentage errors can be incorporated to account for systematics in the models and instrumental effects (as in e.g. \cite{leja2019measure} and \cite{doore2021impact}). Including the systematic effects can mitigate the significant under-fitting in some bands for even the best-fitting models. Furthermore, the much larger statistical weight of optical-radio data in the SEDs may dilute the contribution of the X-rays to the fit. To assess the potential consequences of the previous, we modified the X-ray uncertainties to equally weight the less energetic band by following the \cite{zou2022spectral} prescription and we performed again the SED fittings. Although the change in the likelihood values was expected due to its dependence on the photometric errors, no significant differences were found in the inferred physical parameters or the contribution of the AGN and host galaxy emission components. Then, the potential main drivers of the underfitting of our sample are systematic effects and AGN variability.

\subsection{Current advances and limitations on the modelling of the torus}
\label{subsec:toruscomparison}

Previously, we discussed the clear bump we found in the residuals at $1.5-5$ $\mu \rm m$  and how the CAT3D model reduces most of that feature, meaning that dusty winds might be a required element for $\sim 70\%$ of the sources. Beyond the quality of the fits achieved by the different models, we also wanted to evaluate the reliability and consistency of the inferred physical parameters.

As a first test, we compare the general trend of the inferred inclination angle parameter found by fitting different torus models to the SEDs of our sources with spectroscopic classifications \citep{veron2010catalogue}. For simplicity, we group Blazars, Sy1, Sy1.2, Sy1.5 and Sy1n as Seyfert 1; and Sy1i, Sy2, Sy1.9 and Sy3 as Seyfert 2. 

In Fig. \ref{fig: TorusM_theta} are shown the overall distributions of the inclination angle obtained for the $31$ AGNs in our sample classified as Seyfert 1 and the $5$ as Seyfert 2.
For this, we evaluate the performance of the CAT3D models and the simplified and complex versions of NK08 and SKIRTOR.
For all the 3-parameter models, there is a difference of more than $22 ^\circ$  in the median values of Sy1 and Sy2 distributions, which means that in principle we can use the inclination angle as a classifier of AGNs. As the distributions overlap, some Sy1 with large inclination angles and a few Sy2 with small angles, this classification would have to be done with caution. The Sy2 distributions for 3-parameter models look quite wide probably due to the low number of AGNs. % and 
In particular, this distribution looks almost flat for the SKIRTOR model. 
Furthermore, it can be noted that SKIRTOR tends to find smaller angles for both Sy1 and Sy2 compared to the other models.

When using the 1-parameter version of the  NK08 and SKIRTOR models, the estimates of the inclination angle for Sy1 present narrower distributions centred at significantly smaller angles than the 3-parameter models. This shows how the simplified version of SKIRTOR manages to find the most physically consistent angles for type 1 AGNs, perhaps by avoiding parameter degeneracies that may be occurring using the more complex version of this torus model. Instead, for the AGN type 2 population, the picture does not improve, as simplified versions of the models result in distribution histograms with very pronounced bimodalities with extreme peaks at around $10$ and $80$ degrees. Hence, the choice of the most suitable torus model to use depends on the properties of the AGNs as nuclear hydrogen column density, in accordance with what was reported in \cite{garcia2022torus}.

Next, we analyse the self-consistency of models from a geometrical point of view, similar to a comparison done by \cite{cerqueira2023coronal}. 
As proposed by the unified model \citep{antonucci1985spectropolarimetry}, the key spectral features of the diverse population of AGNs could be explained by the interaction of the light emitted by the active nuclei with the distribution of hot nuclear dust. Thus, the angle between the observer's line of sight and the axis of the torus would determine the classification, being Type I and Type II AGNs the two endpoints of unobscured and highly obscured spectra. Comparing the inclination angle with the opening angle of the torus allows us to see whether indeed we reconstruct an edge-on or face-on orientation. 

\begin{figure}[ht!]
    \centering
    \includegraphics[width = 9.0 cm]{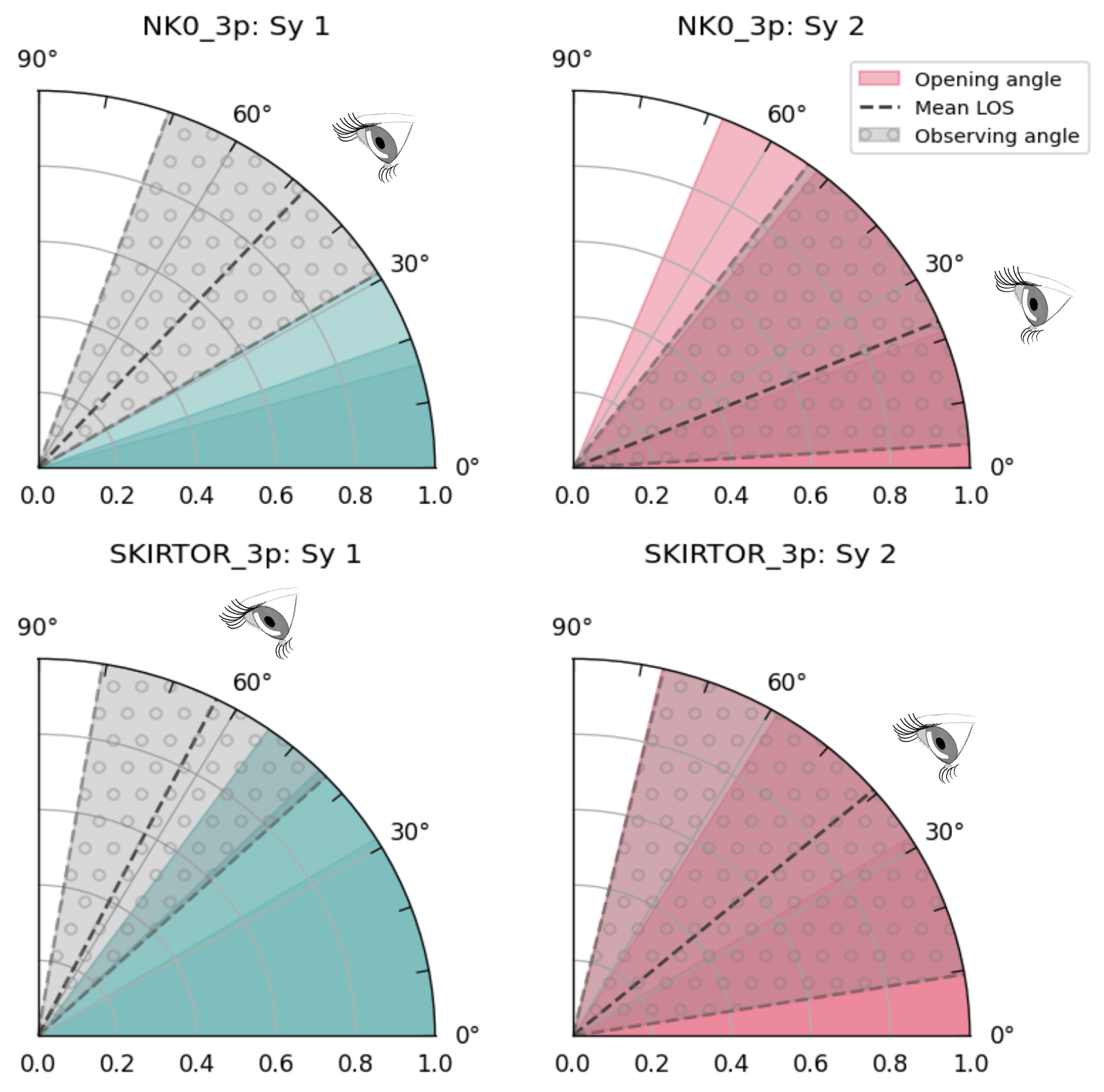}
    \caption{Representation of the upper right quadrant of the cross-section of the toroidal distribution of hot nuclear dust in NK08 (upper panels) and SKIRTOR (lower panels) models. The shaded color areas with decreasing transparency represent the 16th, 50th and 84th percentiles of the torus opening angle in Seyfert 1 (blue) and Seyfert 2 (magenta) AGNs. While the black dotted lines represent the median values of the viewing angle and the dotted gray shaded areas are the 16th and 84th percentiles.}
    \label{fig: TorusM_theta_oa}
\end{figure}

As shown in Fig. \ref{fig: TorusM_theta_oa}, the line of sights in Sy1 AGNs, estimated by the inclination angle parameter,  are outside the toroidal distribution of hot dust for both the NK08 and SKIRTOR models. Even the 16th and 84th percentiles in the distribution of viewing angles do not intercept or have little overlap with dust distribution given by the torus opening angle. Conversely, Sy2 AGNs present distributions of the line of sight that completely intercept the torus for both models. It is outstanding how the SED fitting recovers torus aperture angles and uncertainties in the estimation of the observing angle that is consistently larger for Sy2 than for Sy1, which may be potentially associated with lower luminosity AGN \citep[][]{wada2012radiation}. SKIRTOR is more likely to result in smaller inclinations and larger aperture angles compared to NK08.

Another key physical feature which can be inferred from SED fitting is the estimation of the AGN contribution to the total luminosity in a certain energy range (AGN fraction), such as IR from $8$ to $1000$ $\mu \rm m$ or optical from $400$ to $500$ nm. In Fig. \ref{fig: TOfrac_IR_violin} are presented the distributions of IR AGN fractions found by using each of the torus models in the fitting for each of the sources. With gray backgrounds, the galaxies lack SPIRE and PACS bands, crucial to constrain the emission of the galaxy star-forming regions. 
It can be seen in Fig. \ref{fig: TOfrac_IR_violin} how the uncertainties in the AGN fraction increase without FIR data.
In these cases, \textsc{AGNfitter-rx} can fit high and negligible contributions from cold dust affecting the estimation of the AGN fraction regardless of the choice of torus model. As a consequence, the violin plots of the mentioned galaxies are extended and are out of alignment, with different distributions of AGN fraction values. Since we used a flexible energy balance prior to running the fits, the attenuated stellar emission sets only a lower limit to the cold dust emission luminosity. Nevertheless, a more restrictive energy balance would have raised probabilities of models with a cold dust luminosity equal to the attenuated stellar component, largely constraining the IR AGN fraction and isolating the effect of the torus models.
As previously stated, there are special cases in which the energy balance is broken so the decision of whether to include it or not, as well as the severity of the prior, is left to the user.
For the active galaxies with available IR data, the change of hot dust model doesn't affect the AGN IR luminosity.

\begin{figure}%[ht]
    \centering
    \includegraphics[width = 9.0 cm]{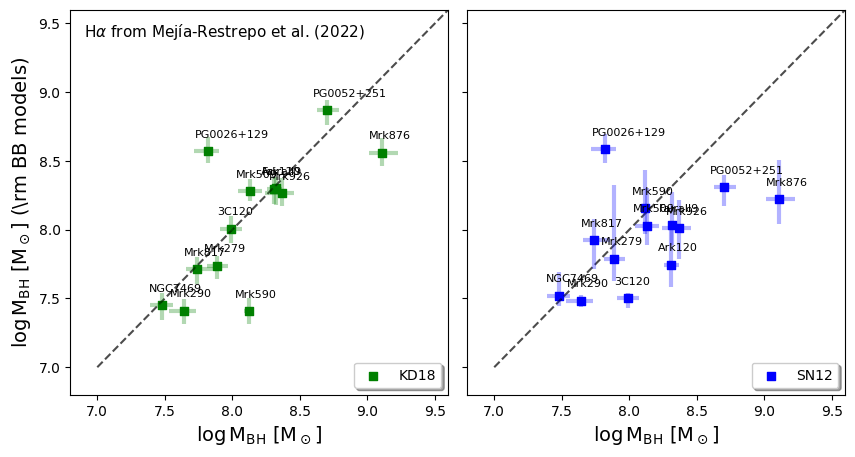}
    \caption{Comparison of the log-scale black hole mass estimates made via \textsc{AGNfitter-rx} using KD18 and SN12 accretion disk models (y-axis) and broad H$\alpha$ emission line (x-axis) for $17$ AGNs in our sample with available information in BASS DR2 catalog. The dotted line corresponds to the 1-to-1 relation.}
    \label{fig: BB_masses}
\end{figure}

\subsection{Current advances and limitations on accretion disk models} \label{subsec:bhcomparison}

%\gaby{Checkear Davis\&Laor+11}
An important decision when choosing the accretion disk model for the SED fitting is whether to opt for semi-empirical or theoretical models.
While the semi-empirical models have likelihood values potentially better than the latter (as THB21 in Figure \ref{fig: BB_likelihood}), theoretical models can provide estimates of important physical parameters such as the black hole mass and accretion rate.
Using \textsc{AGNfitter-rx}, black hole masses can be estimated from the rest-frame UV region of the observed SED only by applying SN12 or KD18 models. 
To evaluate the confidence of the values, in Figure \ref{fig: BB_masses} we compare black hole masses from SED-fitting with virial estimates of black hole masses based on broad emission lines available for $17$ AGNs from the BASS DR2 sample \citep{mejia2022bass}. Since those spectroscopic estimates are reliable, we evaluate how the SED fitting estimates using KD18 (left) and SN12 (right) deviate from the 1-1 line.

It can be easily noticed how the KD18 model is in good agreement with the values of the BASS DR2 catalogue.
Out of the $13$ sources, only $3$ are outliers, while $6$ are within the 1 to 1 ratio taking into account uncertainties and $4$ others are off by only a factor of about 0.1 dex.
These $13$ sources are characterised by X-ray information, which is important for fitting the KD18 model extending up to hard X-rays (100 keV). However, as could be seen in Fig. \ref{fig: BB_likelihood}, the likelihood histogram for the KD18 model has a peak around -log(like) $\sim -40$ but a very spread out distribution reaching values of $-200$. A poor fit to stacked SEDs in the optical regime and unpredictable X-ray fluxes by up to a factor of $2$ were also reported in \cite{mitchell2023soux} for the $\log \text{M}_{\text{BH}}$ regime between $7.5-9 \text{ M}_{\odot}$. One possible explanation for this is that the continuum optical-to-X-rays templates require an observed SED in a narrow time window that ensures the variability is negligible compared to the photometric errors. As this is not guaranteed, sometimes the KD18 model fits well in the UV while producing high errors in the X-rays. Another alternative could be the X-ray obscuration producing changes in the observed spectral slope which does not correspond to the slope variations driven by the accretion rate in the KD18 templates. Precisely, AGNs with a correct estimate of the black hole mass are those that have a good X-ray fit with errors smaller than $2\sigma$. 

In the evaluation of the SN12 model, we included sources available in BASS DR2 even though their SEDs in our catalogue did not have X-ray photometric bands, which increased our set from $13$ to $17$. The use of the $\alpha_{\rm ox}$ relation combined with a power law with the dispersion of the relation ($\Delta \alpha_{\rm ox}$) and the photon index ($\Gamma$) as free parameters allow great flexibility in fitting X-rays. Therefore, X-ray photometry has a mild impact on the performance of specific accretion disk models. In this case, $8$ sources are clear outliers, $6$ are successful estimates and $3$ are off of the 1-1 relation by a factor of about 0.1 dex. 
The clustered sources around $10^{8} M_\odot$ on the right side of the plot show how SN12 faces difficulties in estimating black hole masses for AGNs with masses greater than $10^{8.5} M_\odot$.

\begin{figure*}[t]%[ht!]
    \centering
    \includegraphics[width = 18.0 cm]{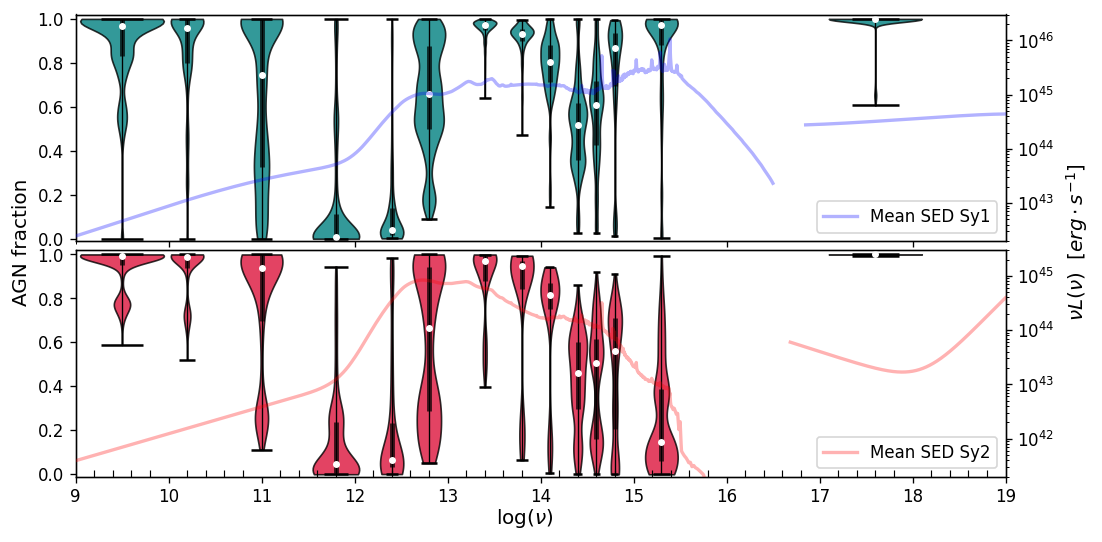}
    \caption{AGN fraction distribution as a function of frequency for Sy1 (upper panel) and Sy2 (lower panel) populations. The white dots within the violin plots highlight the median of the distribution while the black line is the range between the $25$th and $75$th percentile. The blue and red curves present the average radio-to-X-ray SED including BC03_metal, S17, CAT3D, THB21 and combinations of SPL and DPL radio models, for type 1 and type 2 AGNs; respectively.} 
    \label{fig: violin_plot_AGNfrac}
\end{figure*}

It is worth noting that mass measurements from emission lines also have their assumptions and errors as the accretion disk emission models discussed above. The method assumed the dynamic of the gas in the broad line region is virialized, meaning the motion is produced by the gravitational field of the black hole. As mentioned in \cite{mejia2022bass}, the existence of winds can introduce biases by changing the observed gas dynamics.  Additionally, the geometrical factor assumed to estimate the broad line region size in the empirical relations is a source of systematic errors. The specific value according to the chosen geometry or even the inclination angle, if a disk is assumed, increases the uncertainty.

Theoretical models yield estimates of the accretion rate as well, allowing the calculation of bolometric luminosities with the equation \ref{eq:bolometric}. In Fig. \ref{fig: Lbol_BB} we contrast the bolometric luminosities computed via BH mass and accretion rate (BB model product) with the photometry value obtained by adding the integrated luminosities of the AGN-synchrotron, the torus, the accretion disk and the hot corona components (SED-fitting estimation). KD18 matches the AGN components estimates within the errors whereas SN12 systematically overestimates the bolometric luminosities by a factor $~8$. As our sources have BH masses dominantly in the range $\log \text{M}_{\text{BH}}[\text{M}_{\odot}] = 7.5-9$, we are in the parameter space where KD18 templates have demonstrated to be successful in predicting BH mass and $\textit{L}_{2500}$ across different accretion rates \citep{mitchell2023soux}. It can be noticed that: BBB models significantly reddened ($\text{EBV}_{\text{BBB}} > 0.6$) resulting in larger bolometric luminosities than the SED-fitting estimations, in particular with the SN12 model; and the SED-fitting estimations are robust for both models despite of the individual normalization of the AGN components.

Finally, we again assess the effects of the different models, in this case of the accretion disk, on the estimates of the AGN fraction to the optical/UV emission. In Figure \ref{fig: BBfrac_opt_violin2} are presented the distributions of optical AGN fractions (from $4000$ to $5000$ $\AA$) found by using the R06, THB21, SN12 and KD18 models in the fitting. In most of the AGNs, the violin plots of all the models are aligned and even have a similar distribution width. However, within the observed outliers, the most frequent are distributions given by the theoretical models: SN12 and KD18. As shown in \ref{fig:AGNtemplates}, opposite to the empirical models, the theoretical templates have a smooth shape in the big blue bump that hardly fits well the data in the most energetic region of the rest frame UV. The previous might be increasing the emission of the stellar population during the fits to reduce residuals and therefore, raising the contribution of the host galaxy. Also, the availability and shape of X-ray emission influence the fit quality due to the high and low flexibility of SN12 and KD18 while fitting X-rays, respectively.  As discussed previously Fig.\ref{fig: BB_alphas} provides an indicator that the choice of free parameters in KD18 is not sufficient to produce diversity in the templates, especially in X-rays. Further research on the inclusion of relevant parameters such as spin, the spectral index of the warm Comptonisation component shaping the soft X-rays or the dissipated luminosity of the corona is needed. On the semi-empirical side, THB21 predicts a systematically larger host galaxy contribution to the SED of type 1 AGNs than R06.

\subsection{The AGN contribution to the SED from the radio to X-rays}\label{subsec:AGNcontr radio-X-rays}

A key feature of SED fitting is the ability to disentangle the contributions of the AGN and the host galaxies to the overall emission. This contribution can significantly vary across wavelengths. Therefore, defining appropriate frequency regions where AGN fractions should be estimated is crucial. To provide information on this we use the flexibility of \textsc{AGNfitter-rx} in delivering any customary range of AGN fraction provided by the user, and the unique coverage from the radio to the X-ray to compute a spectral AGN fraction distribution as presented in Fig. \ref{fig: violin_plot_AGNfrac}. 

The AGN fraction within a frequency range is estimated as the ratio of the luminosity of the AGN model components, divided by the total observed luminosity (AGN + host galaxy), where the luminosities are estimated by integrating the model templates within the given frequency range.  
The frequency ranges used for the AGN fractions in Fig. \ref{fig: violin_plot_AGNfrac} were chosen qualitatively, based on the photometric coverage in our data set.
The width of the frequency ranges chosen in each case is represented by the maximal width of the violins.
The violin plots in the upper and lower panel of Fig. \ref{fig: violin_plot_AGNfrac} show the density distributions for the combined Sy1 or Sy2 populations, respectively, and illustrate a general trend of how the AGN contribution to the total emission changes as an function of frequency. 
The density distributions were obtained for each source by drawing $100$ realizations from the posterior distributions of the Bayesian fit.
In all cases, the long and thin tails in the distributions (e.g., the 4, 5, 7, 8, 9th violins from the left) are caused by a few outlier realizations from the posterior distributions that encountered models resulting in an AGN fraction distinct from that of the bulk of the realizations for a given source. 

The blue and red curves represent the mean Sy1 and Sy2 SEDs, computed as the average best fit for each population, and are shown in the background for reference with an arbitrary normalization. 
The emission lines of the silicates and the broad line region can be clearly identified, as well as the big blue bump and a steep X-ray spectral slope characteristic of type 1 AGNs. IR-dominated dusty spectra with a mild UV nuclear emission, and a combination of steep and flat X-ray power laws characterize our average type 2 AGNs. Given the size of the Sy2 set, AGN fractions and the average SED might not be representative of the overall type 2 AGN population.

Both Sy1s and Sy2s exhibit dominant AGN emission in radio, MIR and X-ray bands, while emission in FIR bands appears galaxy-dominated. Indeed, this is in agreement with the most reliable methods for AGN selection, which rely on information from these energy ranges: IR-to-radio luminosity ratios \citep{condon1995radio}, radio/FIR correlation \citep{helou1985thermal,donley2005unveiling}, radio excess \citep{yun2001radio}, radio power versus absolute K-band (infrared) magnitude \citep{hardcastle2019radio}, MIR colors \citep{lacy04, stern05, donley12}, X-ray spectral slope \citep{alexander2005x, nandra1994ginga}, X-ray luminosities and X-ray to optical flux ratio \citep{szokoly2004chandra, lehmann2001rosat}, among others. By construction, optical-UV-based selection methods as UV excess are successful in unobscured AGNs \citep{boyle1990catalogue, schmidt1983quasar, wisotzki1996hamburg, verbeek2012first}, and incomplete for dust-obscured AGNs where the UV-optical is dominated by the host galaxy such as the type 2 ones. Consistently, only Sy1s in Fig. \ref{fig: violin_plot_AGNfrac} have a high AGN fraction in UV bands.

Nonetheless, the extension and width of the violin plots evidence the diversity within the AGN populations and also, the degree of uncertainty in the estimate of the AGN fraction. In Sy1s the long and nearly equally wide violin in the FIR (3rd position from the left) is explained by the transition from AGN radio to galactic cold dust as the dominant emission component which occurs at different frequencies depending on the dust temperature and AGN properties. A similar phenomenon takes place at the transition from cold dust to torus (6th violin from the left), and from stellar to accretion disk emissions (4th and 5th violin from the right). 
In the Sy2s, though, the extended UV violin (2nd violin from the right) is a consequence of $3$ of the $5$ AGNs having so reddened accretion disk emission that stellar emission dominates, giving rise to a low AGN fraction, while the remaining sources the contributions are similar or dominated by the disk. We note that we have used the observed accretion disk luminosities (the blue curve in the Figs. \ref{fig: SEDfitting_1}, \ref{fig: SEDfitting2}), not corrected for reddening, into the estimation of the AGN fraction. However, if we use the intrinsic (reddening-corrected) luminosity instead, the UV violin for the Sy2 population would be shifted towards higher values and a shape equivalent to the Sy1 population.

\section{Summary and conclusions}

In this paper, we introduced the new radio-to-X-ray fitting capabilities of the Bayesian Python code \textsc{AGNfitter} \citep{calistrorivera16}, released as \textsc{AGNfitter-rx}.

\textsc{AGNfitter-rx} is designed for consistently fitting the radio-to-X-ray SEDs of active galaxies by sampling the parameter space through MCMC or nested sampling. 
The physical model considers the contributions of the host galaxy, which consist of the emission from stellar populations, cold dust and cosmic ray electrons in star-forming regions; as well as the AGN components, which include the emission of the hot corona, the accretion disk, the hot dust and the radio emission from jets and/or shocks. The code now incorporates a variety of semi-empirical and theoretical emission models, as well as informative priors, including the assumption of the energy balance between dust-attenuated stellar emission and the re-emission by cold dust.

To test the new capabilities of the code, we used \textsc{AGNfitter-rx} to fit 36 radio-to-X-ray SEDs of local active galaxies from the AGN SED ATLAS \citep{brown19} at $z < 0.7$. 
In this study, we selected this sample of well-characterized and diverse AGNs with the main scope of assessing and comparing the performance of state-of-the-art torus and accretion disk emission models. 
We discuss the effects of different models on the fit quality and the most important physical properties as well as the performance of the code by implementing the best model combination for different AGN populations.

\begin{enumerate}
    \item In $\sim 70 \%$ of the studied sources the inclusion of polar winds in torus models produced more than a $3$ order of magnitude advantage in the median of the likelihood distribution, a Bayesian factor higher than $100$ and the lowest AIC values compared to other clumpy torus models.  Considering that the sample is dominated by Sy1 AGNs, this result suggests that the presence of infrared dusty polar outflows is essential in shaping the spectral energy distributions of type 1 AGNs.

    \item By comparing less complex torus template libraries, we find that the inclusion of absorption and emission silicate features likely plays a key role in achieving precise estimates of the inclination angles in most cases.

    \item We analyzed the fitted inclination angle (i.e. the line of sight) and the opening angle parameters of the hot nuclear dusty torus in the Type 1 and 2 AGNs. While the more complex torus models (>3 parameters) revealed overlapping broad inclination angle distributions, their simplified 1-parameter versions yielded more physically consistent results with their spectroscopic classification, potentially alleviating parameter degeneracies inherent in complex torus models. However, the geometrical information obtained from the multiple parameters in the more complex torus models revealed lines of sight outside the toroidal hot dust distribution for Sy1 and lines of sight that fully intercepted the torus for Sy2 AGNs, consistent with the AGN unification model.

    \item In $\approx 67 \%$ of the cases the accretion disk models including emission lines stand out as the optimal choice for the fitting, with its likelihood distribution boasting a median value $>\mathrm 10^{7}$ times higher, a Bayes factor of $10^{5}$ and the lowest AIC values than that of other models. Emission lines are a crucial feature to avoid the narrow peak in the residuals at $0.7 \hspace{1mm}\mu \rm m$, likely stemming from the overlap of the doublet of [\ion{N}{ii}] $\lambda\lambda$ 6549,6585 \AA \hspace{1mm} and  \ion{H}{$\alpha$} $\lambda$ 6563 \AA. 
    While such empirical templates do not offer estimates for key properties such as black hole mass or accretion rate, this result underscores the substantial impact of high-resolution spectral features within broad AGN emission on the photometry, aligning with previous studies. %\citep{schaerer09, shim11}.
    
    \item We conducted a comprehensive comparison between the black hole masses obtained from photometry only via SED fitting and virial estimates based on broad H$\alpha$ emission lines. The theoretical model that incorporates X-ray emission demonstrates strong agreement with virial estimates, aligning within the $0.1$ dex range around the 1-to-1 ratio for $10$ out of $13$ sources. Conversely, models that do not account for X-rays encounter difficulties in estimating black hole masses for AGNs exceeding $10^{8.5}$ $\rm M_\odot$.

    \item We investigate the frequency dependence of estimated AGN fractions across the radio-to-Xray spectrum, finding that Sy1 and Sy2 exhibit dominant AGN emission in radio, MIR and X-ray bands, in agreement with empirical AGN selection strategies. From our probabilistic analysis, we conclude that significant uncertainties in disentangling the AGN and galaxy emission in the IR and optical-UV can arise from the absence of FIR data, degeneracies between the accretion disk and host-galaxy UV emission, and the variable flexibility of models when fitting X-ray data.  

\end{enumerate}

The radio-to-X-ray SED fitting capabilities in \textsc{AGNfitter-rx} are a powerful tool for comprehensive studies of AGN physics. In addition to offering an unsurpassed wavelength coverage, \textsc{AGNfitter-rx} allows the user to easily incorporate customary telescope filters, state-of-the-art theoretical and empirical model templates, as well as informative priors.
\textsc{AGNfitter-rx} is thus a promising and reliable tool for pushing boundaries in the interface between observational and theory communities, allowing them to test, compare and customize models for the increasing data sets of deeper and larger panchromatic surveys.

In particular, \textsc{AGNfitter-rx} was developed to capitalize on the present and future availability of exquisite radio and X-ray data sets, such as wide and deep radio surveys from the Low-Frequency Array \citep[LOFAR;][]{van2013lofar} and the Square Kilometre Array \citep[SKA;][]{mcmullin2020square}, as well as new observational parameter spaces covered by X-ray observatories (e.g., eROSITA, Athena). Crucially, the focus of this investigation is the implementation and testing of infrared AGN models of diverse complexity. While these complex models could be only explored in the nearby Universe till recently, the remarkable sensitivity, resolution and wavelength coverage by the JWST will now allow us to test and use these models at intermediate and high redshifts. 
The rich library for infrared AGN and host galaxy models in \textsc{AGNfitter-rx} is thus an ideal laboratory for modelling JWST/MIRI, JWST/NIRCam, HST and Euclid photometry, to investigate the hot dust emission and physics and geometry of AGN obscuration at high-$z$, including polar dust distributions, and the disentanglement of AGN emission from star formation.

The \textsc{AGNfitter} code \citep{calistrorivera16}, including the \textsc{AGNfitter-rx} code release, are publicly available as open-source code in Python 3 in \url{https://github.com/GabrielaCR/AGNfitter}.

\begin{appendix} %First appendix
\section{Catalog of AGNs fitted}
\begin{table*}
	\centering
	\caption{Sample of 36 local AGNs from SED ATLAS.}
	\label{table:agnsummary}
	\begin{tabular}{lrrcll} 
		\hline
Name & RA (J2000) & DEC (J2000) & $z$ & AGN type & Available bands \\
		\hline
2MASXJ13000533+1632151  & 195.0223 &  16.5374 & 0.0799 & S1i   &  $41$ \\ 
             3C 120   &  68.2962 &   5.3543 & 0.0330 & S1.5  & $47$ \\ 
             3C 273   & 187.2779 &   2.0524 & 0.1583 & S1.0  & $49$ \\ 
             3C 351   & 256.1724 &  60.7418 & 0.3719 & S1.5  & $42$ \\ 
           3C 390.3   & 280.5375 &  79.7714 & 0.0561 & S1.5  &  $42$ \\ 
            Ark 120   &  79.0475 &  -0.1498 & 0.0327 & S1.0  &  $49$ \\ 
            Ark 564   & 340.6639 &  29.7254 & 0.0247 & S3    &  $41$ \\ 
          Fairall 9   &  20.9407 & -58.8058 & 0.0470 & S1.2  &  $43$ \\ 
       F2M1113+1244   & 168.4777 &  12.7443 & 0.6812 & S1     & $57$ \\ 
         H 1821+643   & 275.4888 &  64.3434 & 0.2968 & S1.2  & $42$ \\ 
    IRAS 11119+3257   & 168.6620 &  32.6926 & 0.1876 & S1n   &  $42$ \\ 
   IRAS F16156+0146   & 244.5392 &   1.6559 & 0.1320 & S2    & $56$ \\ 
            Mrk 110   & 141.3035 &  52.2862 & 0.0353 & S1n   & $57$ \\ 
           Mrk 1502   &  13.3956 &  12.6934 & 0.0589 & S1n   & $49$ \\ 
            Mrk 231   & 194.0593 &  56.8737 & 0.0422 & S1.0  & $48$ \\ 
            Mrk 279   & 208.2644 &  69.3082 & 0.0305 & S1.0  & $48$ \\ 
            Mrk 290   & 233.9683 &  57.9026 & 0.0302 & S1.5  & $42$ \\ 
            Mrk 421   & 166.1138 &  38.2088 & 0.0300 & HP    & $43$ \\ 
            Mrk 493   & 239.7901 &  35.0299 & 0.0310 & S1n   &  $41 $ \\ 
            Mrk 509   & 311.0406 & -10.7235 & 0.0344 & S1.5  & $49$ \\ 
            Mrk 590   &  33.6398 &  -0.7667 & 0.0261 & S1.0  & $48$ \\ 
            Mrk 817   & 219.0919 &  58.7943 & 0.0315 & S1.5  & $49$ \\ 
            Mrk 876   & 243.4882 &  65.7193 & 0.1290 & S1.0  & $49$ \\ 
            Mrk 926   & 346.1812 &  -8.6857 & 0.0469 & S1.5  & $47$ \\ 
           NGC 5728   & 220.5996 & -17.2531 & 0.0094 & S1.9  & $43$ \\ 
           NGC 7469   & 345.8151 &   8.8739 & 0.0163 & S1.5  &  $49$ \\ 
             OQ 530   & 214.9441 &  54.3874 & 0.1525 & HP   & $43$ \\ 
        PG 0026+129   &   7.3071 &  13.2677 & 0.1420 & S1.2  &  $41$ \\ 
        PG 0052+251   &  13.7172 &  25.4275 & 0.1545 & S1.2  &  $47$ \\ 
        PG 1211+143   & 183.5736 &  14.0536 & 0.0809 & S1n   &  $43$ \\ 
        PG 1307+085   & 197.4454 &   8.3302 & 0.1538 & S1.2  & $47$ \\ 
        PG 1415+451   & 214.2534 &  44.9351 & 0.1137 & S1.0  & $41$ \\ 
        PG 2349-014   & 357.9838 &  -1.1537 & 0.1738 & S1.2  & $43$ \\ 
        PKS 1345+12   & 206.8892 &  12.2900 & 0.1205 & S2    &  $41$ \\ 
            Ton 951   & 131.9269 &  34.7512 & 0.0640 & S1.0  & $43$ \\ 
              W Com   & 185.3820 &  28.2329 & 0.1020 & BL    &$43$ \\ 
 		\hline
	\end{tabular}
~\\   
$\dagger$ Spectral classifications by \citet{veron2010catalogue}.    
\end{table*}

\section{SED fitting of the remaining sample}

\begin{figure*}[ht!]
\centering
    \includegraphics[trim={0 1.57cm 0 0},clip, width = 0.8\linewidth]{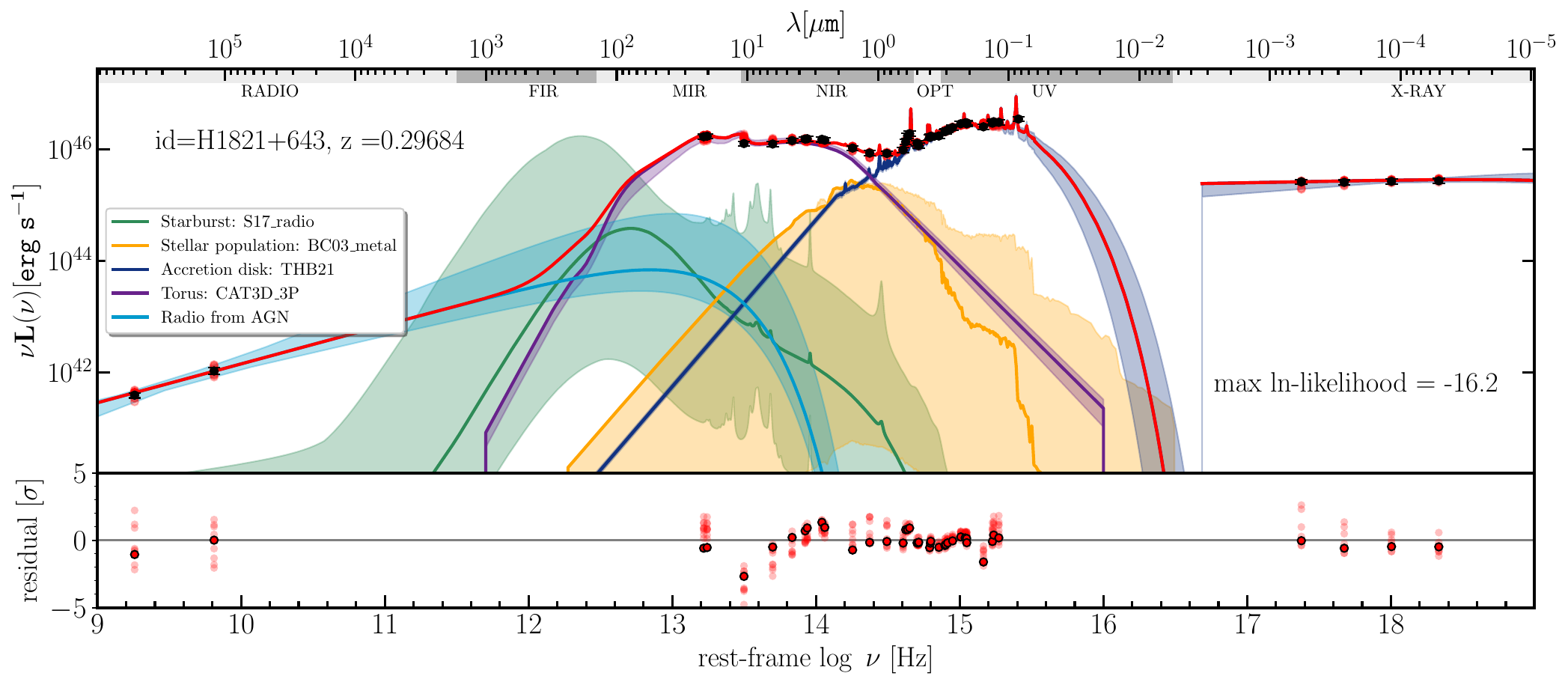}
    \includegraphics[trim={0 1.57cm 0 1.55cm},clip, width = 0.8\linewidth]{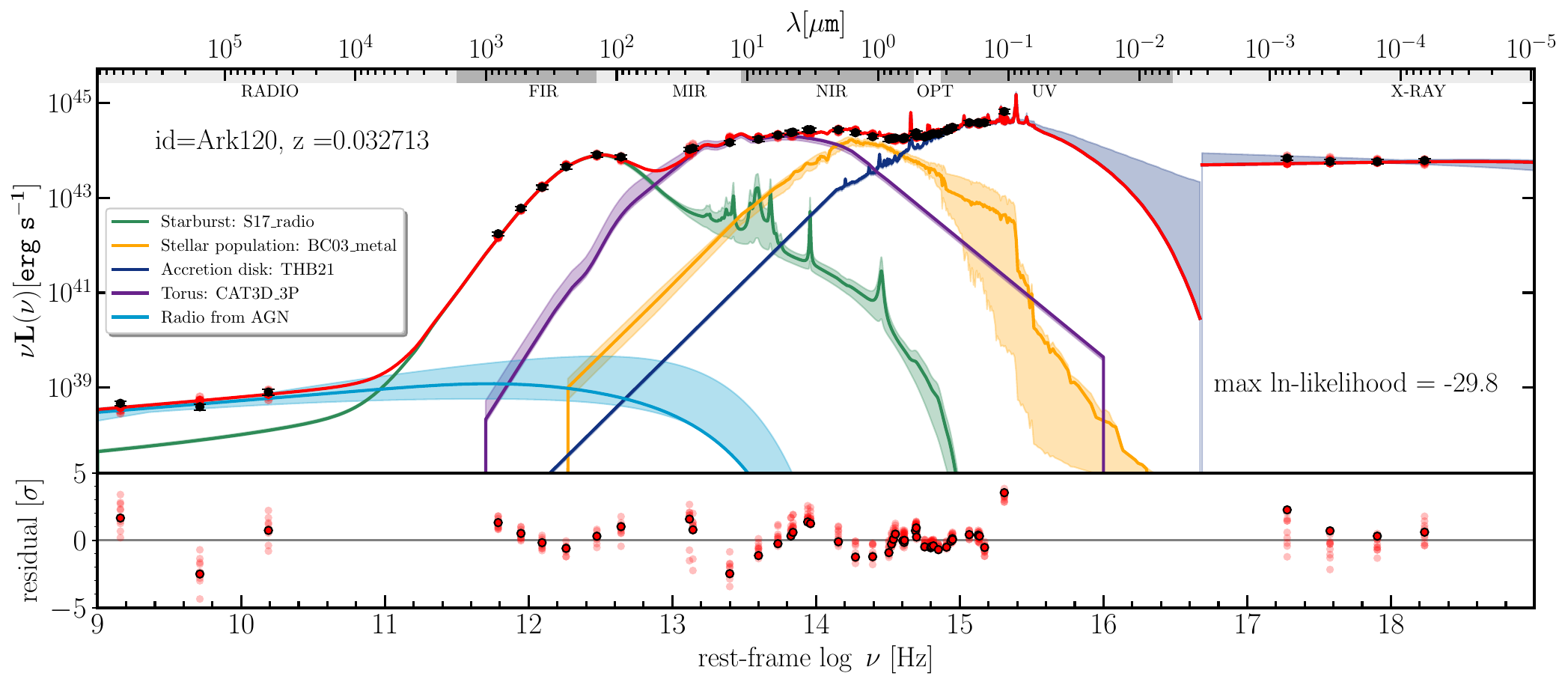}
    \includegraphics[trim={0 1.57cm 0 1.55cm},clip, width = 0.8\linewidth]{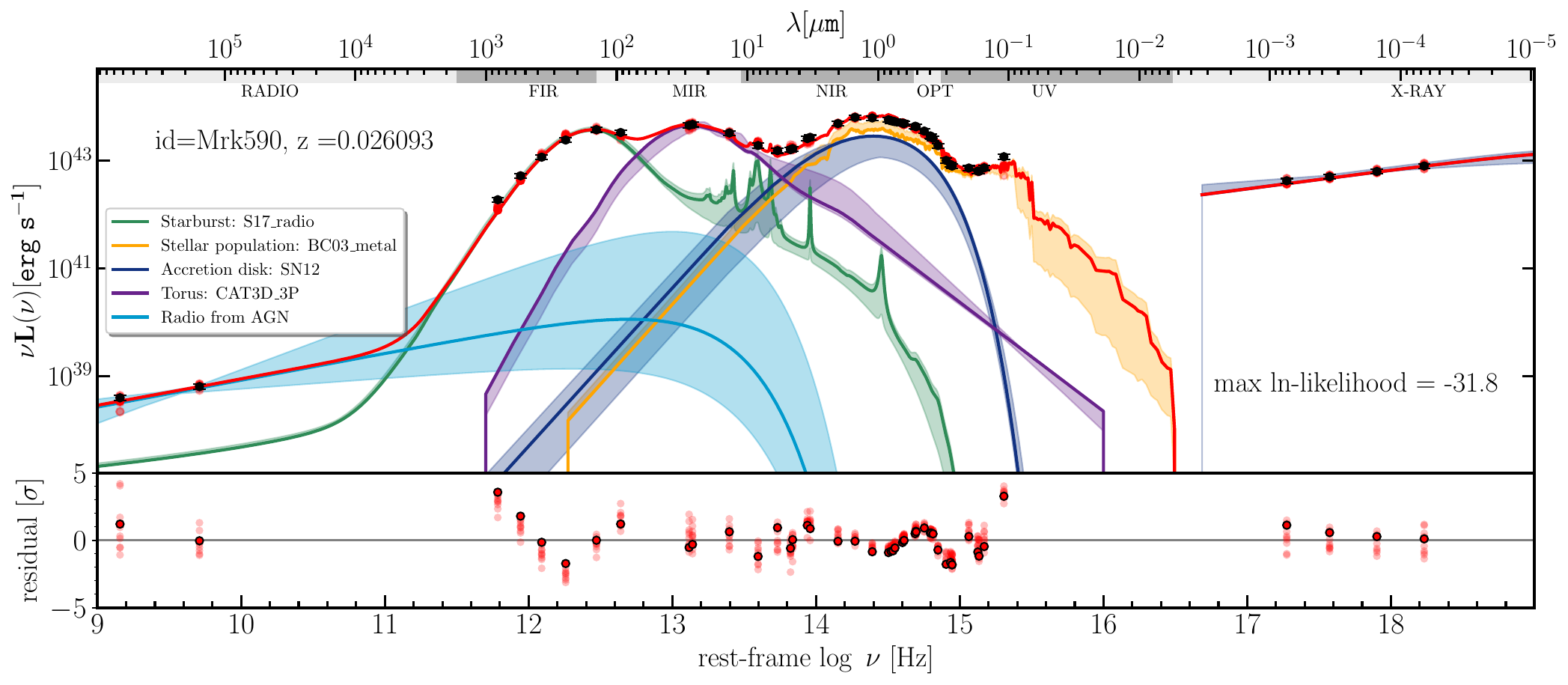}
    \includegraphics[trim={0 0 0 1.55cm},clip, width = 0.8\linewidth]{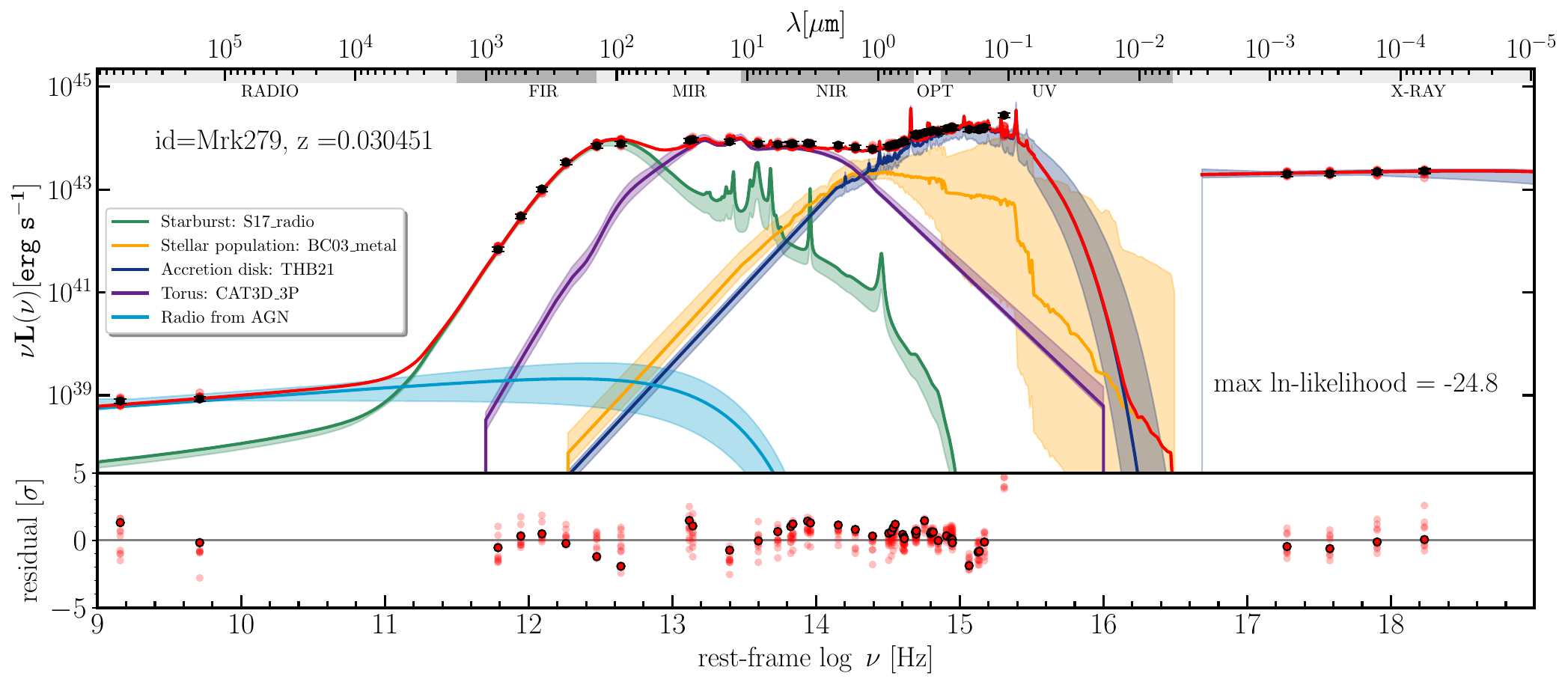}
    \caption{Continued from Figure \ref{fig: SEDfitting_1}. Examples of the best SED-fittings for H 1821+643, Ark 120, Mrk590 and Mrk 279.}
    \label{fig: SEDfitting3}
\end{figure*}

\begin{figure*}[ht!]
\centering
    \includegraphics[trim={0 1.57cm 0 0},clip, width = 0.8\linewidth]{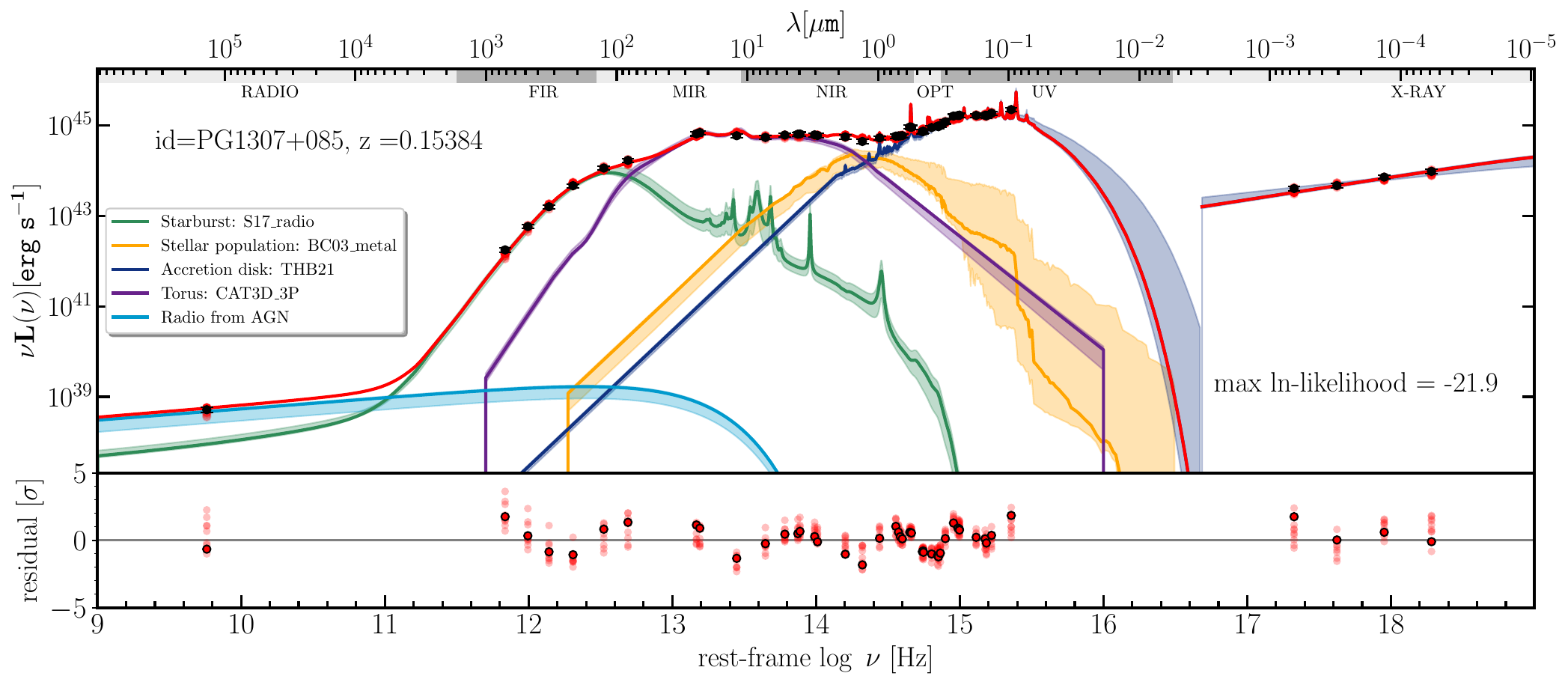}
    \includegraphics[trim={0 1.57cm 0 1.55cm},clip, width = 0.8\linewidth]{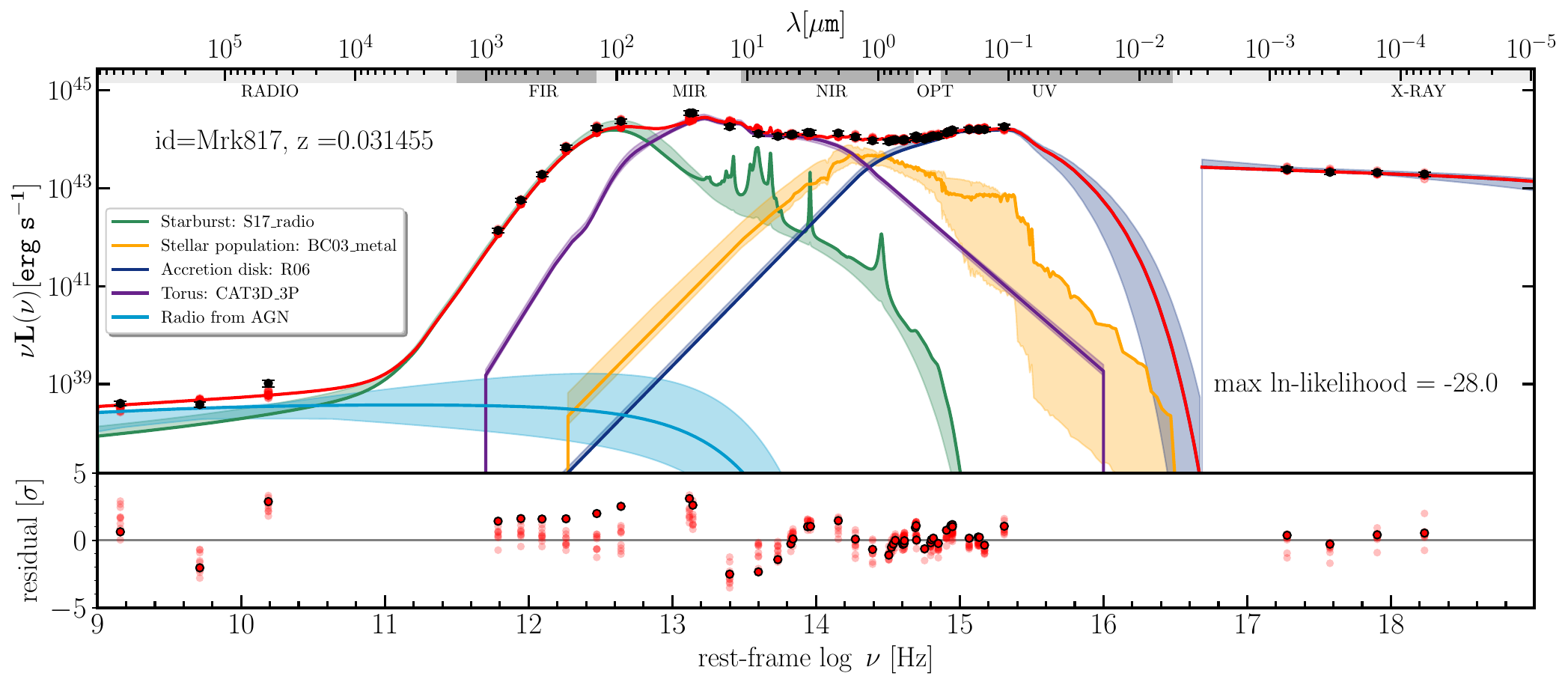}
    \includegraphics[trim={0 1.57cm 0 1.55cm},clip, width = 0.8\linewidth]{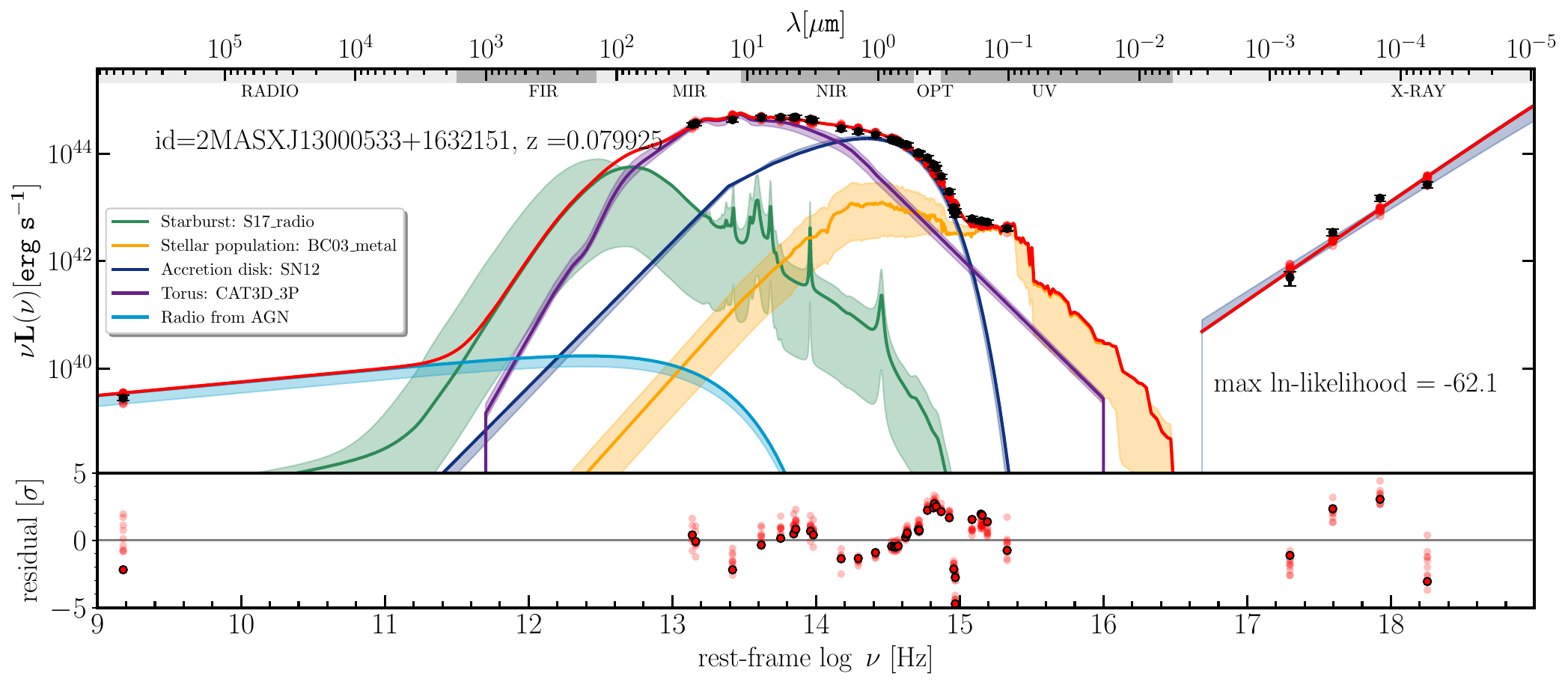}
    \includegraphics[trim={0 0 0 1.55cm},clip, width = 0.8\linewidth]{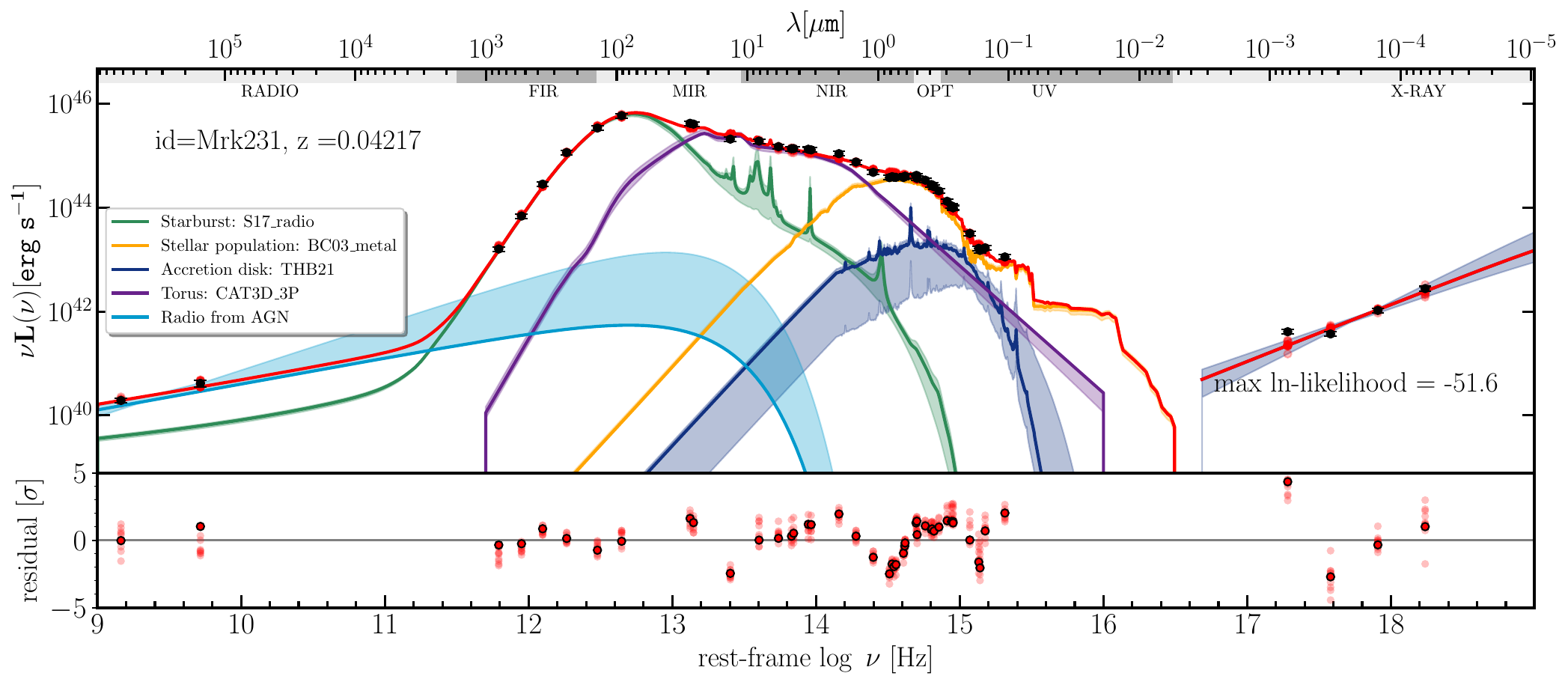}
    \caption{Continued from Figure \ref{fig: SEDfitting_1}. Examples of the best SED-fittings for PG 1307+085, Mrk 817, 2 MASXJ13000533+1632151 and Mrk 231.}
    \label{fig: SEDfitting4}
\end{figure*}

\begin{figure*}[ht!]
\centering
    \includegraphics[trim={0 1.57cm 0 0},clip, width = 0.8\linewidth]{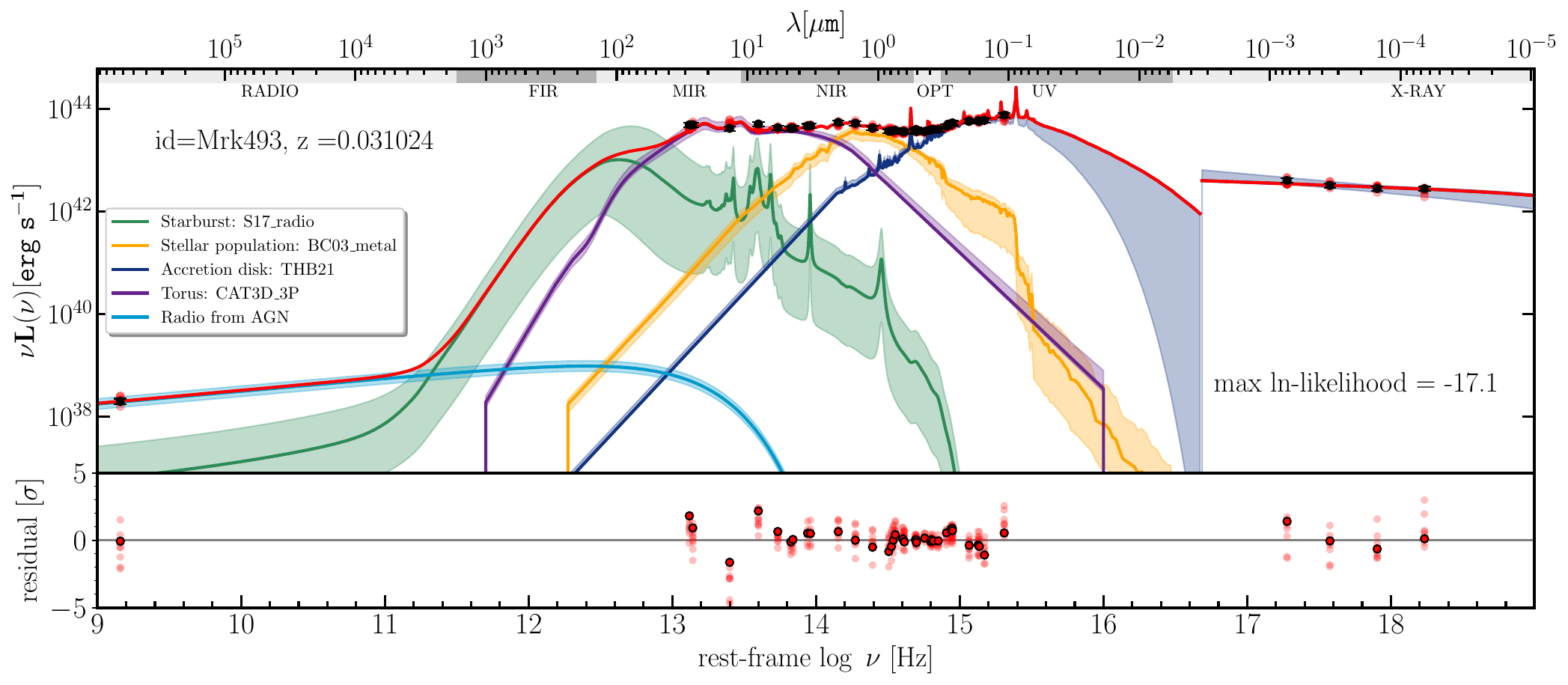}
    \includegraphics[trim={0 1.57cm 0 1.55cm},clip, width = 0.8\linewidth]{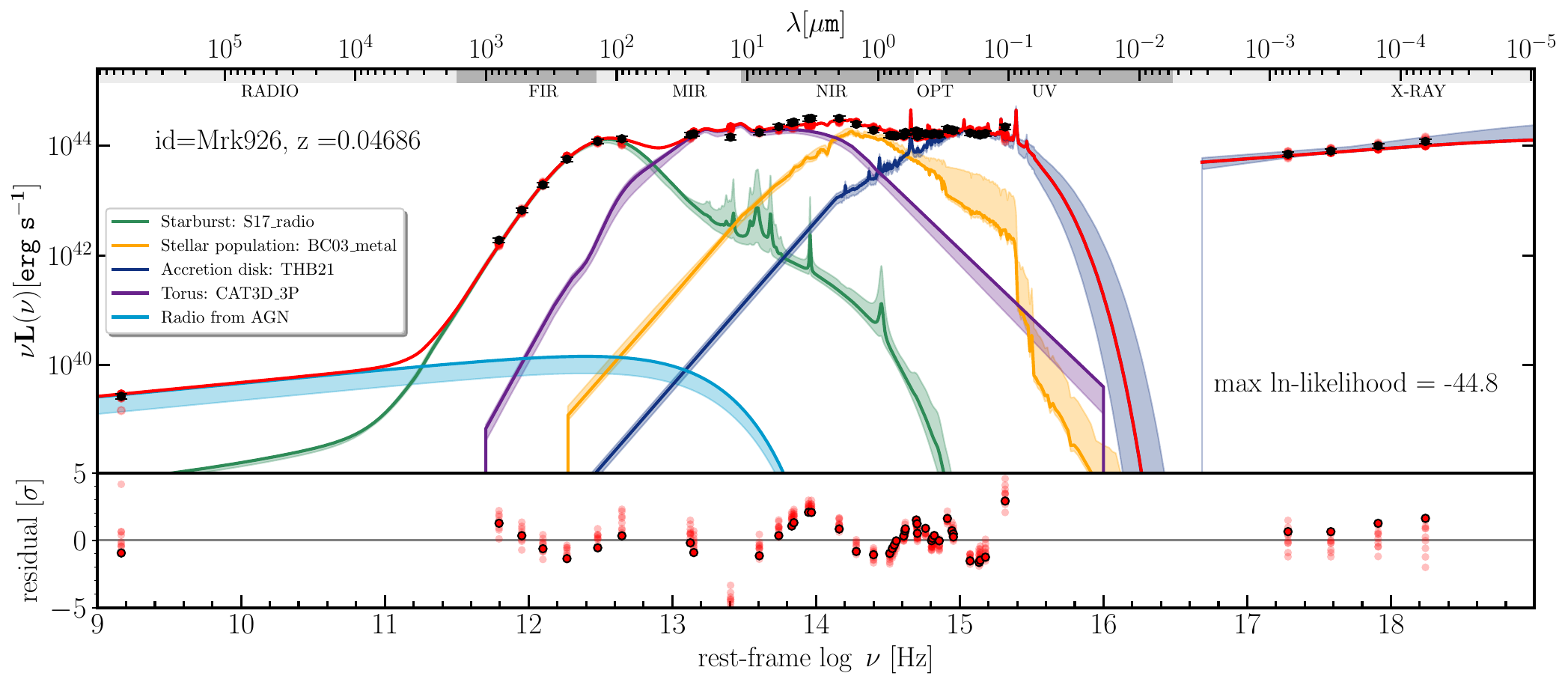}
    \includegraphics[trim={0 1.57cm 0 1.55cm},clip, width = 0.8\linewidth]{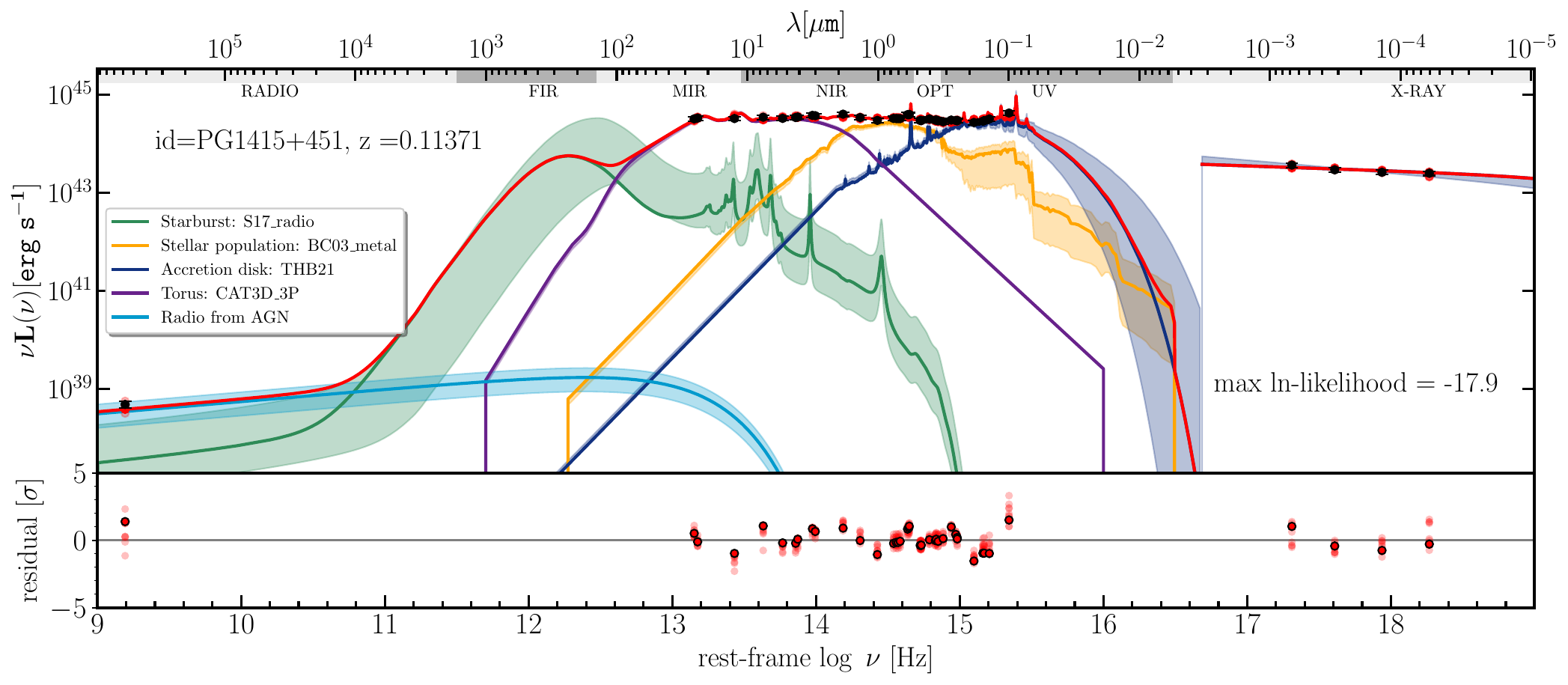}
    \includegraphics[trim={0 0 0 1.55cm},clip, width = 0.8\linewidth]{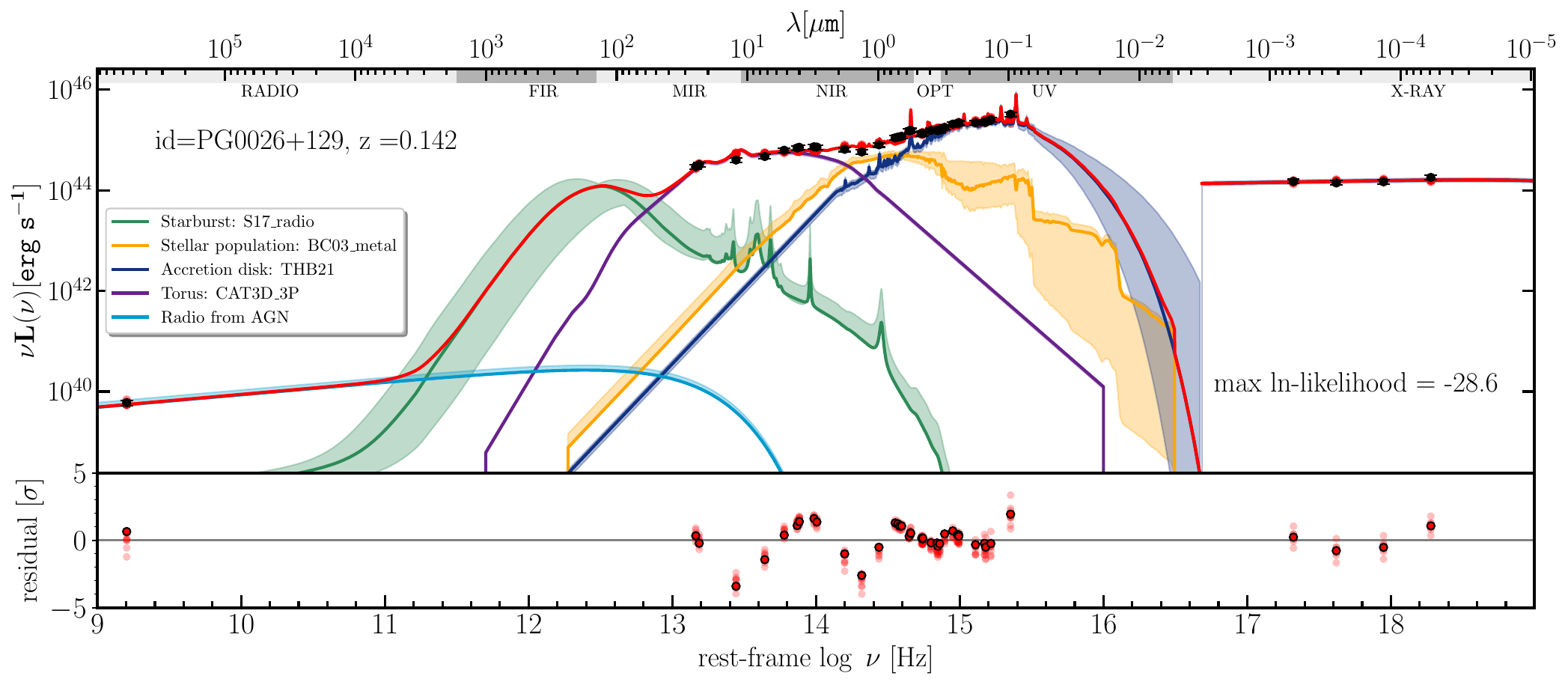}
    \caption{Continued from Figure \ref{fig: SEDfitting_1}. Examples of the best SED-fittings for Mrk 493, Mrk 926, PG 1415+451 and PG 026+129.}
    \label{fig: SEDfitting5}
\end{figure*}

\begin{figure*}[ht!]
\centering
    \includegraphics[trim={0 1.57cm 0 0},clip, width = 0.8\linewidth]{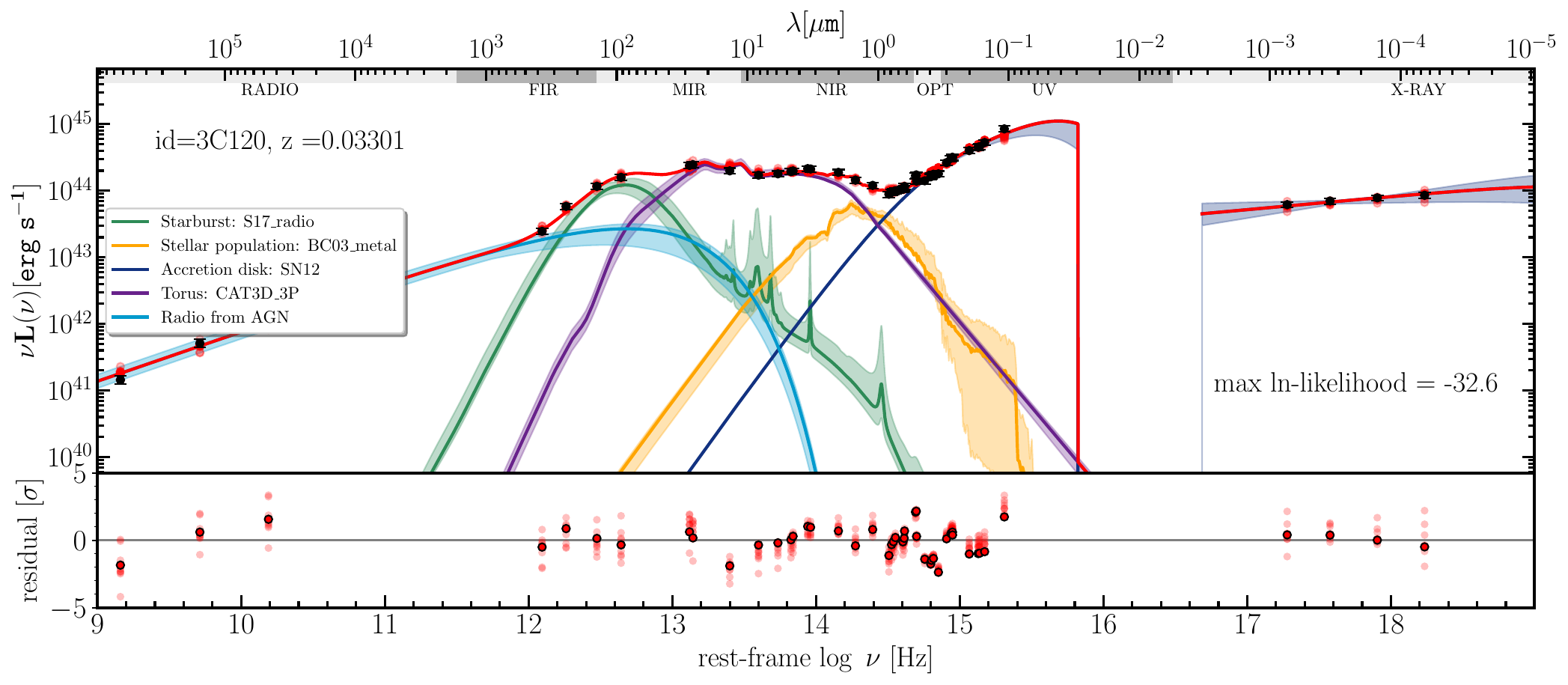}
    \includegraphics[trim={0 1.57cm 0 1.55cm},clip, width = 0.8\linewidth]{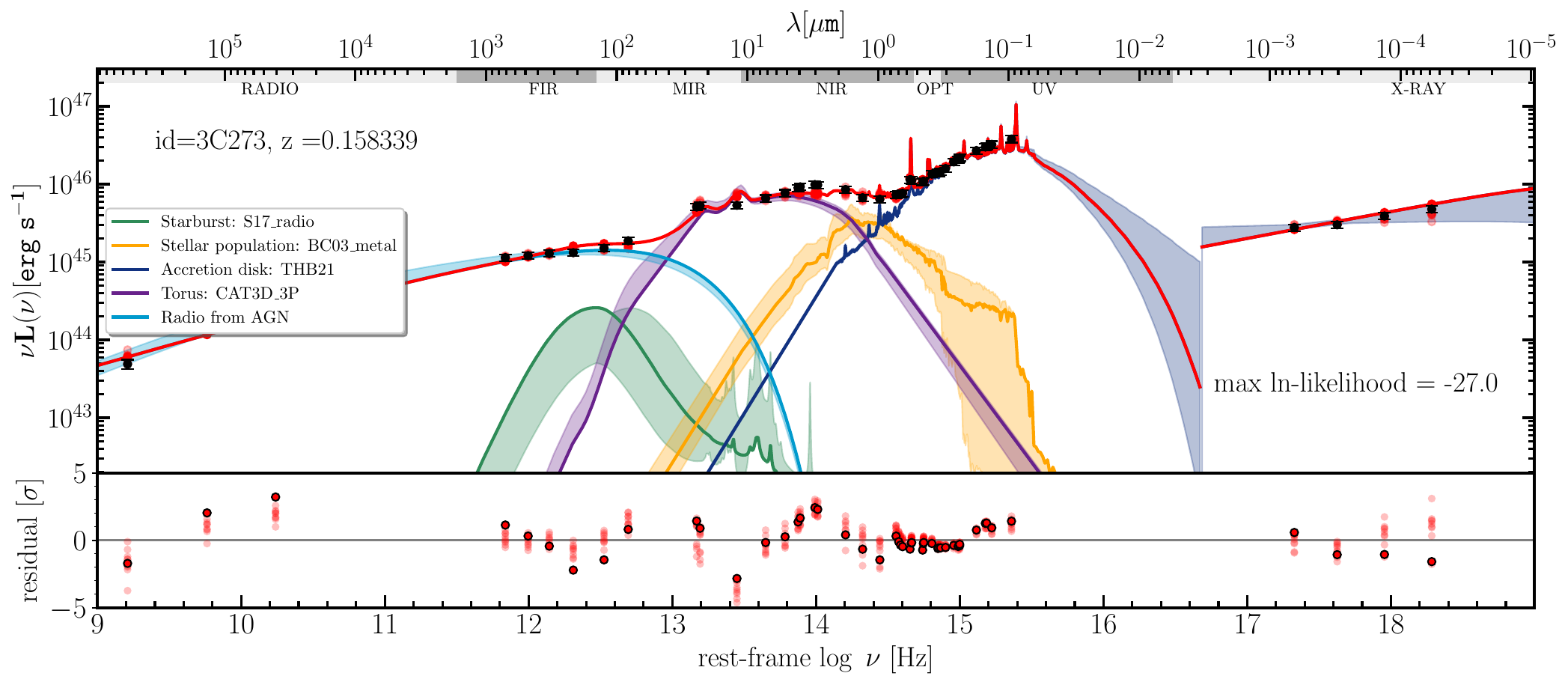}
    \includegraphics[trim={0 1.57cm 0 1.55cm},clip, width = 0.8\linewidth]{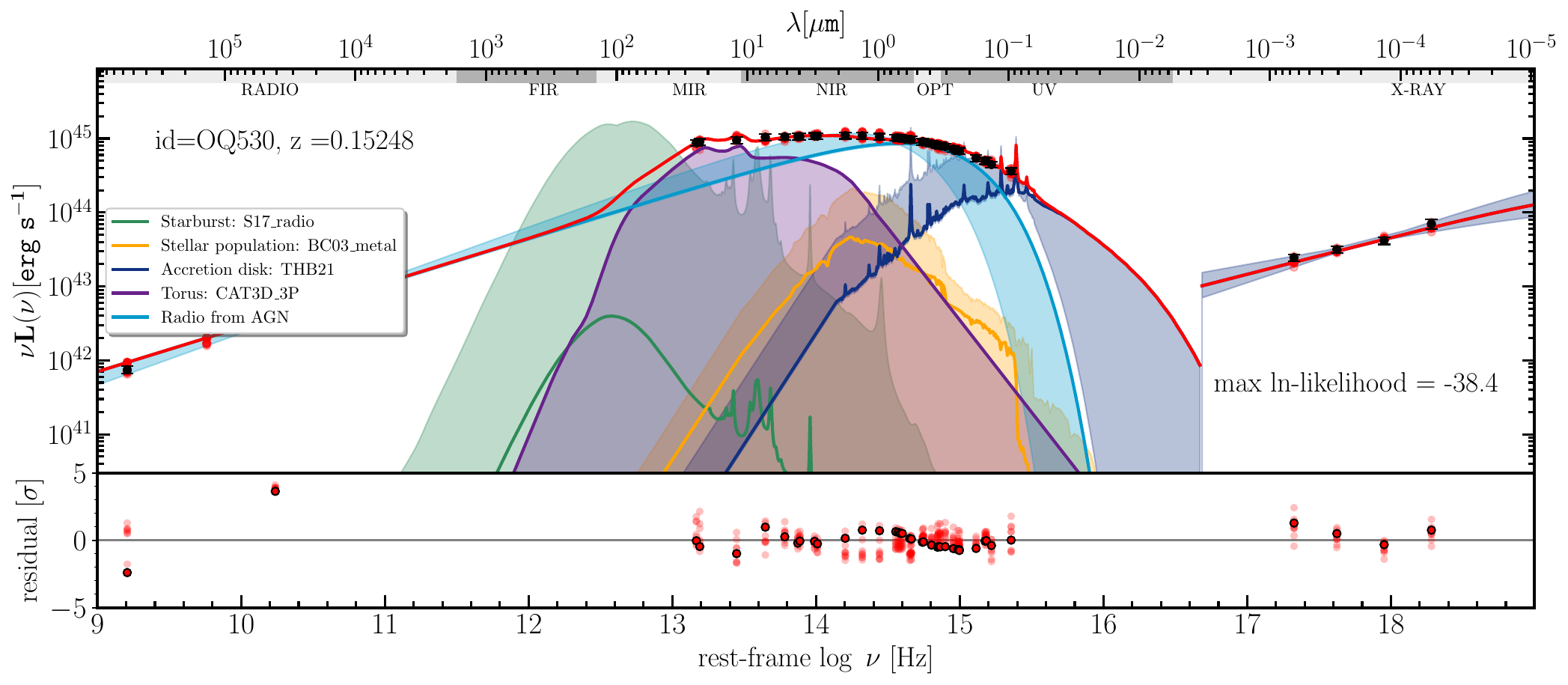}
    \includegraphics[trim={0 0 0 1.55cm},clip, width = 0.8\linewidth]{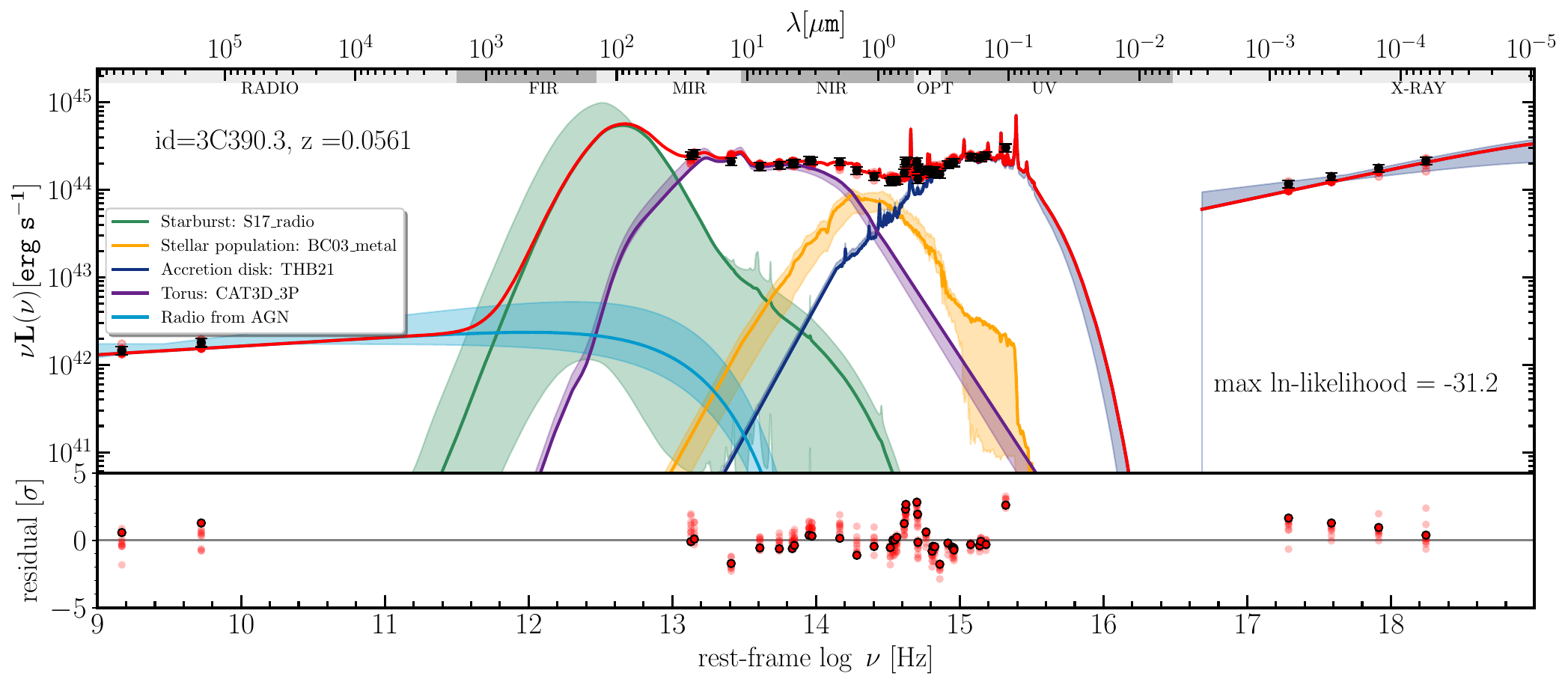}
    \caption{Continued from Figure \ref{fig: SEDfitting_1}. Examples of the best SED-fittings for 3C 120, 3C 273, OQ530 and 3C 390.3.}
    \label{fig: SEDfitting6}
\end{figure*}

\begin{figure*}[ht!]
\centering
    \includegraphics[trim={0 0 0 0},clip, width = 0.8\linewidth]{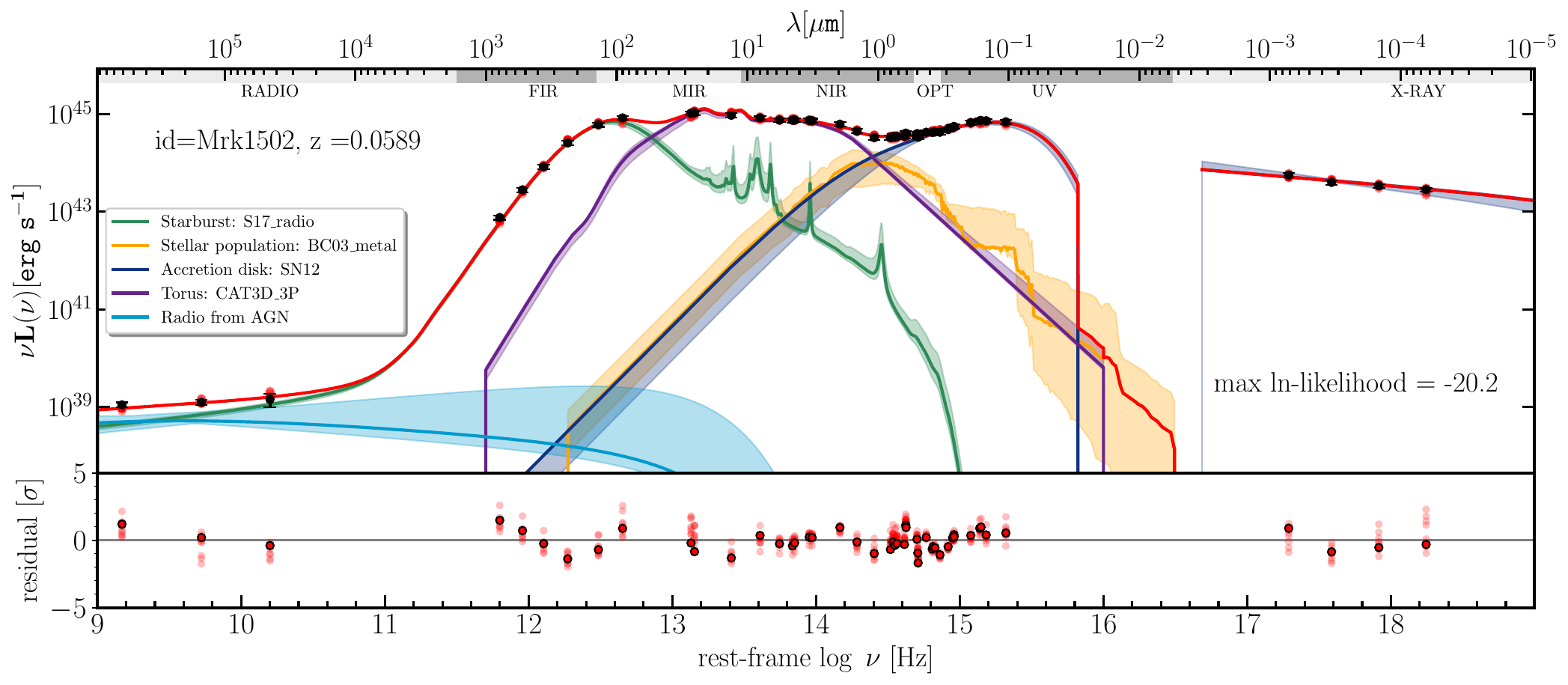}
    \includegraphics[trim={0 1.57cm 0 0},clip, width = 0.8\linewidth]{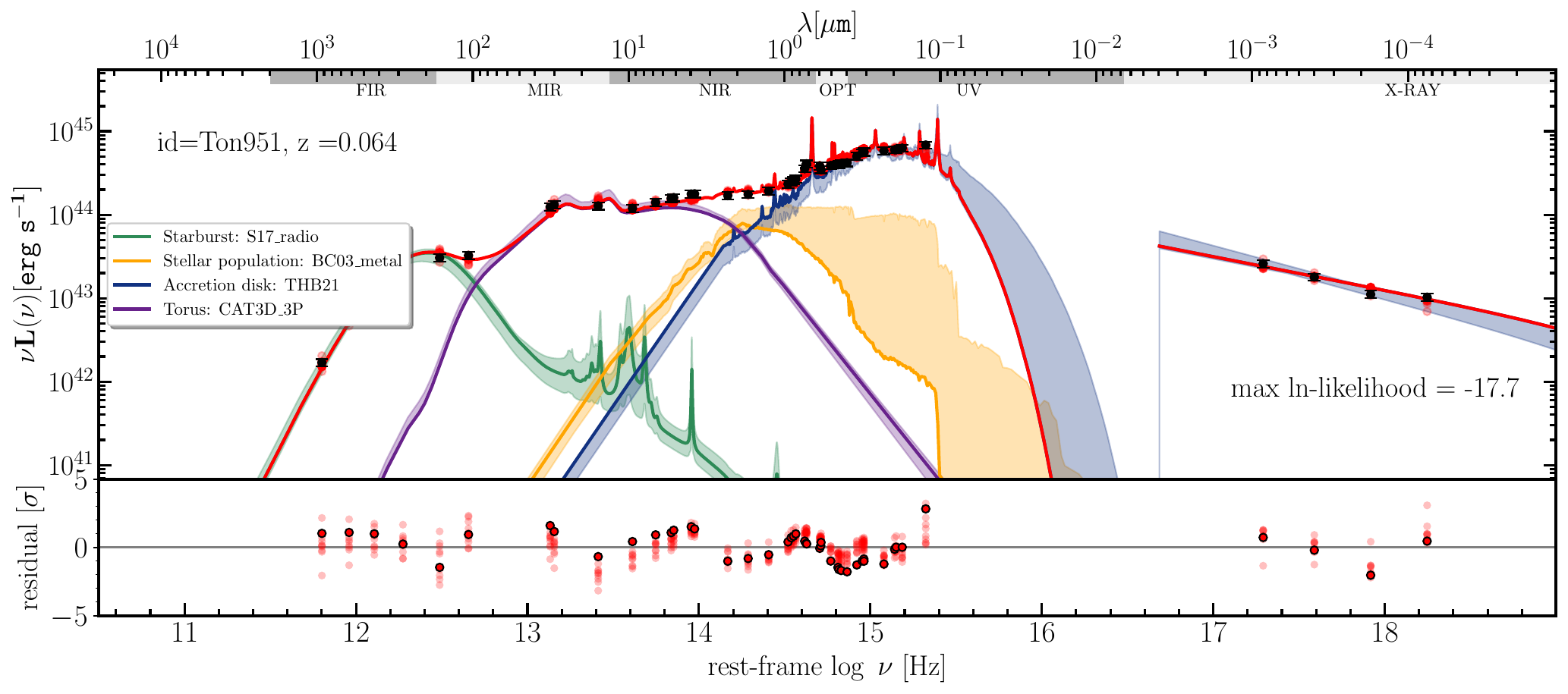}
    \includegraphics[trim={0 1.57cm 0 1.55cm},clip, width = 0.8\linewidth]{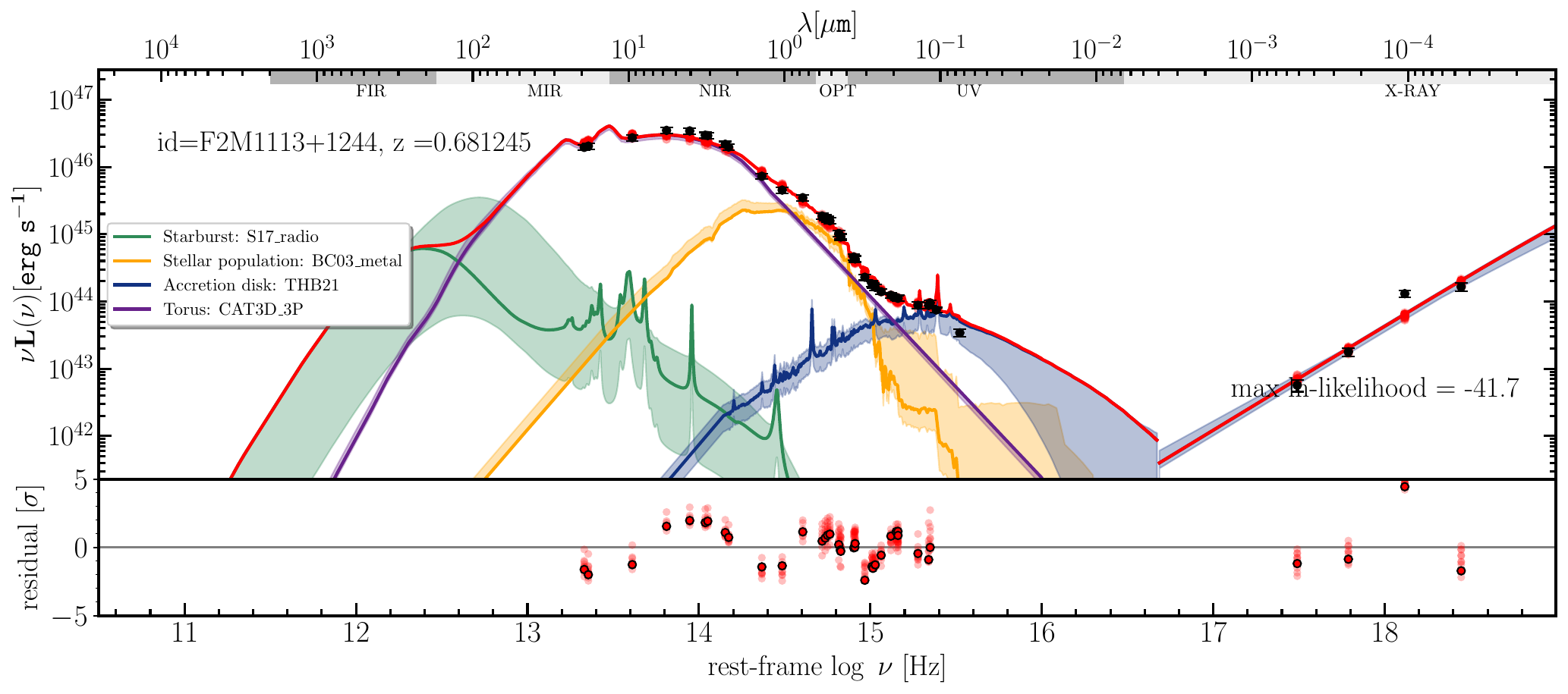}
    \includegraphics[trim={0 0 0 1.55cm},clip, width = 0.8\linewidth]{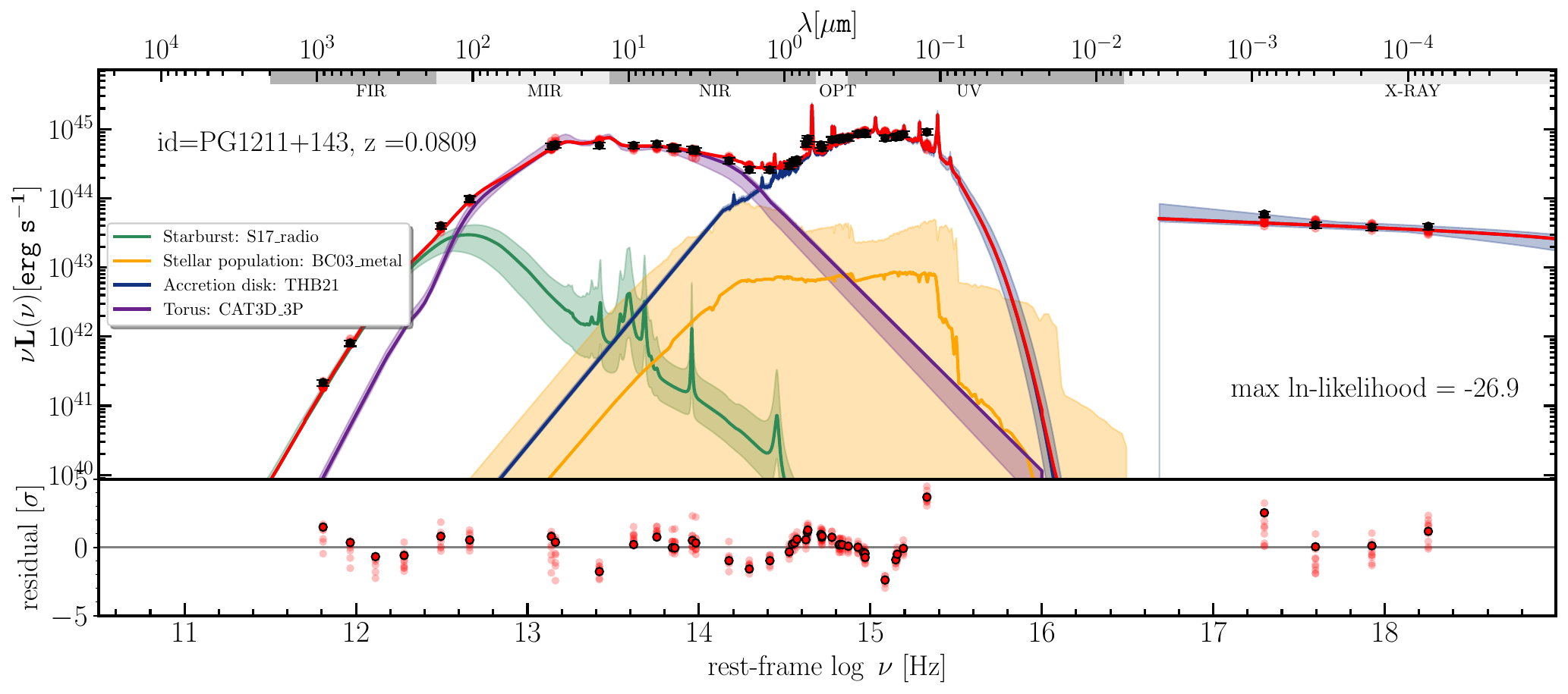}
    \caption{Continued from Figure \ref{fig: SEDfitting_1}. Examples of the best SED-fittings for Mrk 1502, Ton 951, F2M1113+1244 and PG 1211+143.}
    \label{fig: SEDfitting7}
\end{figure*}

\begin{figure*}[ht!]
\centering
    \includegraphics[trim={0 0 0 0},clip, width = 0.8\linewidth]{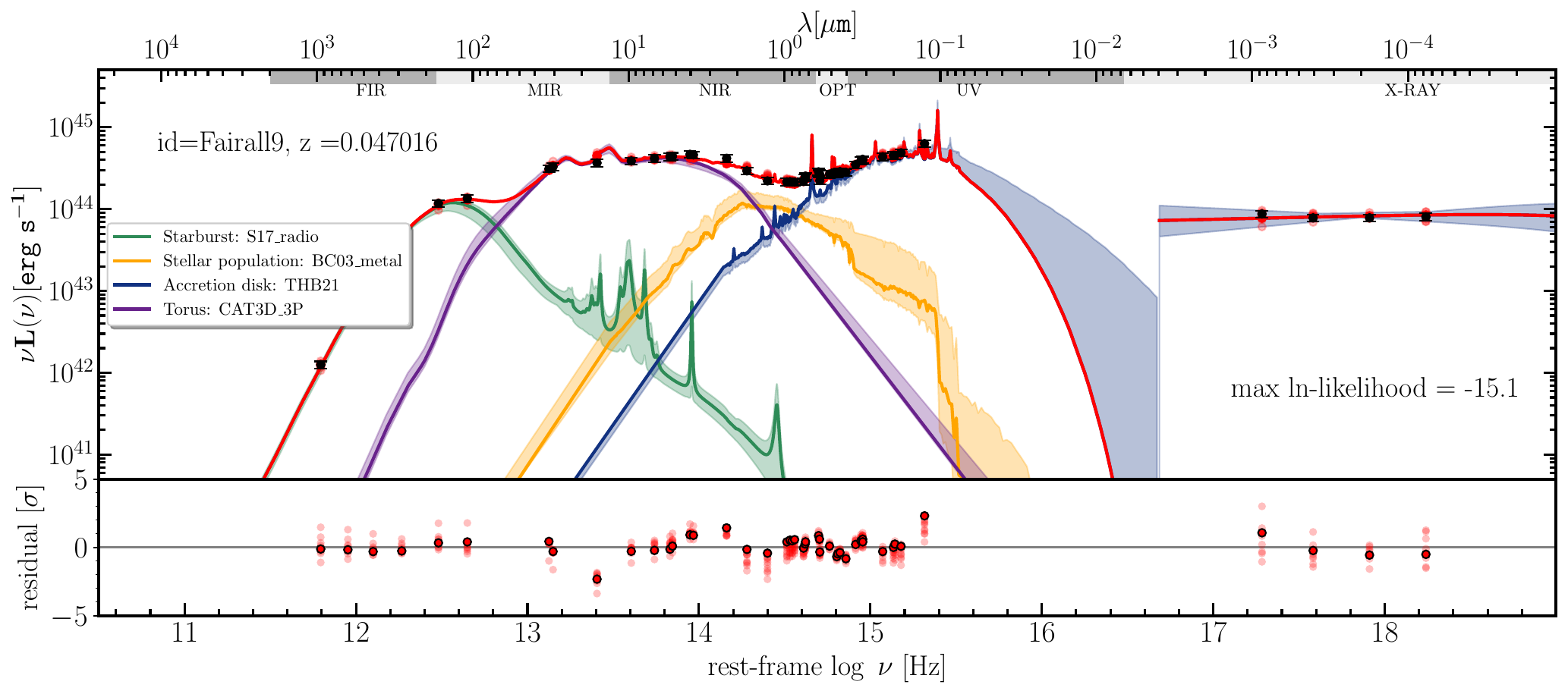}
    \includegraphics[trim={0 1.57cm 0 0},clip, width = 0.8\linewidth]{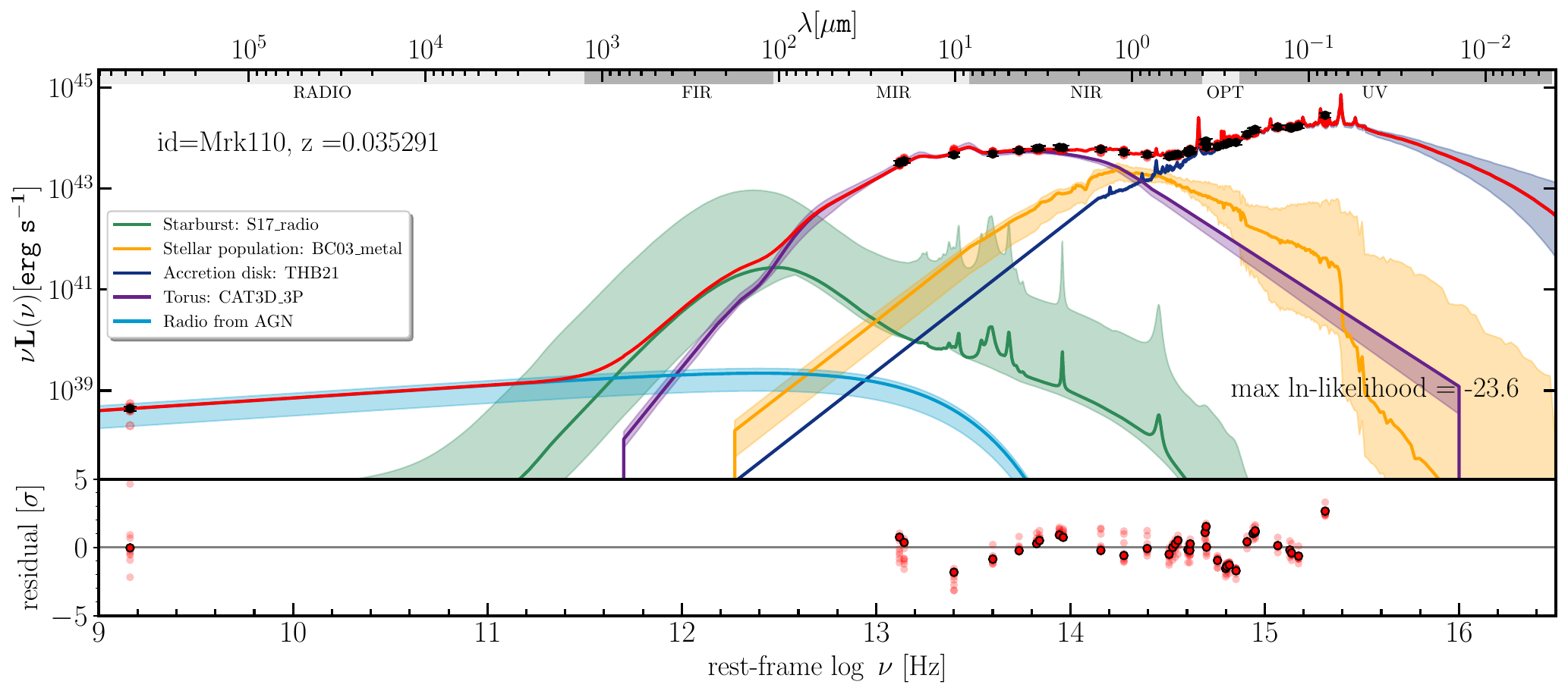}
    \includegraphics[trim={0 1.57cm 0 1.55cm},clip, width = 0.8\linewidth]{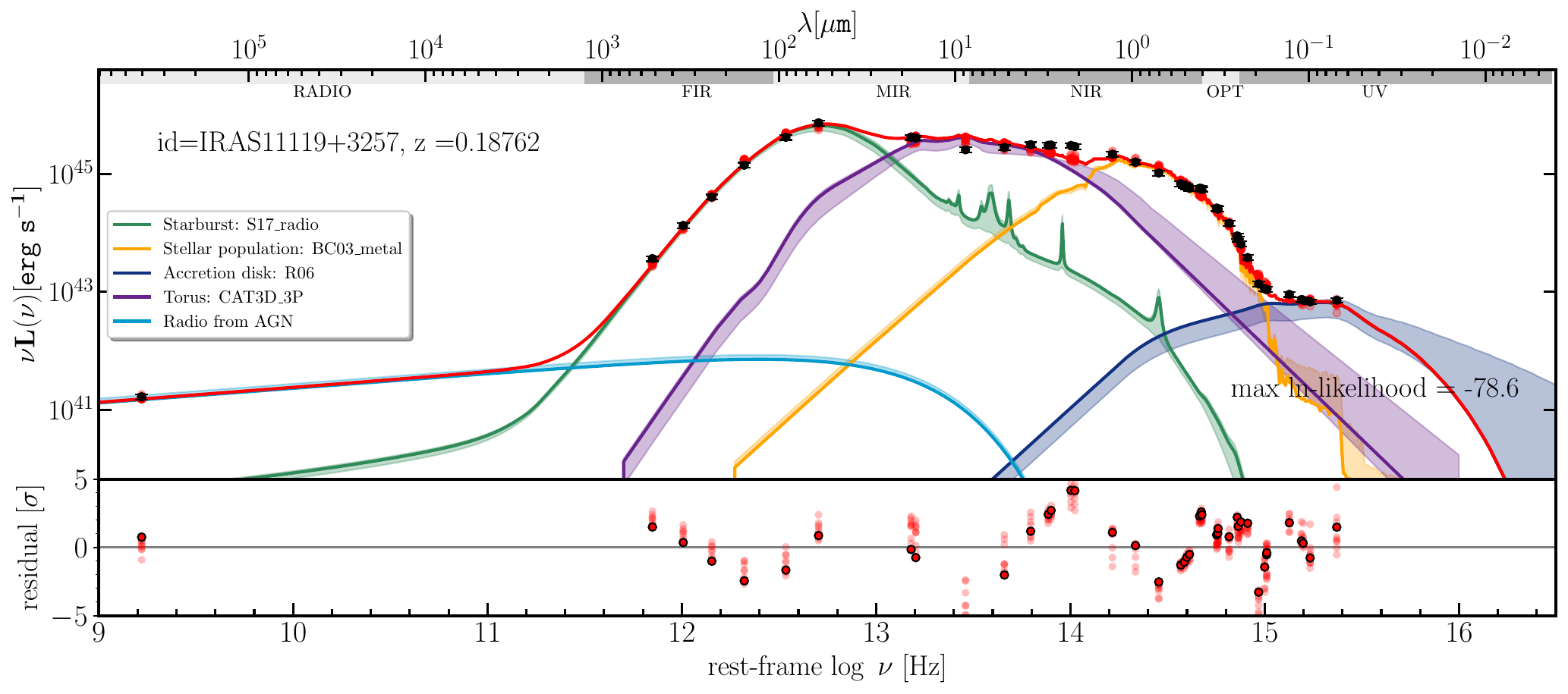}
    \includegraphics[trim={0 0 0 1.55cm},clip, width = 0.8\linewidth]{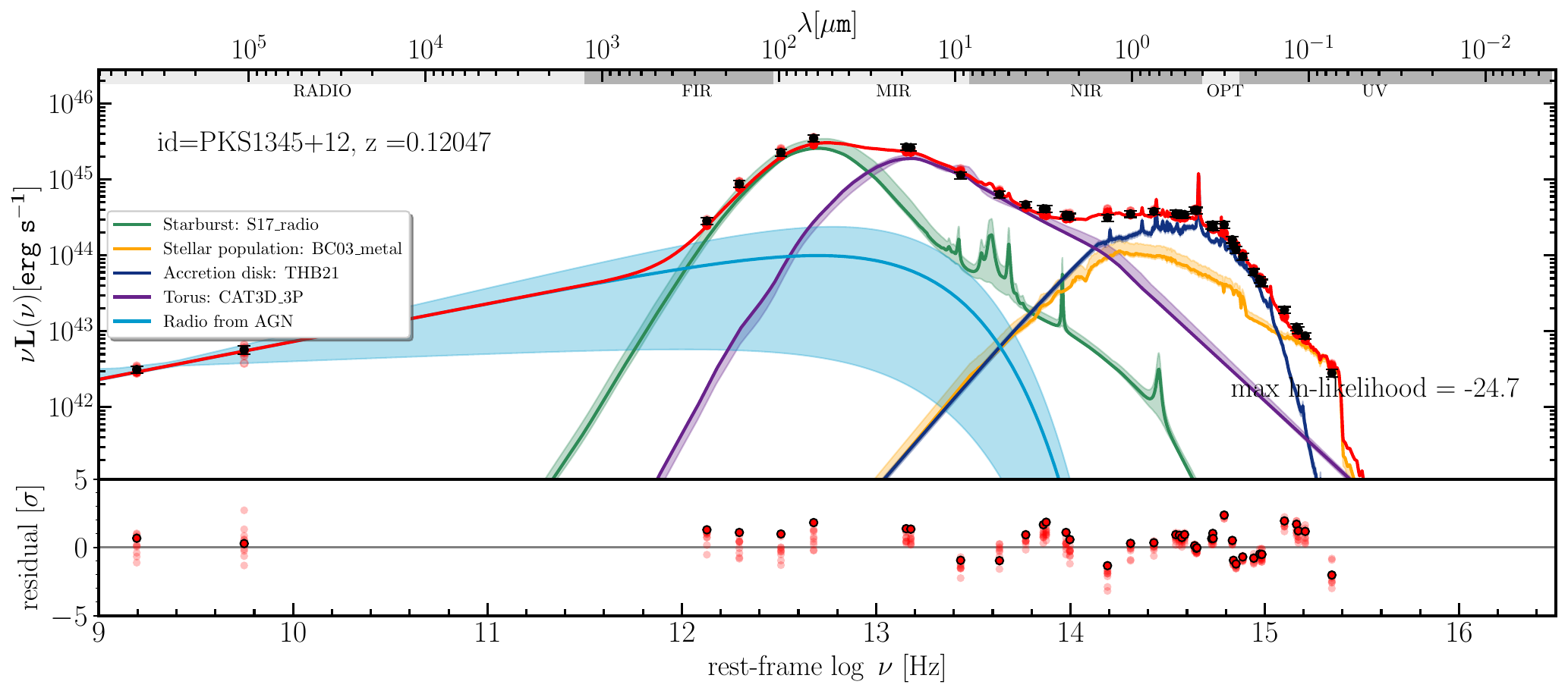}
    \caption{Continued from Figure \ref{fig: SEDfitting_1}. Examples of the best SED-fittings for Fairall9, Mrk110, IRAS 11119+3257 and PKS1345+12.}
    \label{fig: SEDfitting8}
\end{figure*}

\begin{figure*}[ht!]
\centering
    \includegraphics[trim={0 1.57cm 0 0},clip, width = 0.8\linewidth]{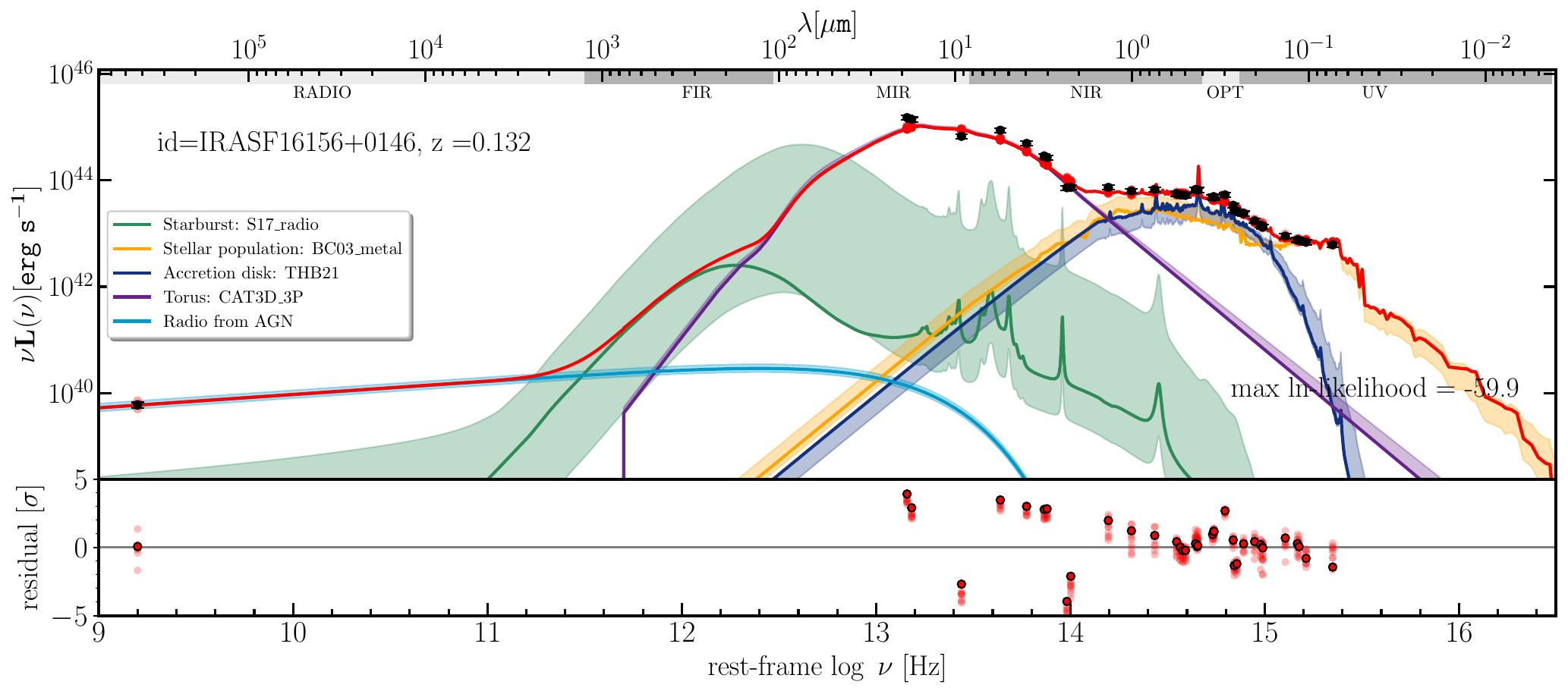}
    \includegraphics[trim={0 1.57cm 0 1.55cm},clip, width = 0.8\linewidth]{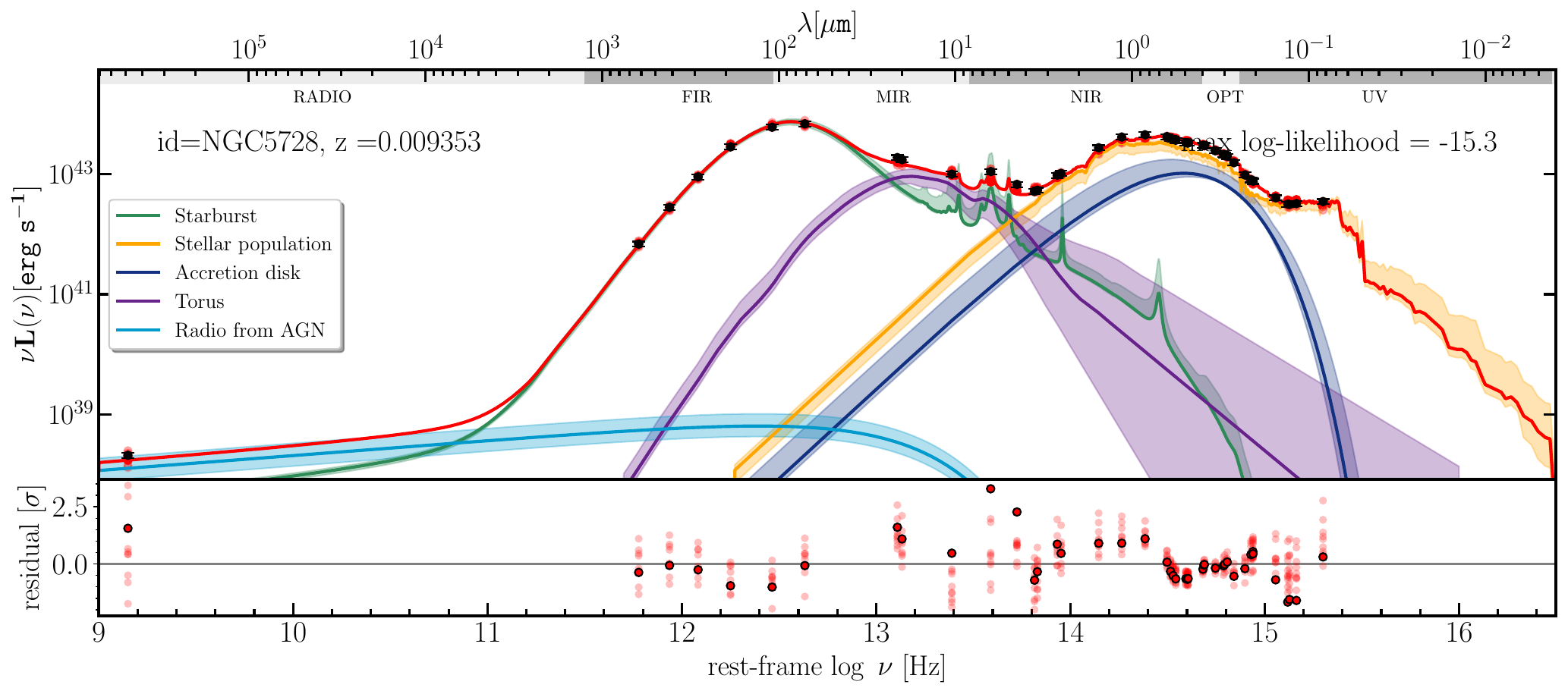}
    \includegraphics[trim={0 1.57cm 0 1.55cm},clip, width = 0.8\linewidth]{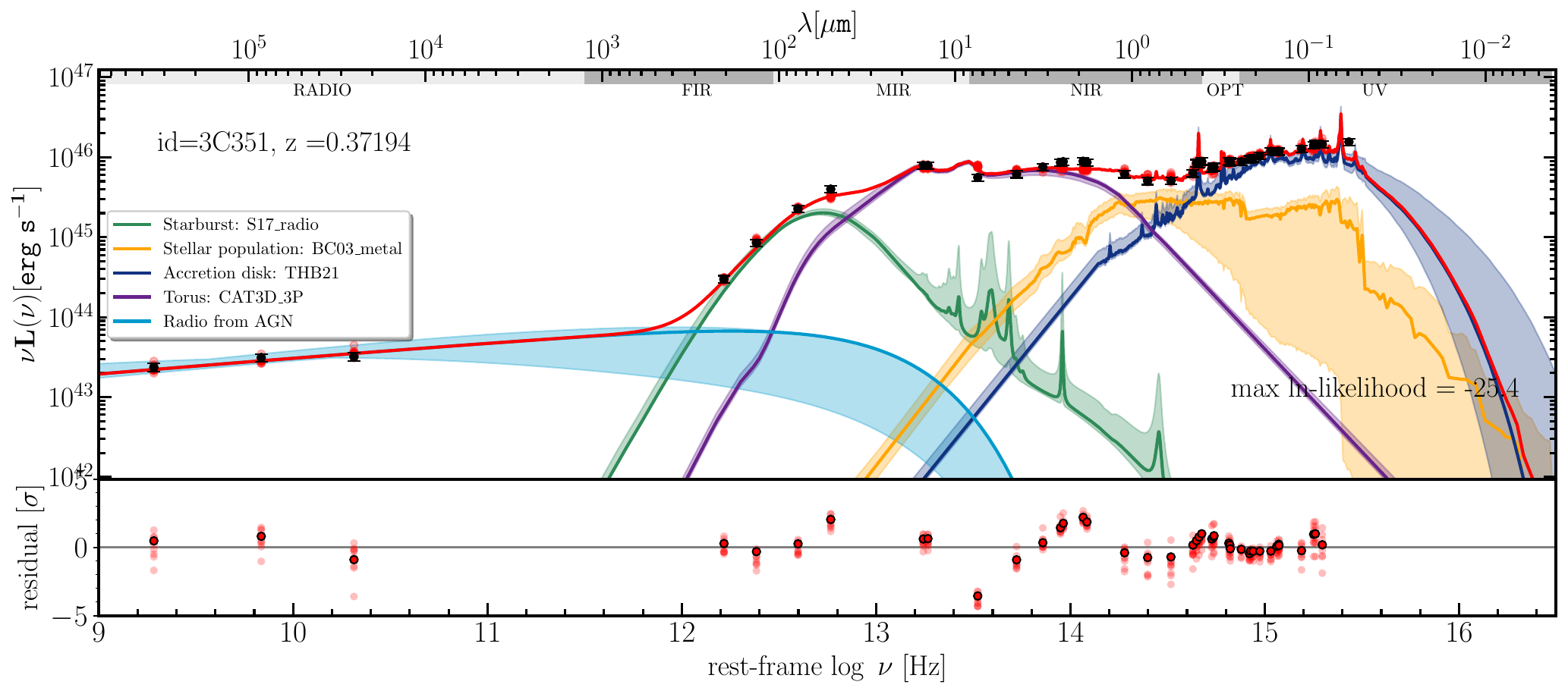}
    \includegraphics[trim={0 0 0 1.55cm},clip, width = 0.8\linewidth]{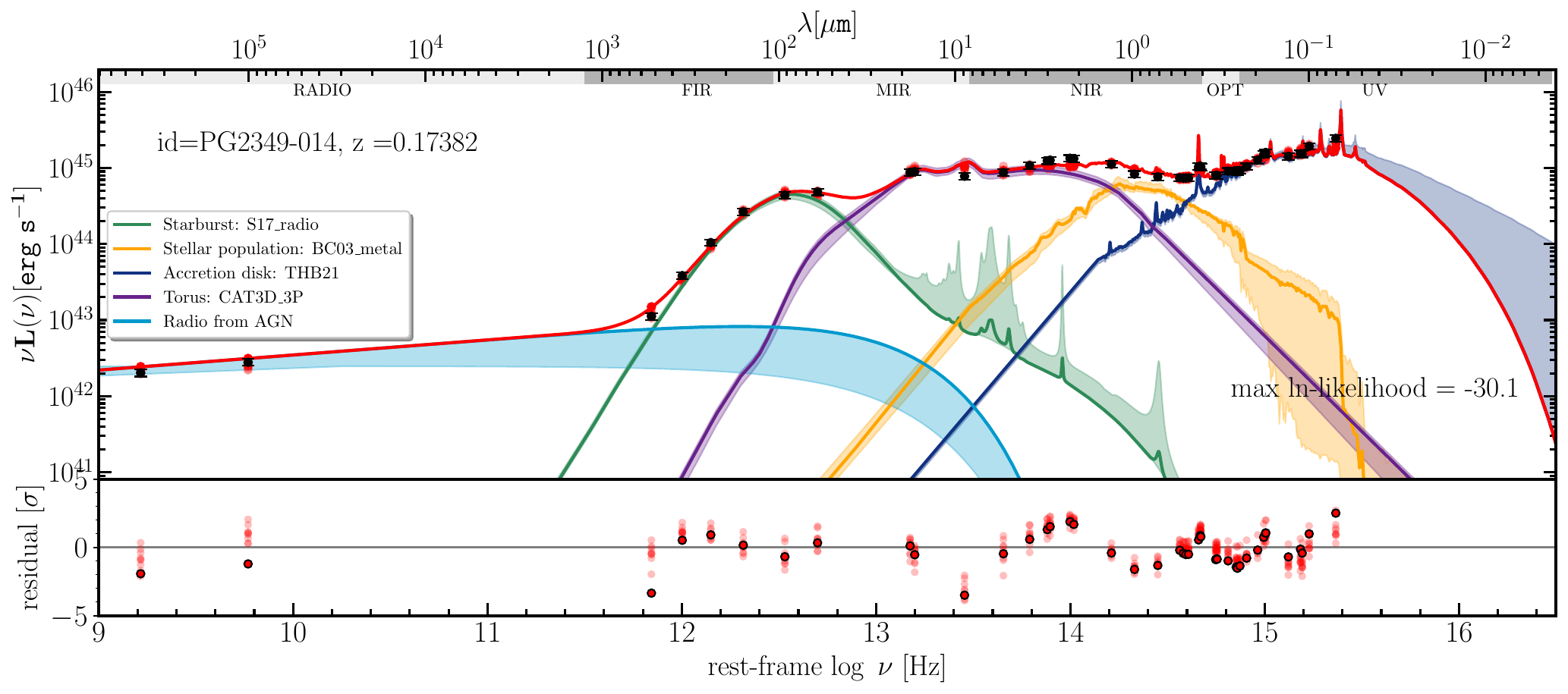}
    \caption{Continued from Figure \ref{fig: SEDfitting_1}. Examples of the best SED-fittings for IRAS F16156+0146, NGC5728, 3C351 and PG 2349-014.}
    \label{fig: SEDfitting9}
\end{figure*}

\section{Parameter space covered by the models}
Spectral slope space covered by the torus models included in this study, given by $\alpha = -\log(F_\nu (\lambda_2)/F_\nu(\lambda_1))/\log(\lambda_2 / \lambda_1)$ with $\alpha_{\text{MIR}}$ and $\alpha_{\text{NIR}}$ corresponding to $[\lambda_2, \lambda_1]$ equal to $[14, 8]$ and $[6, 3]$ $\mu \rm m$. The estimates for the AGNs of our sample were based on WISE band 3 at $\lambda = 12$ $\mu$m and IRAC band 4 at $\lambda = 8$ $\mu$m for $\alpha_{\text{MIR}}$; and IRAC band 3 at $\lambda = 5.8$ $\mu$m and WISE band 1 at $\lambda = 3.4$ $\mu$m for $\alpha_{\text{NIR}}$.

\begin{figure}[ht]
    \centering
    \includegraphics[width = 9.0 cm]{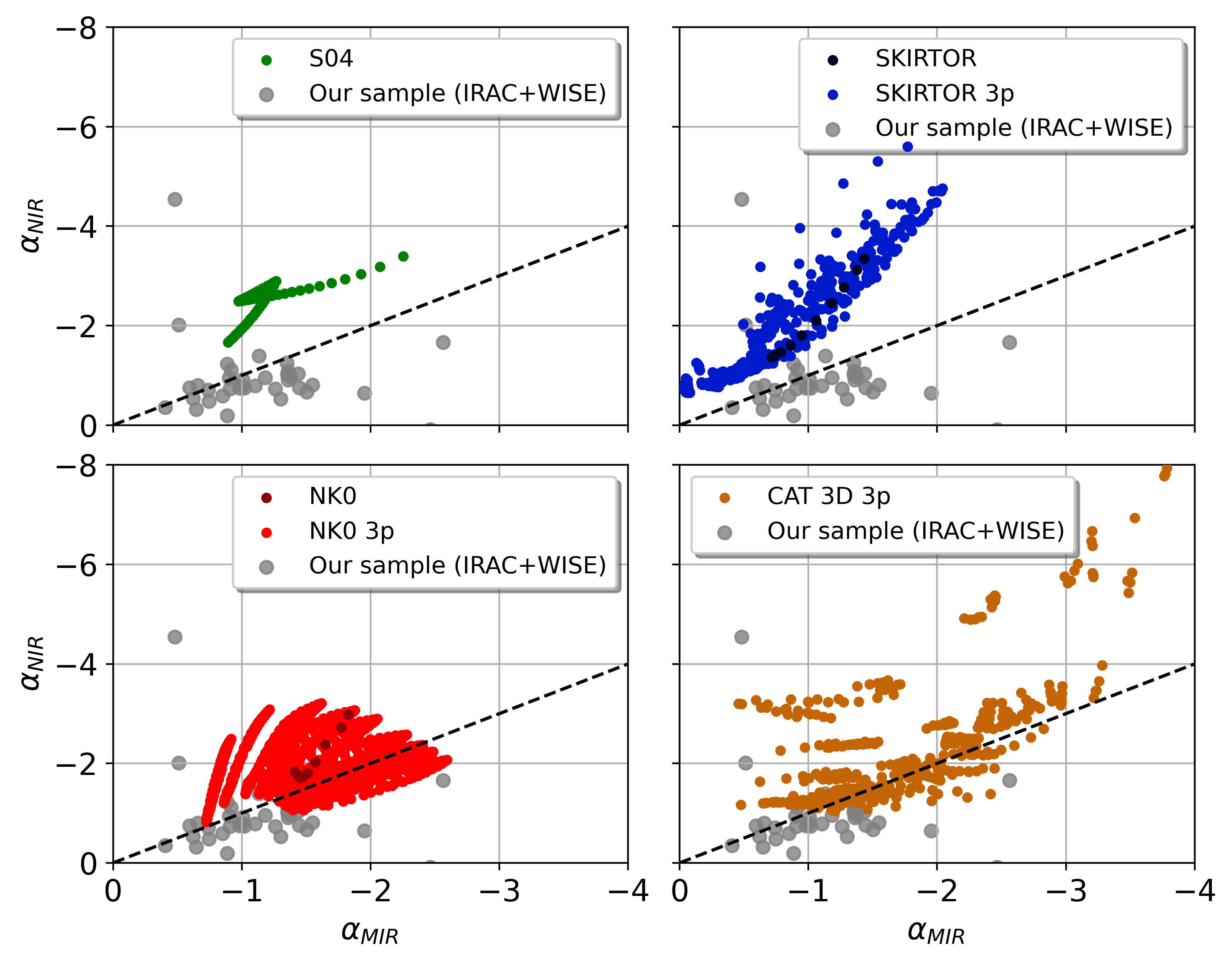}%{FIGURES/SEDs_allmodelsTEST.png}
    \caption{$\alpha_{\text{MIR}}$-$\alpha_{\text{NIR}}$ spectral slope space covered by the different torus models included in \textsc{AGNfitter-rx} with different complexity levels. The gray circles show estimates of the spectral slope covered by our sample of AGN, computed with Spitzer and WISE observed photometry.}
    \label{fig: TO_alphas}
\end{figure}

Similarly, the spectral slope space covered by the accretion disk models was computed with $\alpha_{\text{B-V}}$ and $\alpha_{\text{ox}}$ corresponding to $[\lambda_2, \lambda_1]$ equal to $[4420, 5400]$ and $[2500, 8.21]$ $\AA$. The slopes of our sample were based on SWIFT band B at $\lambda = 4392$ $\AA$ and SWIFT band V at $\lambda = 5468$ $\AA$ for $\alpha_{\text{NIR}}$, and GALEX NUV channel at $\lambda = 2500$ $\AA$ and XMM-Newton band 3 for $\alpha_{\text{ox}}$.

\begin{figure}[ht]
    \centering
    \includegraphics[width = 8.0 cm]{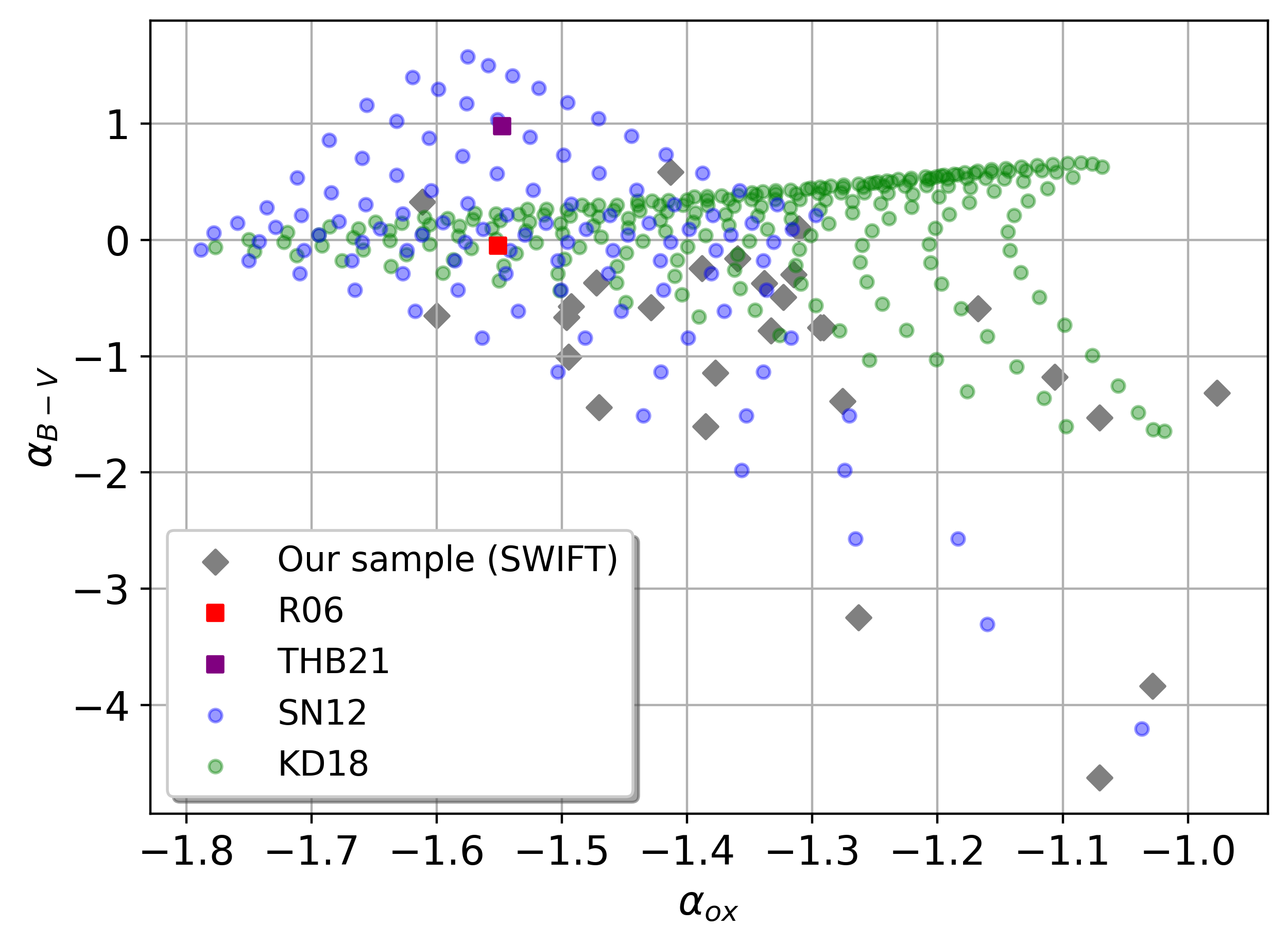}
    \caption{$\alpha_{\text{ox}}$-$\alpha_{\text{B-V}}$ spectral slope space covered by the different accretion disk models included in \textsc{AGNfitter-rx}. The gray circles show estimates of the spectral slope covered by our sample of AGN, computed with SWIFT, GALEX and XMM-Newton observed photometry.}
    \label{fig: BB_alphas}
\end{figure}

\section{Bolometric luminosity from accretion disk models}
 We compute bolometric luminosities based on the black hole mass and accretion rate estimates from the theoretical accretion disk models SN12 and KD18 as follows:

\begin{equation}
    \hspace{8mm} \textit{L}_{\rm bol} = \left( \frac{\dot{m}}{\dot{m}_{\rm edd}} \right) \times 1.26 \times 10^{38} \times \rm M_{\rm BH}.
\label{eq:bolometric}
\end{equation}
and compare with the values obtained by adding the integrated luminosities of the AGN-synchrotron, the torus, the accretion disk and the hot corona components.

\begin{figure}[ht]
    \centering
    \includegraphics[width = 8.0 cm]{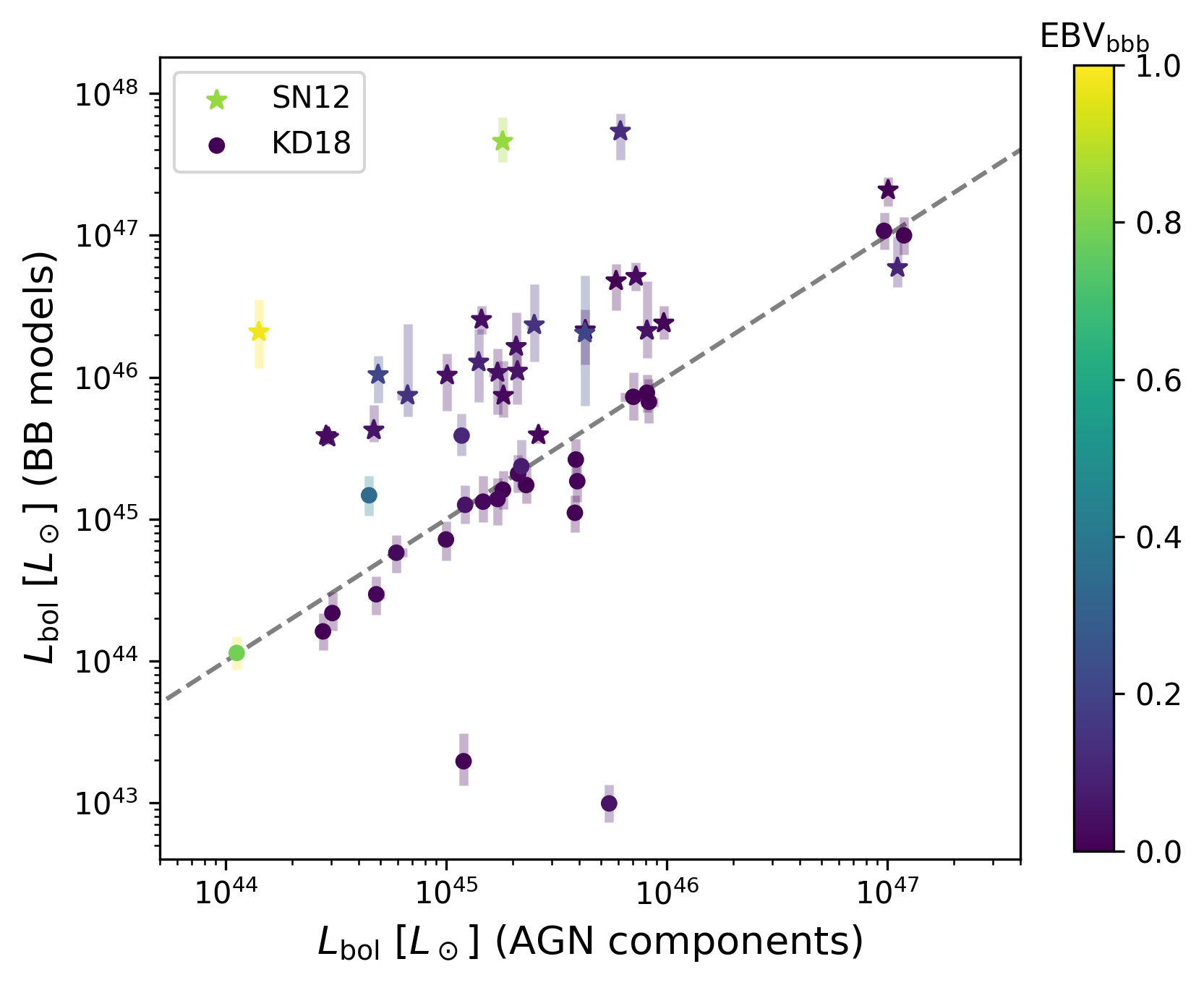}
    \caption{Bolometric luminosities are computed by adding integrated luminosities of the AGN components vs bolometric luminosities based on the inferred black hole mass and accretion rate by the accretion disk models. The gray dashed line indicates the 1-1 ratio and the color bar is the accretion disk reddening parameter.}
    \label{fig: Lbol_BB}
\end{figure}

\section{IR and optical AGN fraction}

\begin{figure*}[ht]
    \centering
    \includegraphics[width = 16 cm]{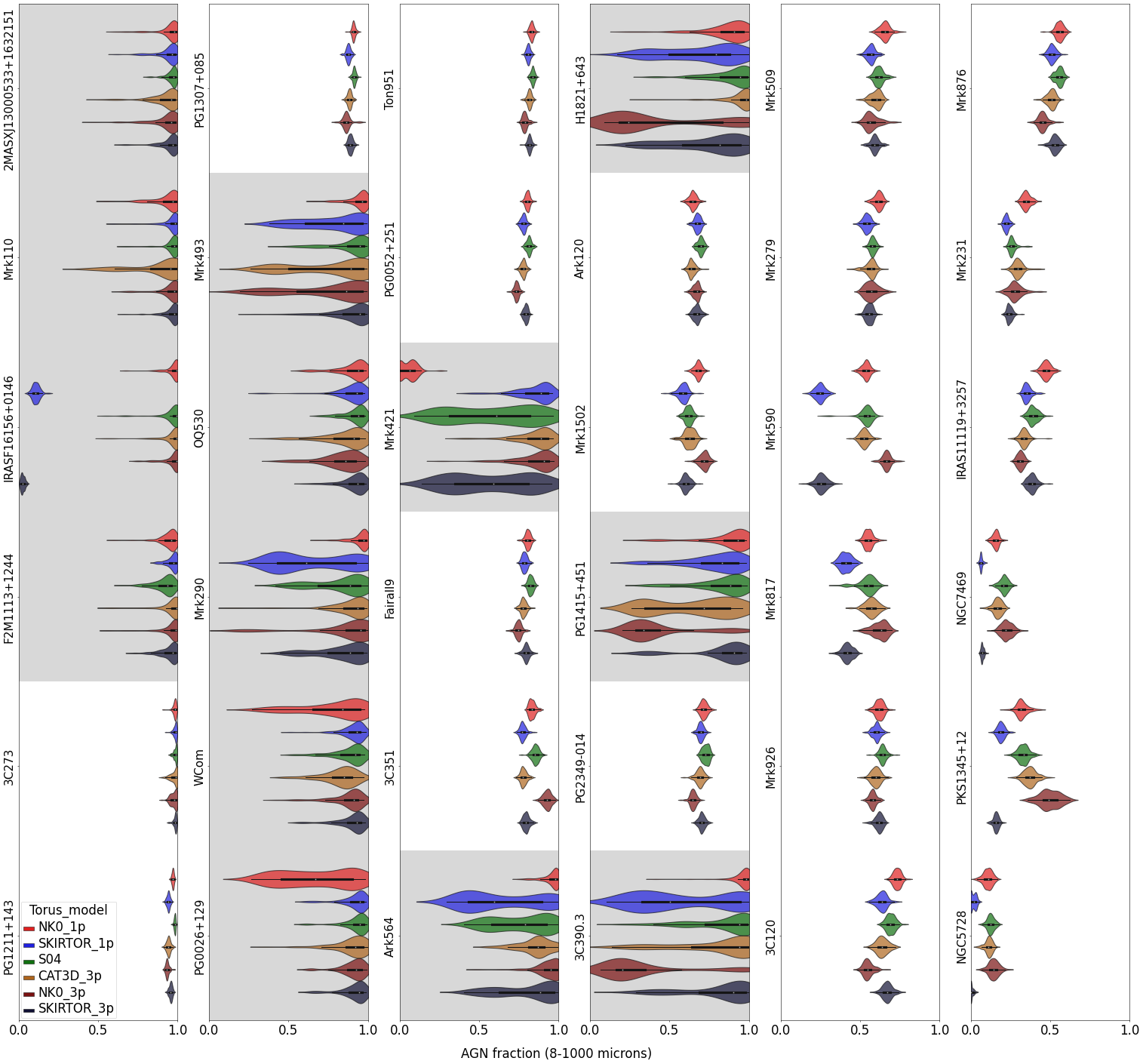}
    \caption{AGN fraction contributing to the total IR luminosity ($8-1000\, \mu$m) for the 36 galaxies of the sample and each torus model tested. The gray area highlights the SEDs with missing data in the FIR. The different models used do not have a large effect on the overall AGN fraction in this range.}
    \label{fig: TOfrac_IR_violin}
\end{figure*}

\begin{figure*}[ht]
    \centering
    \includegraphics[width = 18.5cm]{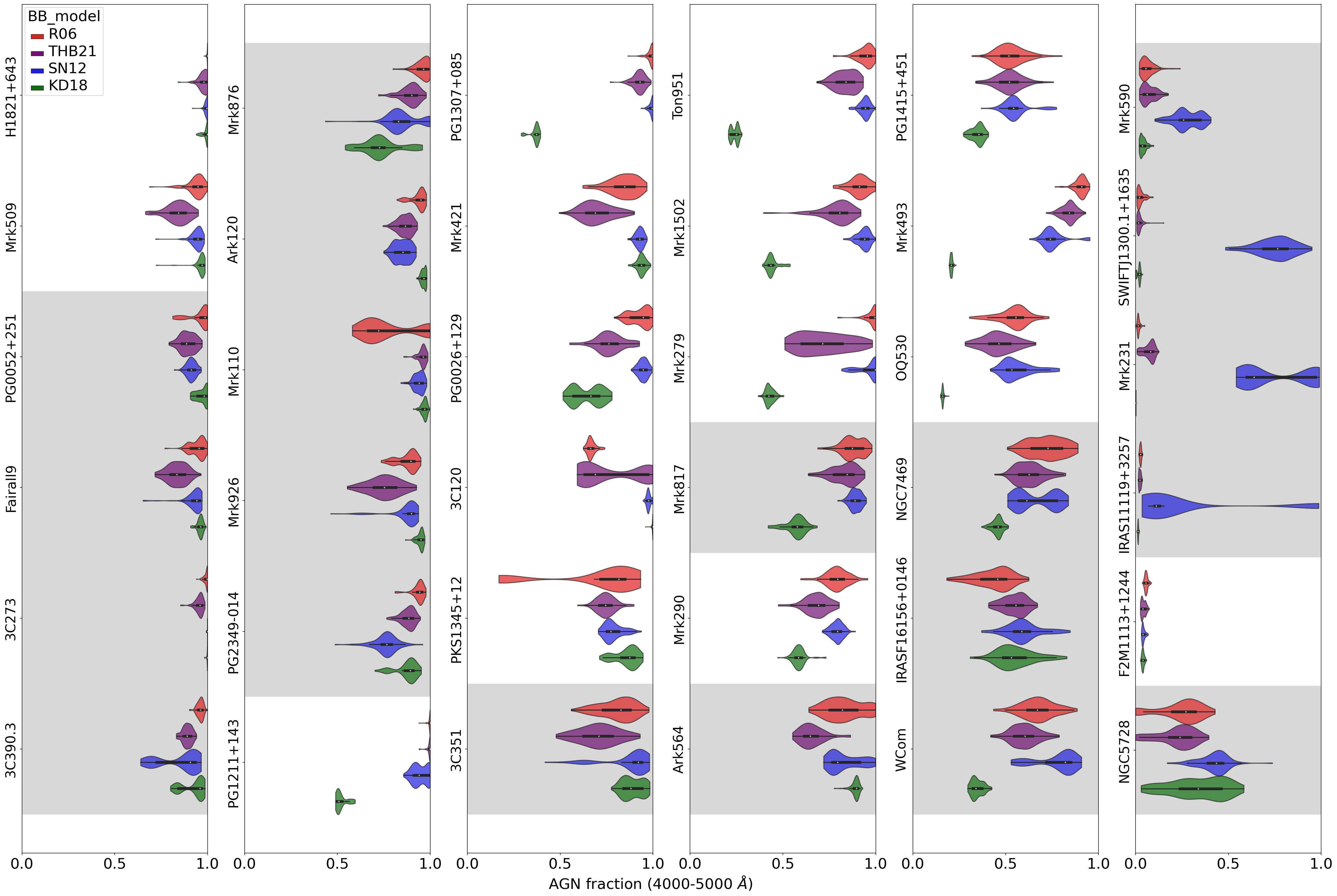}
    \caption{AGN fraction contributing to the total optical luminosity ($400-500$nm) for the 36 galaxies of the sample and each accretion disk model tested. The flexibility in modelling X-rays and the effects of UV degeneracy causes outliers more often with the KD18 accretion disk model.}
    \label{fig: BBfrac_opt_violin2}
\end{figure*}

\end{appendix}

\begin{acknowledgements}
We gratefully acknowledge support from the National Agency for Research and Development (ANID) under the fellowship ANID Becas/Doctorado Nacional, \#21220337 (LNMR-R), Millennium Science Initiative Program - ICN12\_009 (LNM-R, FEB), CATA-BASAL - FB210003 (LNM-R, FEB), and FONDECYT Regular - \#1200495 (LNM-R, FEB); the Vicerrectoría de Investigación of Pontificia Universidad Católica de Chile under the fellowship Stay of Doctoral Co-tutelage Abroad, leading to double degree; the Vicerrectoría de Investigación y Extensión de la Universidad Industrial de Santander under project 2494 (JCBP); and the Ministry of Science, Technological Development and Innovation of the Republic of Serbia, through contract No. 451-03-9/2024-14/200002 (MS).
%Juan C. B. Pineda acknowledges financial support of Vicerrectoría de Investigación y Extensión de la Universidad Industrial de Santander under project 2494. M.S. acknowledges support by the Ministry of Science, Technological Development and Innovation of the Republic of Serbia, through contract No. 451-03-9/2024-14/200002. 
We thank Sebastian H\"oning, Paolo Padovani and Christine Done for useful comments, valuable discussions and suggestions on the implementation of CAT3D-WIND, KD18 and synchrotron emission models.
We thank David W Hogg and Luis A. Núñez for their support during the initial phases of the project.
\end{acknowledgements}

\bibliographystyle{aa}
\bibliography{AGNfitter_lowz_ref.bib}

\end{document}